\documentclass[useAMS,usenatbib,usegraphicx]{mn2e}

\begin{document}

\newcommand{\aj}{AJ}
\newcommand{\apj}{ApJ}
\newcommand{\apjl}{ApJL}
\newcommand{\apjs}{ApJS}
\newcommand{\mnras}{MNRAS}
\newcommand{\nat}{Nature}
\newcommand{\araa}{ARAA}
\newcommand{\aap}{A{\&}A}
\newcommand{\aaps}{A{\&}AS}

\newcommand{\kms}{\,km\,s\ensuremath{^{-1}}}
\newcommand{\vej}{v_{\mathrm{ej}}}
\newcommand{\vlsr}{v_{\mathrm{LSR}}}
\newcommand{\acd}{N_{\mathrm{a}}}
\newcommand{\zabs}{z_{\mathrm{abs}}}
\newcommand{\zem}{z_{\mathrm{em}}}
\newcommand{\fnurest}{f^\mathrm{rest}_\nu}
\newcommand{\fnuobs}{f^\mathrm{obs}_\nu}

\newcommand{\lesssim}{\la}
\newcommand{\gtrsim}{\ga}
\newcommand{\nodata}{\ldots}

\newcommand{\hi}{{H\,{\sc i}}}
\newcommand{\cii}{{C\,{\sc ii}}}
\newcommand{\ciii}{{C\,{\sc iii}}}
\newcommand{\civ}{{C\,{\sc iv}}}
\newcommand{\niii}{{N\,{\sc iii}}}
\newcommand{\niv}{{N\,{\sc iv}}}
\newcommand{\nv}{{N\,{\sc v}}}
\newcommand{\oi}{{O\,{\sc i}}}
\newcommand{\oii}{{O\,{\sc ii}}}
\newcommand{\oiii}{{O\,{\sc iii}}}
\newcommand{\oiv}{{O\,{\sc iv}}}
\newcommand{\ovi}{{O\,{\sc vi}}}
\newcommand{\ov}{{O\,{\sc v}}}
\newcommand{\ovii}{{O\,{\sc vii}}}
\newcommand{\oviii}{{O\,{\sc viii}}}
\newcommand{\neviii}{{Ne\,{\sc viii}}}
\newcommand{\mgii}{{Mg\,{\sc ii}}}
\newcommand{\silii}{{Si\,{\sc ii}}}
\newcommand{\siliv}{{Si\,{\sc iv}}}
\newcommand{\suliv}{{S\,{\sc iv}}}
\newcommand{\feii}{{Fe\,{\sc ii}}}

\title[QSO-Intrinsic Absorbers in HST/STIS Spectra]{A Census of Quasar-Intrinsic Absorption in the Hubble Space Telescope Archive: Systems from High Resolution Echelle Spectra\thanks{Based on observations made with the NASA/ESA Hubble Space Telescope, which is operated by the Association of Universities for Research in Astronomy, Inc., under NASA contract NAS 5-26555.}}

\author[Ganguly et al.]{Rajib Ganguly$^1$,
        Ryan S. Lynch$^2$,
        Jane C. Charlton$^3$,
        Michael Eracleous$^3$,
        \newauthor
        Todd M. Tripp$^4$,
        Christopher Palma$^3$,
        Kenneth R. Sembach$^5$,
        Toru Misawa$^6$,
        \newauthor
        Joseph R. Masiero$^7$,
        Nikola Milutinovic$^8$,
        Benjamin D. Lackey$^9$,
        Therese M. Jones$^{10}$\\
        $^1$Department of Computer Science, Engineering, \& Physics, The University of Michigan-Flint, 213 Murchie Science Building, \\ 303 East Keasley St., Flint, MI, 48503\\
        $^2$Department of Physics, McGill University, 3600 rue University, Montreal, QC Canada H3A 2T8\\
        $^3$Department of Astronomy \& Astrophysics, The Pennsylvania State University, University Park, PA 16802\\
        $^4$Department of Astronomy, University of Massachusetts, Amherst, MA 01003\\
        $^5$Space Telescope Science Institute, 3700 San Martin Drive, Baltimore, MD  21218\\
        $^6$School of General Education, Shinshu University, 3-1-1 Asahi, Matsumoto, Nagano 390-8621, Japan\\
        $^7$Jet Propulsion Laboratory, California Institute of Technology, 4800 Oak Grove Dr., MS 321-520, Pasadena, CA 91109\\
        $^8$Dept. of Physics \& Astronomy, University of Victoria, Elliott Building, 3800 Finnerty Rd, Victoria, BC, V8P 5C2 Canada\\
        $^9$Department of Physics, University of Wisconsin-Milwaukee, Milwaukee, WI, 53201\\
        $^{10}$Department of Astronomy, University of California-Berkeley, Berkeley, CA, 94720
        }

\maketitle
\begin{abstract}
We present a census of $\zabs \lesssim 2$\, intrinsic (those showing partial coverage) and associated ($\zabs \sim \zem$) quasar absorption-line systems detected in the Hubble Space Telescope archive of Space Telescope Imaging Spectrograph echelle spectra. This work complements the \citet{misawa07} survey of $2 < \zem < 4$\ quasars that selects systems using similar techniques. We confirm the existence of so-called ``strong {\nv}'' intrinsic systems (where the equivalent width of {\hi} Ly\,$\alpha$\ is small compared to {\nv} $\lambda$1238) presented in that work, but find no convincing cases of ``strong {\civ}'' intrinsic systems at low redshift/luminosity. Moreover, we also report on the existence of ``strong \ovi'' systems. From a comparison of partial coverage results as a function of ion, we conclude that systems selected by the {\nv} ion have the highest probability of being intrinsic. By contrast, the {\civ} and {\ovi} ions are poor selectors. Of the 30 {\ovi} systems tested, only two of the systems in the spectrum on 3C\,351 show convincing evidence for partial coverage. However, there is a $\sim3\sigma$\ excess in the number of absorbers near the quasar redshift ($|\Delta v| \le 5000$\,\kms) over absorbers at large redshift differences. In at least two cases, the associated {\ovi} systems are known not to arise close to the accretion disk of the quasar.
\end{abstract}


\section{Introduction}

It is well accepted now that quasars are powered by the accretion of matter onto a central supermassive ($10^{7-9}$\,M$_\odot$) black hole. The detailed physics of how angular momentum is carried away allowing accretion and black hole growth to proceed, however, is still unclear. Alongside the photon bubble and magneto-rotational instabilities \citep[e.g.][]{blaes2011,begelman06}, mass outflows may play an important role in this process. Outflows are also at the heart of understanding the overall structure of quasars \citep[e.g.,][]{elvis00,gan01a} and other active galactic nuclei (AGN) and have been invoked to explain many observed properties of quasars \citep[e.g., single-peaked broad emission lines, broad absorption lines;][]{mur95,mur97,psk00,pk04,everett05}.

More recently, outflows from AGN have become an integral component in understanding the effects that the accreting black hole has on the surrounding galaxy and intergalactic medium (i.e., AGN feedback). AGN feedback is invoked to prevent an excess of bright/massive objects, and to reproduce the colors of massive galaxies. Models can explain this in two ways. In high accretion rate (hence, high luminosity) objects, a small fraction of the power output is deposited as thermal energy in the interstellar gas \citep*[e.g.,][]{tdm05,he10}. This added heat source prevents the cooling necessary to further form stars. Furthermore, mechanical energy from outflows can affect both the interstellar and intergalactic material though the increase of entropy and the blow-out of material \citep*[e.g.,][]{so04,soc10}. These effects potentially combine in galaxy evolution scenarios to produce giant, red elliptical galaxies \citep*[e.g.,][]{hopkins07e,tdm08}.

Understanding the physics of outflows, then, is a crucial component in understanding accretion/black hole growth, as well as galaxy evolution. There are many theoretical flavors of mass outflows, with essentially three different mechanisms for driving the gas: gas pressure, radiation pressure, and magnetocentrifugal forces. In the first case, gas is thought to be ablated off the dense, dusty torus. The gas is then ionized beyond the point where radiative cooling is an efficient means of energy transport. This causes a thermal runaway, increasing the gas pressure. The gas expands violently in a multi-temperature, multi-density outflow \citep[e.g.,][]{kk01}. In the second case, ultraviolet light produced in the inner regions of the accretion disk radially drives gas that is not too ionized \citep*[via electron scattering, and absorption and re-emission of lines and edges, e.g.,][]{alb94,mur95,psk00}. Outside the dust sublimation radius, radiative driving of dust may also be important \citep*[e.g.,][]{egk09} Finally, in the presence of magnetic fields, ionized gas can spiral away from the accretion disk along open field lines. \citep[e.g.,][]{kk94,everett05}.

Observations of outflows are primarily carried out in absorption against the central compact UV/X-ray continuum. Detailed studies of absorption lines provide a powerful means of diagnosing the ionization conditions and metal abundances of the outflows. Large surveys provide a statistical means of gauging the frequency with which outflows are observed. This frequency is vitally important in statistically understanding the geometric structure of outflows and its relative importance to galaxy evolution. \citet{gb08} provide a review of the recent literature regarding the frequency of outflows, finding that, largely independent of luminosity, about 60\% of AGN show outflows in absorption, though there are differences if one only considers certain classes of outflows.

This paper is a companion to \citet[][hereafter M07]{misawa07} which presents a survey of narrow absorption-line systems ($v_\mathrm{FWHM} \lesssim 500$\,\kms) intrinsic to a sample of $z \sim 2.5$\ quasars. These are separate from the less frequent, but more dramatic {\it broad} absorption lines (BALs) that often have $v_\mathrm{FWHM} \gtrsim 5000$\,\kms\ depending on the choice of definition \citep[e.g.,][]{weymann91,hallai}. An intermediate class of intrinsic absorption-line systems, called ``mini-BALs,'' is often invoked to refer to the remaining unclassified systems that fall in the $\sim 1$\ order of magnitude gap in the velocity widths spanned by the former two classes. Furthermore, the quasars in the M07 sample were observed originally with Keck I/HIRES for the purpose of deuterium abundance measurements in high redshift damped Ly\,$\alpha$\ or Lyman-limit systems. In the M07 survey, intrinsic absorption-line systems were selected without explicit velocity bias using the partial coverage method.

In this paper, we supplement the M07 survey with a similar survey using high spectral-resolution observations from the Space Telescope Imaging Spectrograph (STIS) onboard the {\it Hubble Space Telescope} (HST). Consequently, we add two parts of parameter space which were not available to the M07 survey: (1) an extension to lower redshift and luminosity; and (2) a survey of associated\footnote{Here, the term ``associated'' is used to indicate absorption systems that appear near the quasar redshift. Following \citet{foltz86}, we adopt the cut-off of 5000\,\kms\ for this selection.} {\ovi} systems over the quasar redshift range $0.116 \leq z \leq 1.9$. By extending our survey to include associated {\ovi} systems, we are equipped to study high ionization absorption associated with the quasars. It is important to do so to bridge the gap in ionization potential between studies using the {\civ} $\lambda\lambda$1548.204,1550.781 doublet, and the X-ray warm absorbers. Our survey extends the redshift range over which associated {\ovi} absorption has already been studied using FUSE observations of AGN \citep[e.g., ][]{kriss02,dunn07}, which cover the redshift range $z \lesssim 0.15$.

To summarize, our goals are to: (1) bridge the luminosity gap between the \citet{dunn07} catalogs and the M07 survey; (2) catalog intrinsic absorbers at low-redshift independent of velocity; (3) refine selection criteria that are used to create large catalogs of intrinsic absorbers using lower-resolution data (e.g., with the Sloan Digital Sky Survey); and (4) characterize the range of kinematic and ionization properties spanned by intrinsic absorbers.

In \S\ref{sec:data}, we outline the data acquisition and reduction used in this study and characterize the general broadband continuum and emission line properties of the quasars. Our methodology for selecting intrinsic systems and our classification scheme is laid out in \S\ref{sec:pcsys}; the presentation of the sample is in the Appendix. We note some basic statistical results in \S\ref{sec:stats}. Finally, we discuss the implications of our results within the framework of the M07 survey and compare the combined results of these two surveys with other surveys in \S\ref{sec:discussion}. Our conclusions are summarized in \S\ref{sec:summary}.

\section{Data}
\label{sec:data}

\begin{table*}
\begin{minipage}{165mm}
\caption{Journal of HST/STIS Observations}
\label{tab:sample}
\begin{tabular}{lccccclcrrl}
\hline\hline
& \multicolumn{2}{c}{E140M}      & \multicolumn{3}{c}{E230M}      & & & & & \\
& \multicolumn{2}{c}{\hrulefill} & \multicolumn{3}{c}{\hrulefill} & & & & & \\
{Target} & {T$_{\mathrm{exp}}$} & {PID} & {T$_{\mathrm{exp}}$} & {PID} & {$\lambda_{\mathrm{c}}^\mathrm{a}$} & {$\zem^\mathrm{b}$} & {Ref.$^\mathrm{c}$} & {R.A.} & {Dec.} & {Mag$^\mathrm{d}$} \\
& {(ks)} & & {(ks)} & & {(\AA)} & & & {(J2000.0)} & {(J2000.0)} & {(band)}\\ \hline
PG 0117+213     & \nodata & \nodata & 42      & 8673    & 2707    & 1.493:              & (1)  & 01:20:17.3 & $+$21:33:46.2 & 16.05   \\
Ton S210        & 22.5    & 9415    &  4.9    & 9415    & 2415    & 0.116$\pm$0.001     & (2)  & 01:21:51.5 & $-$28:20:57.0 & 14.7    \\
HE 0226-4110    & 43.8    & 9184    & \nodata & \nodata & \nodata & 0.495$\pm$0.001     & (2)  & 02:28:15.2 & $-$40:57:15.6 & 15.2    \\
PKS 0232-04     & \nodata & \nodata & 28      & 8673    & 2707    & 1.4398$\pm$0.0001   & (3)  & 02:35:07.3 & $-$04:02:06.0 & 16.46   \\
PKS 0312-77     & 37.9    & 8651    &  6      & 8651    & 2561    & 0.223$\pm$0.001     & (4)  & 03:11:55.4 & $-$76:51:50.8 & 16.1    \\
PKS 0405-123    & 27.2    & 7576    & \nodata & \nodata & \nodata & 0.5726$\pm$0.0002   & (5)  & 04:07:48.4 & $-$12:11:36.0 & 14.82   \\
PKS 0454-22     & \nodata & \nodata & 10.6    & 8672    & 1978    & 0.533$\pm$0.001     & (5)  & 04:56:08.9 & $-$21:59:09.4 & 16.1    \\
HE 0515-4414    & \nodata & \nodata & 31.5    & 8288    & 2707    & 1.71:               & (6)  & 05:17:07.5 & $-$44:10:55.3 & 14.9    \\
HS 0624+6907    & 61.9    & 9184    & \nodata & \nodata & \nodata & 0.37:               & (7)  & 06:30:02.7 & $+$69:05:03.7 & 14.2    \\
HS 0747+4259    & \nodata & \nodata & 54.2    & 9040    & 2561    & 1.9:                & (6)  & 07:50:54.6 & $+$42:52:19.6 & 15.8    \\
HS 0810+2554    & \nodata & \nodata & 51      & 9040    & 2561    & 1.510$\pm$0.002     & (11) & 08:13:31.2 & $+$25:45:03.0 & 16.0(g) \\
PG 0953+415     & 24.5    & 7747    & \nodata & \nodata & \nodata & 0.2341$\pm$0.0004   & (5)  & 09:56:52.4 & $+$41:15:22.0 & 14.5    \\
Mrk 132         & \nodata & \nodata & 68.9    & 9186    & 2707    & 1.757$\pm$0.002     & (11) & 10:01:29.7 & $+$54:54:38.1 & 16.0(g) \\
Ton 28          & 48.4    & 9184    & \nodata & \nodata & \nodata & 0.3297$\pm$0.0004   & (5)  & 10:04:02.6 & $+$28:55:35.2 & 15.5    \\
3C 249.1        & 24.3    & 9184    & \nodata & \nodata & \nodata & 0.31150$\pm$0.00005 & (5)  & 11:04:13.7 & $+$76:58:58.0 & 15.72   \\
PG 1116+215     & 19.9    & 8097    &  5.6    & 8097    & 2415    & 0.1765$\pm$0.0004   & (5)  & 11:19:08.7 & $+$21:19:18.2 & 15.17   \\
                & 19.9    & 8165    & \nodata & \nodata & \nodata &                     &      &            &              &         \\
PKS 1127-145    & \nodata & \nodata & 52.4    & 9173    & 2707    & 1.184:              & (14) & 11:30:07.1 & $-$14:49:27.0 & 16.9    \\
PG 1206+459     & \nodata & \nodata & 17.3    & 8672    & 2707    & 1.163$\pm$0.003     & (11) & 12:08:58.0 & $+$45:40:35.6 & 15.79   \\
PG 1211+143     & 67.4    & 8571    & \nodata & \nodata & \nodata & 0.0809$\pm$0.0005   & (5)  & 12:14:17.7 & $+$14:03:13.0 & 14.63   \\
PG 1216+069     & 23.2    & 9184    & \nodata & \nodata & \nodata & 0.3313$\pm$0.0003   & (5)  & 12:19:21.0 & $+$06:38:38.4 & 15.68   \\
Mark 205        & 78.3    & 8625    & \nodata & \nodata & \nodata & 0.0708$\pm$0.0002   & (8) & 12:21:44.0 & $+$75:18:38.2 & 14.5    \\
3C 273          & 18.7    & 8017    & \nodata & \nodata & \nodata & 0.15834$\pm$0.00007 & (9) & 12:29:06.7 & $+$02:03:08.2 & 12.86   \\
RX J1230.8-0115 & 27.2    & 7737    & \nodata & \nodata & \nodata & 0.117$\pm$0.001     & (2)  & 12:30:50.0 & $+$01:15:21.7 & 14.42   \\
PG 1241+176     & \nodata & \nodata & 19.2    & 8672    & 2707    & 1.273:              & (1)  & 12:44:10.8 & $+$17:21:04.0 & 15.38   \\
PG 1248+401     & \nodata & \nodata & 25.2    & 8672    & 2707    & 1.03:               & (1)  & 12:50:48.4 & $+$39:51:40.0 & 16.06   \\
PG 1259+593     & 95.8    & 8695    & \nodata & \nodata & \nodata & 0.4778$\pm$0.0004   & (5)  & 13:01:12.9 & $+$59:02:06.8 & 15.6(g) \\
PKS 1302-102    & 22.1    & 8306    & \nodata & \nodata & \nodata & 0.2784$\pm$0.0004   & (5)  & 13:05:33.0 & $-$10:33:20.5 & 14.92   \\
CSO 873         & \nodata & \nodata & 13.6    & 8672    & 2707    & 1.022:              & (10) & 13:19:56.3 & $+$27:28:08.5 & 15.98   \\
PG 1444+407     & 48.6    & 9184    & \nodata & \nodata & \nodata & 0.2673$\pm$0.0004   & (5)  & 14:46:45.9 & $+$40:35:06.4 & 15.95   \\
PG 1630+377     & \nodata & \nodata & 34.1    & 8673    & 2707    & 1.476$\pm$0.002     & (11) & 16:32:01.1 & $+$37:37:49.4 & 16.1(g) \\
PG 1634+706     & \nodata & \nodata & 75.3    & 8312    & 2707    & 1.334:              & (1)  & 16:34:28.9 & $+$70:31:33.0 & 14.9    \\
                & \nodata & \nodata & \nodata & 7292    & 2269    &                     &      &            &              &         \\
3C 351.0        & 74.5    & 8015    & \nodata & \nodata & \nodata & 0.3719$\pm$0.0001   & (11) & 17:04:41.6 & $+$60:44:28.5 & 15.5(g) \\
PG 1718+481     & \nodata & \nodata & 14.1    & 7292    & 2269    & 1.084:              & (1)  & 17:19:38.2 & $+$48:04:12.3 & 15.33   \\
H 1821+643      & 50.9    & 8165    & \nodata & \nodata & \nodata & 0.2970$\pm$0.0003   & (12) & 18:21:57.1 & $+$64:20:36.7 & 14.1    \\
PHL 1811        & 33.9    & 9418    & \nodata & \nodata & \nodata & 0.192:              & (15) & 21:55:01.5 & $-$09:22:25.0 & 14.13   \\
PKS 2155-304    & 28.4    & 8125    & \nodata & \nodata & \nodata & 0.116$\pm$0.002     & (13) & 21:58:52.0 & $-$30:13:32.0 & 14      \\ \hline
\end{tabular}

$^\mathrm{a}$We list the central wavelength of the spectrum since there are several tilts available for observations with the E230M echelle.
The wavelength coverage corresponding to each central wavelength is as follows \citep{stis}:
{\bf 1978}: 1574--2382\,\AA, {\bf 2269}: 1865--2673\,\AA, {\bf 2415}: 2011--2819\,\AA, {\bf 2561}: 2157--2965\,\AA, {\bf 2707}: 2303--3111\,\AA.\\
$^\mathrm{b}$We mark redshifts without reported uncertainties with a colon (:).\\
$^\mathrm{c}$References: (1) \citet{sg83};
(2) \citet{wis00};
(3) \citet{6dfdr2};
(4) \citet{jauncey78};
(5) \citet{mar96};
(6) \citet{rei98};
(7) \citet{bowen94};
(8) \citet{huchra90};
(9) \citet{strauss92};
(10) \citet{hb89};
(11) \citet{sdssqso5};
(12) \citet{degrijp92};
(13) \citet{fpt93};
(14) \citet{wilkes86};
(15) \citet{fbqs3}. \\
$^\mathrm{d}$Unless otherwise noted, the quoted magnitudes are Johnson V.
\end{minipage}
\end{table*}

\subsection{HST/STIS Spectra}

In our search for absorption-line systems (and subsequent assessment of a possible intrinsic origin), we select the medium resolution STIS echelle modes for two reasons. First, the spectral resolution is sufficiently high such that a partial coverage analysis can be meaningfully applied to profiles as a test of an intrinsic origin. Second, the wavelength coverage is higher than the first-order grating modes, which yields an interesting redshift/velocity path for a given transition.

In Table~\ref{tab:sample}, we list all quasars and active galaxies that have been observed with either the E140M or E230M echelle along with their exposure times (column 2 and 4, respectively) and the proposal identification numbers (columns 3 and 5, respectively). The spectra were obtained using either the $0.\!''2 \times 0.\!''2$\ or the $0.\!''2 \times 0.\!''06$\ slit and the MAMA detectors. According to \citet{stis}, these modes provide spectral resolutions of 6.5\,\kms\ and 10\,\kms\ for the E140M and E230M echelle gratings, respectively, with a sampling 2-3 pixels per resolution element. While there are occasionally somewhat larger errors in the wavelength calibration \citep[e.g.,][]{tripp05}, the relative wavelength calibration is usually excellent (0.25 - 0.5 pixels), while the absolute calibration is good to about 0.5-1.0 pixels. (These correspond to $\sim 0.01$\,\AA, and $\sim0.03$\,\AA\ uncertainties in the E140M and E230M spectra, respectively.) Similarly, the relative flux calibration is good to about 5\%, while the absolute calibration is good to about 8\%. Since we are looking for the signature of partial coverage, background subtraction can be important. For the FUV-MAMAs there is a glow that been growing (currently at the level of $1.6 \times 10^{-5}$\,counts\,s$^{-1}$), and hence can be problematic for the on-the-fly-calibration (OTFC).  However, our objects have source counts rates an order of magnitude larger, so the OTFC are sufficient for our purposes. E140M spectra, covering 1123-1710\,\AA, are available for 21 quasars. E230M spectra are available for 18 quasars. The wavelength coverage for the E230M spectra depends on the grating tilt, which we list in Table~\ref{tab:sample} (column 6), as well as the corresponding wavelength coverage in the footnotes of the table.

\subsection{Redshifts}

It is important to obtain accurate systemic redshifts for the quasars in order to understand the location(s) of the absorbers, especially the associated absorbers. Ideally, one would like to have precise information from narrow, forbidden emission lines as these are thought to originate in the host galaxy, far from the quasar central engine. In Table~\ref{tab:sample}, we list the redshift of each quasar and its statistical uncertainty (column 7) and the reference for that redshift measurement (column 8).

Quasars with $\zem \lesssim 0.9$\ will have narrow emission lines like [\oii] $\lambda$3727 or [\oiii] $\lambda$5007 in the optical bandpass. In these cases, the quoted redshift is likely to be close to the systemic redshift. For higher redshift quasars, the redshift measurement comes from broad ultraviolet emission lines (e.g., {\mgii}$\lambda$2800, {\civ}$\lambda$1549) that are shifted into the optical bandpass. In these cases, the quoted redshift can be blueshifted (or redshifted) relative to the systemic velocity \citep[e.g.,][]{gaskell83,wilkes86,tf92,esp93,bro94,mar96,rich02b}. Hereafter, velocity refers to the putative ejection velocity relative to the quasar redshift from Table~\ref{tab:sample}. A positive velocity indicates a blueshift:
\begin{equation}
\beta = {v \over c} = {{(1+\zem)^2 - (1+\zabs)^2} \over {(1+\zem)^2
+ (1+\zabs)^2}}. \label{eq:beta}
\end{equation}
From a composite spectrum of Sloan Digital Sky Survey quasars, \citet{vdb01} find that the average blueshift from the [\oiii]$\lambda$5007 line is $563\pm27$\,\kms\ for {\civ}$\lambda$1549 and $-160\pm10$\,\kms\ for {\mgii}$\lambda$2800. Typically, the quasar-to-quasar variation in these velocity offsets is under 200\,\kms \citep{tf92}, though there are some systematic differences between radio-loud and radio-quiet quasars \citep[e.g.,][]{tf92,mar96}.

\begin{figure}
\includegraphics[width=0.95\columnwidth,angle=-90]{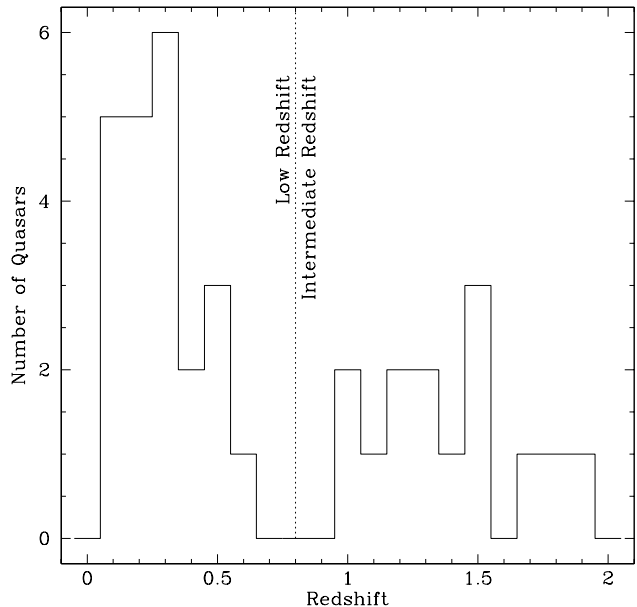}
\protect\caption[redshift distribution]{Redshift distribution of quasars for the HST sample. A vertical dotted line marks our division between low-redshift and intermediate-redshift subsamples.}
\label{fig:zdist}
\end{figure}

Overall, the redshifts of the quasars in this sample range from very low (Mrk 205, $\zem=0.0708$), to moderately high (HS\,0747+4259, $\zem=1.9$). Figure~\ref{fig:zdist} shows the redshift distribution of the quasars. There is a dearth of quasars in the range $0.65 < \zem < 0.95$. Thus, we form two subsamples of quasars, a low redshift sample with $\zem < 0.7$, and an intermediate redshift sample with $0.9 < \zem < 2$. [We refer to this higher-redshift subsample as intermediate redshift and reserve the term high-redshift for the $\zem \gtrsim 2.5$\ quasars in the companion survey (M07).] Our low-redshift subsample contains 22 quasars (median $\zem=0.2729$) and covers about half of cosmological history. Our intermediate-redshift subsample contains 14 quasars (median $\zem=1.3849$), and covers about a fifth of cosmological history.

\subsection{Radio Properties}
\label{sec:radio}

Historically, quasar radio properties have been observationally linked to the presence and form of intrinsic/associated absorption. The radio-loudness of a quasar is parameterized by the ratio, $R^*$, between the rest-frame 5\,GHz flux density and the 2500\,\AA\ (or sometimes 3000\,\AA) flux-density \citep[$R^* = F_\nu(5\,\mathrm{GHz})/F_\nu(2500\,\mathrm{\AA})$, ][]{sw80}. Quasars with $R^* > 10$\ are usually taken to be radio-loud \citep[e.g.][]{kel89}. Older surveys of the most extreme forms of intrinsic absorption found that broad absorption lines only appeared in radio-quiet quasars \citep{weymann85,turn88,turn88b}. An interesting complement to this was the observation that strong ($W_\lambda \gtrsim 1-1.5$\,\AA), associated {\civ} absorption existed preferentially in optically-faint, steep-spectrum radio-loud quasars \citep{wwpt,foltz86,and87,mj87,sbs88,foltz88}, and that the absorption strength was correlated with the observer's viewing angle \citep[e.g,][]{btv97}.

In the past two decades, there have been several observations that have complicated these two simple trends. The {\it Faint Images of the Radio Sky At Twenty-cm} Bright Quasar Survey \citep[FBQS,][]{gregg96,first,second} has shown clearly that BALs do exist in ``radio--loud'' quasars \citep{broth97,bro98,wbl99,becker00,gregg00,gbd06} particularly in the range $3 < R^* < 100$\ where a dearth of quasars was thought to exist \citep[i.e., the apparent radio dichotomy, e.g.,][]{mpm90,stocke92}. Even the simple orientation scheme of a predominantly equatorial wind that produced BALs in quasars viewed at large inclination angles \citep[e.g.,][]{good95,goodrich97} has come under fire with the discovery of BAL quasars that may be viewed nearly face-on \citep[e.g.,][]{gp07,brotherton06}.

For narrow absorption lines, advancements in our ability to separate intrinsic from intervening absorbers and the increased sensitivity of radio surveys have made the connection less clear. \citet{gan01a} and \citet{ves03} conducted surveys for associated high-ionization absorption at two different redshifts and found no overt preference for radio-properties contrary to the previous surveys. \citet{gan01a} showed that, in low-redshift quasars, associated absorption appears (with a frequency that depends on the quasar properties) in quasars with virtually all properties. In addition, several works in the past two decades have shown the clear presence of intrinsic absorption at very high blueshifts from the emission redshift \citep[e.g.,][]{kp96,ham97c,gan01b,rich02,ham11,rhh11}. \citet{rich99,rich01a} and \citet{rich01b} showed that the frequency of absorption as a function of velocity changes with radio property. Radio-loud quasars show a greater frequency of absorbers than radio-quiet quasars, and flat-spectrum radio-loud quasars show a higher frequency than steep-spectrum radio-loud quasars. Hence, there still may be a connection between intrinsic narrow absorption lines and radio properties.

Even though the relationship between quasar radio properties and the intrinsic absorption is, at best, unclear, we report the radio-loudness parameters of the quasars. Following \citet{sw80}, we derive the radio flux density at rest-frame 5\,GHz, $\fnurest(5\,\mathrm{GHz})$, from an observed flux density, $\fnuobs(\nu)$, at observed frequency $\nu$:
\begin{eqnarray}
\log \fnurest(5\,\mathrm{GHz}) & = & \log \fnuobs(\nu) +
\alpha_\mathrm{r} \log (5\,\mathrm{GHz}/\nu) \nonumber \\
 & & - (1 + \alpha_\mathrm{r}) \log (1+\zem), \label{eq:radioflux}
\end{eqnarray}
where $\alpha_\mathrm{r}$\ is the spectral index ($f_\nu \sim \nu^\alpha$). We adopt a value of $\alpha_\mathrm{r} = -0.5$. To derive the $\fnurest(2500\,\mathrm{\AA})$\ flux density, we use the reported Sloan magnitude in the bandpass that covers $\lambda_\mathrm{obs}=2500(1+\zem)$\,\AA. [Note: Per our convention, $\fnurest(2500\,\mathrm{\AA}) = \fnuobs(\lambda_\mathrm{obs})$.] The Sloan magnitude system is an AB system with magnitude zero points calibrated to 3631\,Jy \citep{ugriz1}. [There are zero-point shifts in the u and z bands of -0.04 mag and +0.02 mag, respectively, and these have been taken into account.] While Sloan reports magnitudes based on an asinh system \citep{ugriz2}, our objects are bright enough that the difference between this and the standard Pogson magnitude is negligible. Hence, we convert a magnitude, $m$, to a flux density, $\fnuobs$\ via:
\begin{equation}
m = -2.5 \log \fnuobs(\lambda_\mathrm{c})[Jy] + 8.9,
\end{equation}
where $\lambda_\mathrm{c}$\ is the central wavelength of the bandpass. We extrapolate the flux density at wavelength $\lambda_\mathrm{c}$\ to the flux density at $\lambda_\mathrm{obs}$\
assuming a power-law with a spectral index of $\alpha_\mathrm{o}=-0.44$ \citep[$f_\nu \sim \nu^\alpha$][]{vdb01}:
\begin{equation}
\fnuobs(\lambda_\mathrm{obs}) = \fnuobs(\lambda_\mathrm{c}) \left (
{{\lambda_\mathrm{c}} \over {\lambda_\mathrm{obs}}} \right )^\alpha.
\end{equation}
When the desired wavelength is not covered by the Sloan bandpasses (i.e., if the object's redshift is too low), we use the flux implied by the u magnitude. For objects that are not covered by Sloan, we use the Johnson $V$\ magnitude and transform this to an observed-frame $f_\nu(5500\,\mathrm{\AA})$\ flux \citep{sg83,os70} using the equation:
\begin{equation}
V = -2.5 \log f_\nu(5500\,\mathrm{\AA}) - 48.60,
\end{equation}
and extrapolate to the appropriate wavelength assuming the same power-law as above.

Table~\ref{tab:radio} lists the observed radio flux densities of our quasars from the literature and our estimate of the radio-loudness. Twelve of the quasars are radio-loud, while 22 are radio-quiet. We do not have information for one of the quasars (HE\,0515-4414), and we do not have sufficient constraints on HE\,0226-4110 to make a definitive classification. In addition to our redshift bins as defined above, we also define radio-loud and radio-quiet subsamples.
We exclude HE\,0515-4414 and HE\,0226-4110 from these two subsamples (but not from the total sample or the redshift-based subsamples) due to the lack of information on their radio properties.

\begin{table}
\caption{Radio Properties of Quasars}
\label{tab:radio}
\begin{tabular}{lrcrcc}
\hline\hline
{Target} & {$\fnuobs$} & {$\nu$} & {$\log R^*$} & {Class.} & Ref.$^\mathrm{a}$ \\
         & {(mJy)}     & {(GHz)} & & &\\ \hline
PG 0117+213     &  $<2$    & 1.4     & $<0.08$  & RQ      & (1) \\
Ton S210        &  $<2$    & 1.4     & $<0.00$  & RQ      & (1) \\
HE 0226-4110    &  $<47.3$ & 4.85    & $<1.78$  & \nodata & (2) \\
PKS 0232-04     &  1495.3  & 1.4     &   3.12   & RL      & (1) \\
PKS 0312-77     &   620    & 5       &   3.31   & RL      & (3) \\
PKS 0405-123    &  2940.2  & 1.4     &   3.14   & RL      & (1) \\
PKS 0454-22     &  1900    & 5       &   3.75   & RL      & (3) \\
HE 0515-4414    &  \nodata & \nodata & \nodata  & \nodata & \nodata \\
HS 0624+6907    &  $<2$    & 1.4     & $<-0.24$ & RQ      & (1) \\
HS 0747+4259    &  $<2$    & 1.4     &  $<0.03$ & RQ      & (1) \\
HS 0810+2554    &  $<2$    & 1.4     & $<-0.06$ & RQ      & (1) \\
PG 0953+415     &  $<2$    & 1.4     & $<-0.10$ & RQ      & (1) \\
Mrk 132         &  $<2$    & 1.4     & $<-0.01$ & RQ      & (1) \\
Ton 28          &  $<2$    & 1.4     &  $<0.28$ & RQ      & (1) \\
3C 249.1        &  2339.4  & 1.4     &    3.44  & RL      & (1) \\
PG 1116+215     &     6.1  & 1.4     &    0.66  & RQ      & (1) \\
PKS 1127-145    &  5622.0  & 1.4     &    3.90  & RL      & (1) \\
PG 1206+459     &  $<2$    & 1.4     & $<-0.14$ & RQ      & (1) \\
PG 1211+143     &  $<2$    & 1.4     & $<-0.02$ & RQ      & (1) \\
PG 1216+069     &  $<2$    & 1.4     &  $<0.36$ & RQ      & (1) \\
Mrk 205         &  $<2$    & 1.4     & $<-0.07$ & RQ      & (1) \\
3C273           & 54991.2  & 1.4     &    3.70  & RL      & (1) \\
RX J1230.8-0115 &  $<2$    & 1.4     & $<-0.07$ & RQ      & (1) \\
PG 1241+176     &   378.8  & 1.4     &    2.11  & RL      & (1) \\
PG 1248+401     &  $<2$    & 1.4     &  $<0.24$ & RQ      & (1) \\
PG 1259+593     &  $<2$    & 1.4     &  $<0.06$ & RQ      & (1) \\
PKS 1302-102    &   711.3  & 1.4     &    2.61  & RL      & (1) \\
CSO 873         &     3.8  & 1.4     &    0.50  & RQ      & (1) \\
PG 1444+407     &  $<2$    & 1.4     &  $<0.23$ & RQ      & (1) \\
PG 1630+377     &  $<2$    & 1.4     &  $<0.07$ & RQ      & (1) \\
PG 1634+706     &  $<2$    & 1.4     & $<-0.37$ & RQ      & (1) \\
3C351.0         &  3074.7  & 1.4     &    3.27  & RL      & (1) \\
PG 1718+481     &    63.3  & 1.4     &    1.45  & RL      & (1) \\
H 1821+643      &  $<2$    & 1.4     & $<-0.27$ & RQ      & (1) \\
PHL 1811        &     2.1  & 1.4     &   -0.22  & RQ      & (1) \\
PKS 2155-304    &   489.3  & 1.4     &    2.11  & RL      & (1) \\ \hline
\end{tabular}
$^\mathrm{a}$References: (1) \citet{nvss}; (2) \citet{gw93}; (3) \citet{pkscat}
\end{table}

\section{Method for Surveying Absorption Systems}
\label{sec:pcsys}

To produce a catalog of metal-line systems that may be related to the background quasar, we take the following approach. First we identify extragalactic metal absorption-line systems in the sight line (with criteria as outlined below) on the following resonant ultraviolet doublets: \ovi\,$\lambda\lambda$1031.926, 1037.617, {\nv}\,$\lambda\lambda$1238.821, 1242.804, \siliv\,$\lambda\lambda$1393.760, 1402.773, and {\civ}\,$\lambda\lambda$1548.204, 1550.781. [We adopt the wavelengths and oscillator strengths listed in \citet{morton03}.] We seek two types of systems for inclusion in this catalog: (1) systems that lie within 5000\,\kms\ of the quasar redshift (``associated'' systems); and (2) systems that exhibit the signature of partial coverage. In the following subsections, we outline two aspects of this approach -- finding/identifying absorption-line systems, and testing whether they exhibit partial coverage.

\subsection{Finding and Identifying Absorption-Line Systems}
Our criteria for identifying absorption-line systems are as follows:
\begin{itemize}
\item[(1)] the stronger (i.e., higher $f\lambda$) member of a resonant ultraviolet doublet is detected at $\geq 5\sigma$\ confidence using the unresolved feature detection algorithm of \citet{kpii} and \citet{crcv99};
\item[(2)] the weaker member of the same doublet is detected at $\geq 3\sigma$\ confidence.
\end{itemize}
In addition, the system must satisfy at least one of the following criteria:
\begin{itemize}
\item[(3)] The doublet ratio (defined as the ratio of equivalent widths of the stronger to the weaker transition) is in the range 1--2 with matching kinematic profiles; or
\item[(4)] other transitions (e.g., {\hi} Lyman series, {\ciii} $\lambda$977.020) corroborate the existence of the system.
\end{itemize}
We note that this last criterion is especially useful for the identification of line-locked pairs of systems. Briefly, two systems are apparently line-locked if their velocity separation matches that of a pair of lines. Examples of line pairs are the UV resonant doublets, {\ovi} $\lambda1031.926$ - {\hi} Ly$\beta$, and {\nv} $\lambda1238.821$ - {\hi} Ly$\alpha$ \citep[e.g.,][]{gan03b,korista93}. When line-locking is physical, it is typically attributed to situations where both systems are radiatively driven, as in a quasar outflow \citep[e.g.,][]{milne26,scargle73,bm89}.

Furthermore, we also note that as a result of these criteria, we may not identify all absorption-line systems in the sight-lines. However, a small amount of incompleteness should not adversely affect our results. In particular, our use of the unresolved feature detection method biases our search against systems that are shallow. However, this bias is reasonable considering our goals of using partial coverage as a test. The test is not robust for shallow systems. Moreover, since we are not using integrated equivalent widths, what we consider an ``$n\sigma$'' detection will differ from other works.

The following additional points are worth noting regarding our use of the partial coverage technique to separate intrinsic systems from others. Our ultimate goal is to understand the role of outflows in the structure of quasars, the accretion process, and in feedback scenarios. We restrict the term ``intrinsic'' as meaning having an origin in the central engine. \citet{ck12} have pointed out that, in low redshift (and hence, low luminosity) AGN, the mass outflow rates tend to be very large compared to the accretion rates, indicating that the outflows are driven outside the inner accretion disk, potentially at large distances from continuum source. Hence, it would be unlikely for those outflows to exhibit partial coverage. Furthermore, some investigations of $\zabs \approx \zem$\ absorbers in higher bolometric luminosity objects indicate distances at the several kiloparsec scale \citep{bor12,dunn10,moe09}. The large distances also imply large kinetic luminosities, pointing toward the importance of outflows to feedback processes. We point out, however, that these are conclusions based on excited states of species that have ionization potential $\lesssim 60$\,eV: \feii -- 16.2\,eV, \silii -- 16.3\,eV, \cii -- 24.4\,eV, \suliv -- 47.3\,eV, \ciii -- 47.9\,eV. This is a lower ionization potential than the \civ\ (64.5\,eV), \nv\ (97.9\,eV), and \ovi\ (138\,eV) species that are commonly used to search for outflows. Moreover, the incidence of these excited-state species is decidedly lower than that of the more common high-ionization species (i.e., they are not observed \emph{in every outflow}), suggesting that the densities required for observing excited states may be a special case. Thus, we are uncertain how these should apply to the more general population of high-ionization outflows.

In addition, we note that some studies of associated absorbers have found implied distances that are so large ($\gg 100$\,kpc) that the main conclusion is that it is most likely that the absorption has no connection to the central engine at all \citep[e.g.,][]{mor86,tls96}. Indeed, this provides a strong motivation for our decision to use an alternative criterion to identify intrinsic absorbers: the condition that $\zabs \approx \zem$\ provides no assurance that the absorbing gas is anywhere close to the black hole. \citet{gb08} provide a review of this issue and secondary means to distinguish intrinsic absorbers from those far from the central engine. In particular, the two most successful means have been time variability and partial coverage. In our sample, time variability is by-and-large not of use. It is worth noting that a few objects with associated absorbers \citep[e.g., 3C\,351 -- ][]{yuan02} do have multiple epochs of observations, and variability is usually not evident. However, the time baseline of the observations may not be very useful/interesting, and certainly these are the exceptions, not the rule.

\begin{figure}
\includegraphics[width=0.93\columnwidth,angle=-90]{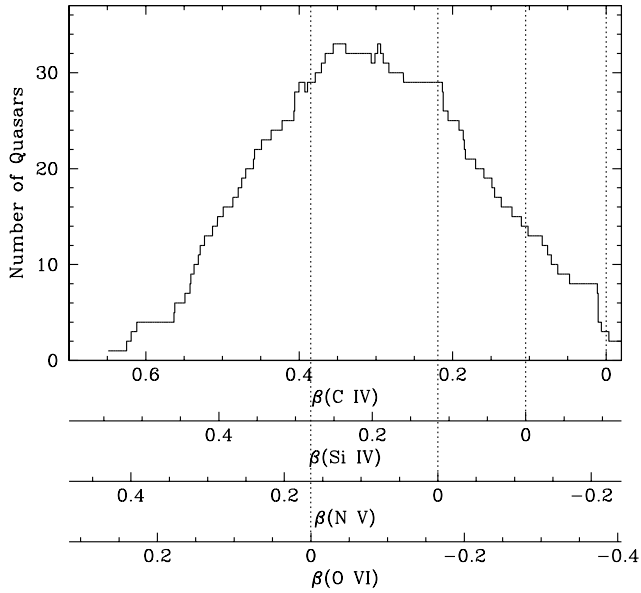}
\protect\caption[ion coverage plot]{Plot of the number of quasars where we have coverage of a particular ion as a function of velocity. Ejection velocity (defined as positive for blueshifts
relative to the quasar redshift) increases to the left; wavelength increases to the right. A normalized version of this curve is used in \S\ref{sec:vejdist} and Fig.~\ref{fig:vejdist} as a sensitivity curve. No equivalent width limit has been applied.}
\label{fig:ioncoverage}
\end{figure}

\begin{figure}
\includegraphics[width=0.95\columnwidth]{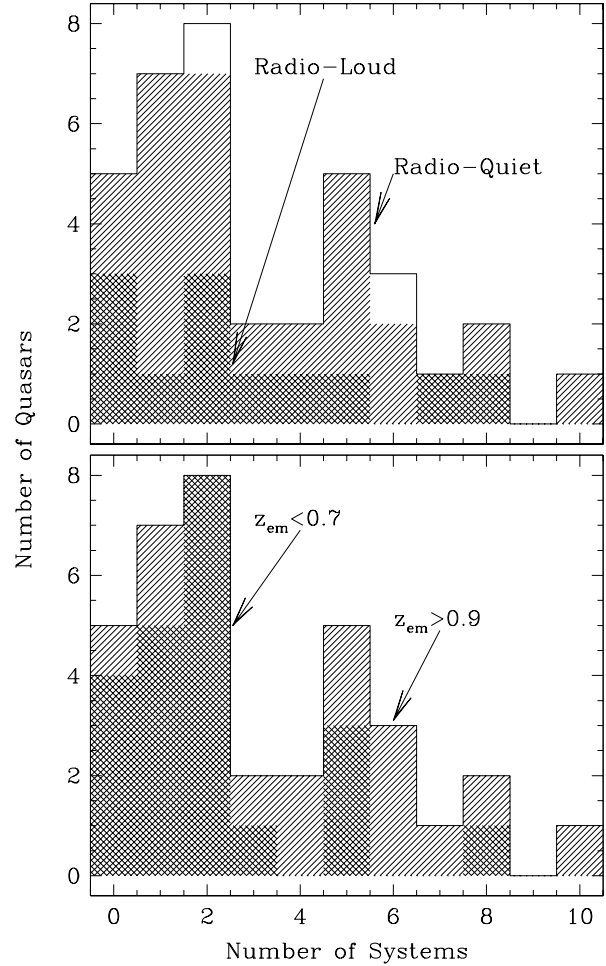}
\protect\caption[system/quasar demographics]{We plot the number of
sight-lines (quasars) as a function of the number of absorption-line
systems observed in those sight-lines. In the top panel,
single-hatched regions indicate sight-lines toward radio-quiet
quasars; double-hatched regions indicate radio-loud quasars. Two
quasars have no radio classification (see \S\ref{sec:radio}. In the
bottom panel, single-hatched regions indicate sight-lines toward
intermediate-redshift quasars; double-hatched regions indicate
low-redshift quasars.}
\label{fig:los}
\end{figure}

Due to the limited wavelength coverage and the different redshifts of the quasars, not all ions will be covered at a given ejection velocity. Figure~\ref{fig:ioncoverage} shows, for a given ion, the number of quasars in which we could detect that ion as a function of velocity. The curve does not take into account variations in the signal-to-noise across a given spectrum, only the wavelength coverage. The overall shape of all the curves is the same for all ions, but shifted due to the different rest-wavelength of the various transitions (e.g., 1548.204\,\AA\ for {\civ}, or 1031.927\,\AA\ for \ovi). This figure clearly shows the regions of parameter-space where our search is useful - associated {\ovi} systems, {\nv} systems at intermediate ejection velocities, and {\civ} systems at high ejection velocities. To illustrate: Of the 36 quasars for which we have STIS echelle data, we could detect the {\civ} doublet at $\vej \sim 0.4c$\ in 26 of them. Similarly, we could detect {\siliv} at $\vej \sim 0.3c$\ in those same quasars (and, indeed, at the same observed wavelength - hence the curve does not need to be recomputed). However, we would only be able to detect the {\siliv} doublet at $\vej \sim 0.4c$\ in 17 of the 36 quasars. Another example is that we only cover $\vej \sim 0$\ in a few of our quasars if we only look at {\civ}. However, in {\ovi}, we are able to detect systems at $\vej \sim 0$\ in $\sim30$\ of the quasars. In passing, we also note that, our stringent criteria are designed to take full advantage of the high spectral resolution in providing  unambiguous identifications for the absorption lines (e.g. high-velocity {\civ} versus low-velocity {\siliv} which may occur at similar observed wavelengths). Figure~\ref{fig:los} further shows a break-down in the number of systems in the spectra of the radio-quiet, radio-loud, low-redshift and intermediate-redshift subsamples.

\subsection{Testing for Partial Coverage}
For the resonant ultraviolet doublets involved in our search, the ratio of $f\lambda$\ values between the two members of the doublet is 2 within measurement uncertainties \citep{morton03}. With the simplifying assumption that both members occult the fraction, $C(v)$, of the background source, we write the following equations for the continuum-normalized flux, $I$, in both transitions without loss of generality:
\begin{eqnarray}
I_1(v) = & \phi(\Delta v) \otimes [(1-C(v)) + C(v) e^{-2 \tau_2(v)}]
\\
I_2(v) = & \phi(\Delta v) \otimes [(1-C(v)) + C(v) e^{-\tau_2(v)}],
\nonumber
\end{eqnarray}
where $\phi(\Delta v)$\ is the instrumental profile, $\tau$\ is the true optical depth and the subscripts 1 and 2 denote the stronger (i.e., higher $f\lambda$\ value) and weaker members of the doublet, respectively. We note here that the true optical depth depends solely on column density per unit velocity of the absorbing species, and the strength of the transition, $f\lambda$, and is independent of saturation. This is not to be confused with more practical definitions of optical depth, such as $-\ln I$, that rely on the observed flux, as these can be affected by saturation. The simplification that each member of the doublet occults the same faction of the background source is affected by the effective size/intensity of the background source (the continuum, and broad emission-line regions), which can be different even for transitions of the same ion that are closely separated in wavelength \citep[see, for example, ][]{gan99,arav99b,ngc3783ii}.

When the instrumental profile is sufficiently narrower than the kinematic components making up a profile, this system of equations can be solved uniquely for the coverage fraction as a function of velocity:
\begin{equation}
C(v) = {{[1-I_{\mathrm{2}}(v)]^2} \over {I_{\mathrm{1}}(v) - 2
I_{\mathrm{2}}(v) + 1}}, \label{eq:covfrac}
\end{equation}
The coverage fraction is unity when $I_1 = I_2^2$, which is true for intervening systems independent of saturation (again, for resolved profiles). Thus, using the UV resonant doublets, we compute the coverage fraction as a test of an intrinsic origin. [Note that when a profile is strongly saturated, setting $I_1 = I_2$\ reduces the equation to $C(v) = 1 - I(v)$. That is, in saturated profiles that are resolved, it is easier to distinguish absorption that does not fully occult the background quasar.] Because noise can affect the calculation of the coverage fraction, we use eq.~\ref{eq:covfrac} only for velocity bins where $I_2^2 < I_1 < I_2$. Outside of this range we use $C(v) = 1-I_2(v)$\ (for $I_1 > I_2$) or $C(v) = 1$\ (for $I_2^2 > I_1$). For the velocity bins, we generally use individual pixels as driven by the data quality. In a few cases (as noted individually in the appendix), we use a velocity bin that is twice the size of a pixel. This results in more statistically significant bins, but fewer bins to adjudicate partial covering. Since there are 2-3 pixels per resolution elements in the spectra, our rebinning does not affect the velocity resolution.

One important aspect in using partial covering as a test of an intrinsic origin is estimating the unabsorbed flux (i.e, the effective continuum). We follow the procedures from \citet{ss92}, fitting a low-order ($n<5$) polynomial to regions adjacent to the absorption profiles that are {\it clearly} unabsorbed based on visual inspection. Furthermore, we incorporate the uncertainties from this continuum placement, as well as the statistical uncertainties in the flux, into our error calculation of the coverage fraction. Hence, we feel that our adjudication below of whether an absorption-line system exhibits partial covering is both conservative and robust.

In applying this test without bias in the velocity of the absorbers relative to the quasar, it is important to note that Eq.~\ref{eq:covfrac} can yield non-unity coverage fraction of truly intervening absorbers {\it in the wings of profiles where the absorption is weak} as a result of instrumental smearing \citep{gan99}. In addition, \citet{misawa05} note that, under certain models of intrinsic absorption, covering fractions derived using Eq.~(\ref{eq:covfrac}) where distinct components are blended are not reliable. Therefore, in judging whether or not a system shows evidence for trough-dilution, we focus on the cores of unblended kinematic components that are wider than the instrumental profile. This also has the benefit of reducing potential sensitivity in the uncertainty in continuum placement. In the shallower wings, small discrepancies in the continuum placement can mimic partial coverage. In the appendix, we tabulate the systems detected by our criteria, our adjudications regarding the partial coverage test, as well as notes on systems (those that show partial coverage and those that are associated) that are discussed (as an ensemble) below. We use the following scheme to classify the confidence with which a system does or does not exhibit partial coverage:

\noindent {\bf Cannot Evaluate (CNE) --} Due to blending with transitions from unrelated absorption-line systems, or to insufficient sampling of the line profile (i.e., if the line is too narrow or weak), we cannot evaluate if the doublet shows evidence for partial coverage. These are analogous to the M07 Class C2 and C3 systems.

\noindent {\bf Consistent With 1:2 Doublet Ratio (CON) --} The partial coverage test shows that the ultraviolet doublet is consistent with a 1:2 true optical depth ratio. The coverage fraction in the cores of components reaches unity. These are analogous to the M07 Class C1 systems.

\noindent {\bf Doublet Not Detected (ND) --} The doublet is covered, but not formally detected by the criteria given above. We list an ellipsis (...) if the transition is not covered by the data.

\noindent {\bf Line-Locked Pair (LLo) --} The weaker member of the doublet is blended with the stronger of the same doublet from an adjacent absorption-line system (or visa versa). Due to the blending, the partial coverage test cannot be applied. These are analogous to the M07 Class B2 systems.

\noindent {\bf Possible Partial Coverage (POS) --} The coverage fraction is inconsistent with unity (i.e., inconsistent with full coverage). In some cases, noise in the bins covering the respective troughs of the two members of the doublet may cause an unphysical optical depth ratio (i.e., larger than 2 or less than unity). These are analogous to the M07 Class B1 systems.

\noindent {\bf Partial Coverage Likely (PC) --} The coverage fraction is inconsistent with unity (to reasonable confidence) over the entire absorption profile. There is no evidence for blending, and the smoothness of the profiles implies little (if any) unresolved saturated structure. These are analogous to the M07 Class A systems.

\section{Survey Results}
\label{sec:stats}

\begin{table}
\caption{Absorption-line System Demographics$^\mathrm{a}$}
\label{tab:demographics}
\begin{tabular}{r@{~~}c@{~~}c@{~~}c@{~~}c@{~~}c@{~~}c@{~~}cr}
\hline\hline
{Species} & {Cov.} & {Det.} & {Assoc.} & {LLo} & {Test} & {PC} & {POS} & {Fraction} \\
{(1)} & {(2)} & {(3)} & {(4)} & {(5)} & {(6)} & {(7)} & {(8)} & {(9)} \\ \hline
\multicolumn{9}{c}{All 36 Quasars} \\ \hline
\\[-7pt]
 \civ  &  44 &  43 &  3 &  0 & 26 &  2 &   1 &  8--12\,\% \\
 \nv   &  81 &  31 & 10 &  4 & 10 &  3 &   1 & 30--40\,\% \\
 \ovi  &  67 &  64 & 27 &  4 & 30 &  2 &   3 &  7--17\,\% \\
\siliv &  58 &  32 &  0 &  0 & 12 &  0 &   1 &  0-- 8\,\% \\
   Any & 113 & 113 & 31 &  8 & 63 &  6 &   6 & 10--19\,\% \\
\hline \multicolumn{9}{c}{12 Radio-loud Quasars} \\ \hline
\\[-7pt]
 \civ  &  11 &  10 &  0 &  0 &  4 &  0 &   1 &  0--25\,\% \\
 \nv   &  23 &  10 &  5 &  2 &  4 &  1 &   1 & 25--50\,\% \\
 \ovi  &  23 &  23 & 10 &  2 &  9 &  2 &   1 & 22--33\,\% \\
\siliv &  16 &   9 &  0 &  0 &  3 &  0 &   1 &  0--33\,\% \\
   Any &  34 &  34 & 10 &  4 & 17 &  3 &   4 & 18--41\,\% \\
\hline \multicolumn{9}{c}{22 Radio-quiet Quasars} \\ \hline
\\[-7pt]
 \civ  &  32 &  32 &  3 &  0 & 21 &  2 &   0 & 10--10\,\% \\
 \nv   &  55 &  21 &  5 &  2 &  6 &  2 &   0 & 33--33\,\% \\
 \ovi  &  38 &  35 & 14 &  2 & 19 &  0 &   2 &  0--11\,\% \\
\siliv &  40 &  21 &  0 &  0 &  9 &  0 &   0 &  0-- 0\,\% \\
   Any &  71 &  71 & 18 &  4 & 43 &  3 &   2 &  7--12\,\% \\
\hline \multicolumn{9}{c}{22 $\zem < 0.7$\ Quasars} \\ \hline
\\[-7pt]
 \civ  &  16 &  15 &  3 &  0 & 10 &  2 &   0 & 20--20\,\% \\
 \nv   &  39 &  18 & 10 &  4 &  6 &  2 &   1 & 33--50\,\% \\
 \ovi  &  30 &  29 & 15 &  2 & 16 &  2 &   1 & 12--19\,\% \\
\siliv &  28 &  14 &  0 &  0 &  6 &  0 &   1 &  0--17\,\% \\
   Any &  47 &  47 & 19 &  6 & 31 &  5 &   3 & 16--26\,\% \\
\hline \multicolumn{9}{c}{14 $0.9 < \zem < 2 $\ Quasars} \\ \hline
\\[-7pt]
 \civ  &  28 &  28 &  0 &  0 & 16 &  0 &   1 &  0-- 6\,\% \\
 \nv   &  42 &  13 &  0 &  0 &  4 &  1 &   0 & 25--25\,\% \\
 \ovi  &  37 &  35 & 12 &  2 & 14 &  0 &   2 &  0--14\,\% \\
\siliv &  30 &  18 &  0 &  0 &  6 &  0 &   0 &  0-- 0\,\% \\
   Any &  66 &  66 & 12 &  2 & 32 &  1 &   3 &  3--12\,\% \\ \hline
\end{tabular}
$^\mathrm{a}$Comments: {For each of the five samples of quasars (all, radio--loud, radio--quiet, low--redshift, high--redshift), we list demographics for each species searched (\civ, \nv, \ovi, \siliv, any/all). In each block, the columns are: (1) the species in question; (2) the number of absorption-line systems in which the species was covered; (3) the number of times the species is detected (see text for detection criteria); (4) the number of times the species is detected in an associated absorption-line system; (5) the number of times the species is detected in an apparently line-locked absorption-line system; (6) the number of times a detected species can be tested for an intrinsic origin using our partial coverage criteria; (7) the number of testable species that clearly evidence for trough-dilution; (8) the number of testable species that possibly show evidence for trough-dilution; (9) the implied range of the incidence of systems showing evidence of an intrinsic origin (PC/Test -- PC+POS/Test). }
\end{table}

\subsection{Basic Statistics}

In summary, we detect 113 extragalactic metal-line systems in the UV doublet of at least one of the high-ionization species searched. Table~\ref{tab:demographics} lists for each quasar subsample (all, radio--loud, radio--quiet, low redshift, intermediate redshift) and each high-ionization species (column 1) the number of absorption-line systems that are covered (column 2) and detected (column 3) in each species. The table also lists how many of the systems are associated (column 4), how many are line-locked (column 5), the number of systems we are able to test for partial coverage (column 6), the number of testable systems that show evidence for diluted troughs by our prescription (PC: column 7, POS: column 8), and implied fraction of intrinsic absorption-line systems (column 9). In the fifth row for each subsample, we list the full demographics of all systems taken together regardless of species.

At face value, the bottom line says that as few as 10\% (6/63, counting only PC cases), and as many as 19\% (12/63, counting PC and POS cases) of absorption-line systems in this sample show evidence of an intrinsic origin. If the same statistics hold for all systems (not just the testable ones), then about 11--22 of the 113 systems are intrinsic. In the most optimistic case, we have found 20 systems (counting PC, POS, and line-locked pairs). There is apparently a difference in intrinsic absorber fraction with radio class, with 18--41\% systems in radio-loud quasars showing an intrinsic origin, while only 7--12\% systems in radio-quiet quasars show an intrinsic origin. We note, however, that the statistics for radio-loud quasars are greatly affected by 3C\,351, which contributes all 3 PC systems. Exclusion of this single quasar brings the intrinsic absorber fraction for radio-loud quasars to 0-29\%, consistent with the radio-quiet sample. In addition, since RX\,J$1230.8+0115$\ contributes 5 systems to our radio-quiet sample, it may also skew the results for that sample. However, only 2 of the systems are actually testable by our technique. Removing these reduces the quoted range for radio-quiet quasars to 3-8\%. In summary, this sample does not show any compelling difference between radio-loud and radio-quiet quasars, but a larger sample is likely required to support a definitive comparison.

With the redshift division, there is a possible difference in the overall statistics, with low-redshift quasars apparently showing a larger fraction of intrinsic systems (16--26\%), compared to the intermediate redshift quasars (3--12\%). However, this is likely the result of small number statistics, since the high redshift quasars from M07 show a fraction of 19--26\% intrinsic systems. This comparison would seem to indicate little evolution in the fraction of quasars showing intrinsic absorption. However, it is important to note that we have not considered the equivalent widths limits of the various subsamples, which may be important. \citet{gan01a} found similar results (i.e., no significant change in the overall percentages) for the fraction of associated absorbers, but did report a dearth of strong [$W_\lambda$({\civ})$\gtrsim1$\,\AA] systems at $z \lesssim 1$.

\subsection{Velocity Distribution of Absorbers}
\label{sec:vejdist}

\begin{figure}
\includegraphics[width=0.95\columnwidth]{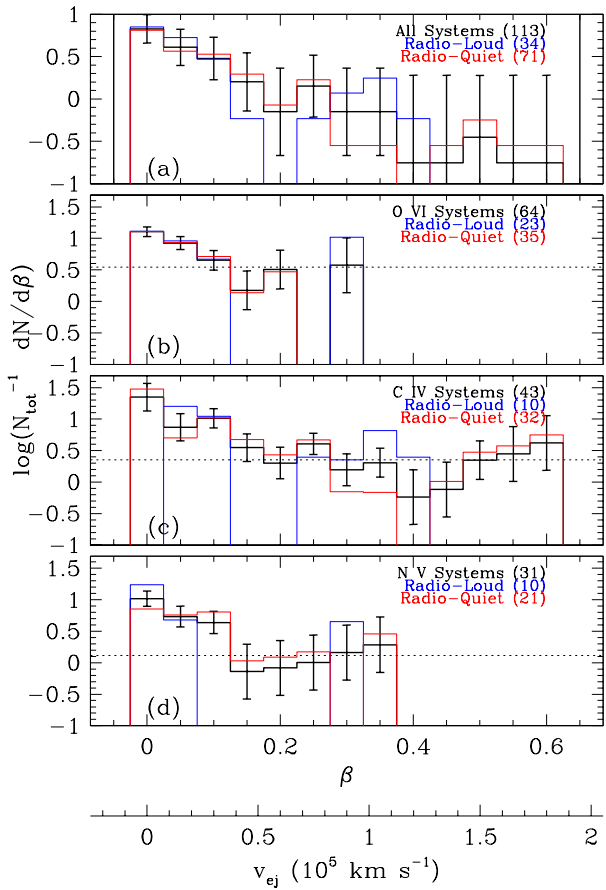}
\protect\caption[velocity distribution]{In panel (a), we show the velocity distribution of all absorption-line systems detected in at least one of the moderate to high ionization doublets searched. Each distribution is normalized separately by the number of systems in the subsample indicated. The numbers in the parentheses give the number of systems used for the normalization. The red (or short-dashed) and blue (or dot-dashed) histograms show the same distribution, but for only radio-quiet and only radio-loud quasars, respectively. In the lower three panels, the velocity distribution is shown for only {\ovi} systems (panel b), only {\civ} systems (panel c), and only {\nv} systems (panel d). The \ovi, \civ, and {\nv} distributions have been corrected by the sensitivity curve from Figure~\ref{fig:ioncoverage}. The error bars reflect both the statistical counting uncertainties and those associated with the sensitivity curve. The horizontal line in each panel is drawn at the average for all $\vej > 5000$\,\kms\ ($\beta > 0.17$) bins to guide the eye as to the consistency of absorbers at large velocity with a flat distribution. The averages computed using the full velocity distributions (black histograms) are $\log (N_\mathrm{tot}^{-1} \frac{dN}{d\beta}) = 0.54$ (\ovi), 0.35 (\civ), and 0.12 (\nv).}
\label{fig:vejdist}
\end{figure}

Panel (a) of Figure~\ref{fig:vejdist} shows the normalized velocity distribution ($N_\mathrm{tot}^{-1} dN/d\beta$), also called the velocity-path density, of absorbers (regardless of ion -- i.e., without double-counting). The normalization is the number of absorption-line systems and is indicated in parentheses in the figure. The panel also shows the normalized velocity distributions for the radio-loud and radio-quiet subsamples. From a similar analysis of absorbers in front of $z\sim2$\ quasars \citet{rich99} found differences between radio-loud and radio-quiet subsamples. For our considerably lower-redshift (and lower-luminosity) quasars, we find no such difference. Even if we placed all systems at velocities larger than 5000\,\kms\ in a single bin, we would not see a difference.

However, we note that, as a consequence of the changing sensitivity to finding absorbers as a function of velocity (Figure~\ref{fig:ioncoverage}, panel a) and the finite wavelength coverage, we see a steady decline in the total number of absorbers with increasing velocity, instead of the expected fall-off to a constant value indicative of intervening systems. In panels b--d of Figure~\ref{fig:vejdist}, we isolate the distributions of \ovi-selected (panel b), {\civ}-selected (panel c), and {\nv}-selected (panel d) systems. We correct these distributions for the sensitivity, $f^{-1}(\beta) N_\mathrm{tot}^{-1} dN/d\beta$, where $f(\beta)$\ is taken from Figure~\ref{fig:ioncoverage}. Inspection of panels (B) and (C) shows that non-associated (i.e., $\beta > 0.17$) systems selected by either {\ovi} or {\civ} have a normalized velocity-path density of $N_\mathrm{tot}^{-1} dN/d\beta \sim 3$. For {\nv}, non-associated systems appear to be detected about a third as frequently, $N_\mathrm{tot}^{-1} dN/d\beta \sim 1$. Naively, this difference is expected given that the relative abundance of nitrogen is lower than carbon or oxygen \citep[$\log(N/O)_\odot = -0.88, \log(N/C)_\odot =-0.62$,][]{ags05}. Moreover, nitrogen can be underabundant compared to the solar N/O and N/C ratios due the different nucleosynthetic origins of nitrogen. There is a statistically significant excess over this value at lower velocities ($\beta \lesssim 0.1$), which is also expected \citep[e.g.,][]{foltz88,gan01a,ves03}, but extends to somewhat larger velocities than the standard 5000\,\kms\ velocity cut-off for ``associated'' systems. [\citet{tripp08} also consider the velocity distribution of individual components instead of statistically-associated {\it systems} (see their Figure 15) and come to similar conclusion - that there is about a factor of three more {\ovi} components near the quasar redshift than far from it.] This excess appears much more significant for {\nv} than for {\civ} or {\ovi} given the dearth of non-associated systems detected in {\nv}. This is consistent with the enhanced nitrogen abundance observed in AGN broad emission-line regions \citep[e.g.,][ and references therein]{hf99} and in some absorption-line systems associated with low-redshift AGN \citep[e.g.,][]{arav07,fields07}. Moreover, the excess occurs in both radio-quiet and radio-loud quasars with no apparent preference.

\subsection{Intrinsic Absorber Statistics as a Function of Ion}

Since one of our goals is to determine additional criteria that can be used to separate intrinsic systems from intervening ones, it is interesting to consider the intrinsic absorber fraction as a function of ion. From the top panel in Table~\ref{tab:demographics}, we see that {\civ} and {\ovi} are detected with the greatest frequencies (with 97.7\%, and 95.5\% detection rates, respectively). This is not surprising given that C and O have the largest relative abundances of all metals and that the UV doublets of their lithium-like ions have fairly high $f\lambda$\ values. [In photoionized gas, with a typical quasar ionizing spectrum, the ionization fraction of these two species can easily be $\approx40$\%\ \citep[e.g.][]{ham97d}. While these are not the dominant ionization stages, the combination of high relative metal abundance and $f\lambda$\ values makes these species readily detectable, even at low column densities.]

For {\civ}, only 26 doublets were testable with our partial coverage criteria, and only 2 show evidence for partial coverage (with a third possible, but not definitive). Thus, only 8--12\% of {\civ}-selected systems show partial coverage. Similarly, only 2 of the 30 testable \ovi-selected systems show partial coverage, with an additional 3 possible, giving an intrinsic fraction of 7--17\%. None of the {\siliv} doublets showed partial coverage definitively, and only 1 is even possible; the intrinsic fraction for {\siliv} is 0--8\%.

The {\nv} ion is detected with the lowest frequency; it is covered for 81 systems, but detected in only 31 (38\%). However, the fraction of these systems that are intrinsic is much higher than {\civ}, \ovi, and \siliv. Of the ten testable systems, three show definite signs of partial coverage, and another one is possible. In addition, four systems show evidence for line-locking [one pair in the spectrum of RX\,J$1230.8+0115$ -- see \citet{gan03b}, and one pair in the spectrum of 3C\,351 00 see \citet{yuan02}], giving an intrinsic fraction of 30--57\%. We note that the spectrum of RX\,J$1230.8+0115$\ contributes five {\nv} systems, two of which are consistent with full coverage \citep[one of these is the system at the redshift of the Virgo cluster, see][]{ros03}, one of which shows partial coverage, and two of which are line-locked. Removal of this quasar spectrum gives intrinsic fractions of 29--56\%. We conclude, then, that {\nv} offers the best statistical means of any transition for finding intrinsic systems. This should be of use in constructing large catalogs of intrinsic systems with lower-resolution and/or lower-S/N data.

\subsection{Associated {\ovi} Absorbers}
\label{sec:assocovi}

The statistics for the incidence of associated absorbers are also listed in Table~\ref{tab:demographics} (column 4). Overall, we find 31 AALs (detected in any species) in the spectra of 36 quasars. Since we are most sensitive to finding AALs with {\ovi} (Figure~\ref{fig:ioncoverage}), it is not surprising that 27 of the 31 AALs are detected in this ion. The two {\ovi} systems that show partial coverage are AALs. These are both in the spectrum of 3C\,351 \citep[$\vej = 1007, 2193$\,\kms, see also][]{yuan02}. One of the three possible cases of partial coverage is also an AAL (PG\,1634+706 $\vej=-962$\,\kms). All four line-locked {\ovi} systems are also AALs. Of the 27 {\ovi} AALs, 14 are actually testable. This implies that the fraction of truly intrinsic systems among {\ovi} AALs is between 14\% (2/14, counting only testable systems definitively showing partial coverage) and 21\% (3/14, adding in the possible partial coverage system). If this fraction applies to all {\ovi} AALs, not just testable ones, then we would expect to find 4--6 truly intrinsic {\ovi} systems at $\vej < 5000$\,\kms. The systems showing evidence for partial coverage and the line-locked systems would account for all of these. However, a further analysis bears consideration.

\begin{figure}
\includegraphics[width=0.75\columnwidth,angle=-90]{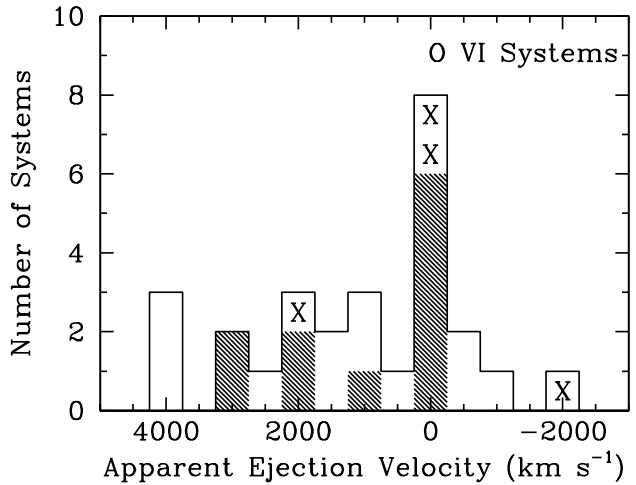}
\protect\caption[ovi velocity distribution]{The figure shows the velocity distribution of the 27 associated \ovi systems. The unshaded histogram shows the distribution of all systems, while the shaded histogram shows the distribution of the subset of 11 systems showing no evidence for partial coverage (i.e., that are flagged as CON). Note that 13 of 27 systems are not testable. 'X' symbols indicate the locations of the four line-locked systems.}
\label{fig:ovivdist}
\end{figure}

Since AALs arise from an eclectic variety of sources, it is useful to estimate the number of AALs that are expected due to intervening structures. We make this estimate using the redshift path density, but we must first consider the velocity distribution of {\ovi} AALs, which is shown in Figure~\ref{fig:ovivdist}. The distribution is sharply peaked at $\vej=0$\ with eight systems. Of the remaining 19 systems, 15 lie toward more positive (that is, more blueshifted) values. There are two systems that have significant redshifts relative to $\zem$, but we note that no {\ovi} system appears more than 2000\,\kms\ redward of the emission redshift. Thus, in our estimate below, we use a velocity path per quasar for {\ovi} AALs of 7000\,\kms\ (i.e, spanning the range $-2000 \leq \vej \leq +5000$\,\kms). There are 29 quasars for which we could detect {\ovi} AALs, so our total velocity and redshift path is $\delta z = \delta \beta = 0.677$.

According to \citet{tripp08}, the redshift path density of low-redshift intervening {\ovi} absorption with rest equivalent widths larger than 30\,m\AA\ is $dN/dz = 15.6_{-2.4}^{+2.9}$. Thus, we statistically expect to find $10.6_{-1.6}^{+2.0}$\ intervening systems in the region searched. We find over twice this number. The Poisson probability of finding 27 AALs when only 10 are expected is $\sim 0.004$\%. Thus, the excess is fairly significant. If (16--17)/27 of the associated {\ovi} systems are intrinsic, then our selection criteria (i.e., partial coverage or line-locking) catch only 12--44\% of intrinsic systems. It is possible that there is an excess of {\ovi} absorbers that are too far from the central engine to cause partial covering, and they may not be line locked. For example, if quasars drive galactic-scale superwinds (e.g., during ``blow out''), the gas could be too far from the compact quasar continuum or line-emitting regions to cause partial coverage. The projected area of those regions would be too small. It might also be possible that quasars tend to be located in regions of higher galaxy density, so there is a greater probability of absorption from the gaseous halo of a galaxy that is distinct and separate from the QSO host galaxy but happens to be close to the QSO redshift \citep[see, for example, ][regarding galaxies in the PHL\,1811 field]{jenkins03}.

\citet{gan01a} noted that the velocity distribution of low-redshift {\civ} AALs was sharply peaked at the emission redshift of the quasar. This suggests a dynamical relationship between AALs and the quasar broad emission-line region, which favors a predominantly intrinsic origin for AALs. Figure~\ref{fig:ovivdist} also seems to show this for {\ovi} AALs [see also Figure~15 of \citep{tripp08}]. Thus it is possible that indeed selection by partial coverage or line-locking does not catch all intrinsic systems. However, \citet{gan06a} and \citet{sembach04b} have shown two examples of {\ovi} AALs that are not related to the quasar central engine. Thus, these statistical arguments regarding the location of AAL gas are not definitive. In addition, \citet{tripp08} note that many of the associated {\ovi} systems do not show {\hi} absorption. These ``{\hi}-free'' {\ovi} absorbers are likely due to a high metal content and high ionization in the presence of quasar radiation field. [The simulations from \citep{ham97d} indicate that at the peak of the {\ovi} ionization fraction, the {\hi} ionization fraction is $f$({\hi})$\sim 10^{-5}$. Under these conditions, a cloud with $N$(\ovi)$\sim10^{14}$\,cm$^{-2}$\ could have an {\hi} column density as low as $N$({\hi})$\sim10^{13}$\,cm$^{-2}$. This would require an equivalent width sensitivity of $\lesssim 40$\,m\AA\ for detecting {\hi} Ly\,$\alpha$. While some of our spectra are not sensitive enough to detect this low column density, some  certainly are.] More information is required in order to distinguish the location (or locations) of this population of absorbers.

\subsection{Quasar Outflow Fraction}

In addition to considering the fraction of systems showing an intrinsic origin, it is also useful to consider the fraction of quasars hosting an intrinsic absorption-line system. In the simplest interpretations, this fraction can be equated either with the solid angle subtended by the outflow from the point of view of the black hole (the ``global covering factor'' of the wind) or with the fraction of the duty cycle of the AGN over which an outflow covering $4\pi$\ str. persists. Of course, this interpretation is complicated by other factors including: (1) We likely have only lower limits on the incidence of intrinsic systems. M07 report that at least 50\% of $\zem=2-4$\ quasars in their sample showed signs of intrinsic absorption. The fraction is reported as a lower limit due to putative intrinsic systems that do not exhibit partial coverage and would, therefore, not be selected by such techniques. (2) Our sample contains no Type 2 AGN. The true outflow fraction would need to take selection effects into account. Nevertheless, we consider the statistics of our survey at ``face value.''

In our sample of 36 quasars (similar in size to that of M07 who considered similar high-resolution Keck spectra of high-redshift quasars), we find that only three quasars (PG\,1206+459, 3C\,351, and RX\,J1230.8+0115) have systems {\it clearly} showing partial coverage and an additional quasar (PG\,0117+213) showing a line-locked pair. Four additional quasars might host systems with partial coverage (PKS\,0312-77, PKS\,0405-123, PG\,1241+176, PG\,1634+706) but we cannot say for certain. The fraction of quasars showing evidence for intrinsic absorption (through partial coverage or line-locking) in this sample is 11-22\%. This fraction is somewhat different between the low redshift subsample (9--18\%) and the intermediate redshift subsample (14--29\%). Taken in conjunction with the M07 result ($>$50\%), this might imply evolution or a luminosity dependence in the fraction of quasars showing intrinsic absorption. For the radio-quiet and radio-loud subsamples, we find 9--18\% and 9--36\% of quasars show intrinsic absorption, respectively.

For $z > 0.15$\ associated {\ovi} absorbers, we cover the doublet in 29 quasars and detect it in 16 (55\%), a fraction comparable to the situation at $z < 0.15$\ \citep{kriss02,dunn07}. However, if we account for AALs that may not be intrinsic by making use of our partial coverage test, line-locking assessments, and the statistical arguments in \S\ref{sec:assocovi}, then the fraction of quasars with intrinsic low-velocity gas is smaller, lying in the range 24--45\%. The range of fractions represents the uncertainty that arises from the fact that we cannot ascertain whether all systems are intrinsic (some intrinsic systems may have full coverage and some of the systems that we were not able to test may also be intrinsic).

\section{Discussion}
\label{sec:discussion}

The overarching interest is to understand quasar outflows, and their role in quasar structure, their relationship to the accretion process, and potentially their relative importance in the galaxy-scale feedback. Toward that goal, we must carefully select absorption-line systems that truly sample outflowing gas. As we have mentioned above, several recent studies have determined that some families of ``intrinsic'' or ``associated'' ($\zabs\sim\zem$) absorption-line systems appear to lie over a range of large distances from the quasar central engine, from galactic-scale \citep[0.1-10\,kpc -- e.g.,][]{bor13,bor12,dunn10,moe09} to extragalactic-scale \citep[$\gg100$\,kpc -- e.g.,][]{mor86,tls96}.  The relationship (if any) between these families of absorbers and the central engine is not certain. While the extragalactic-scale absorbers are unlikely to have a direct relationship with the central engine, it is possible that the galactic-scale absorbers may be a signature of feedback from the outflow. However, this family of large-distance absorbers tends to have (1) low ionization, (2) low velocity, and (3) full coverage of the continuum source (i.e., the accretion disk) or even line-emitting regions. In order to understand the implications of these galactic-scale intrinsic absorbers with regard to the more general population, additional studies are required to constrain their statistical characteristics and significance. In our selection of ``intrinsic'' systems, we have screened for the presence of partial coverage, for which it is difficult to reconcile large distances from the emitting regions. In the following subsections, we discuss two families of absorbers: (1) intrinsic absorbers selected to have ``strong \nv'' absorption and (2) associated absorbers at low velocity selected by \ovi\ absorption. We discuss the former in the context of the accretion-disk/wind paradigm, and the latter in the broader scope appropriate to the likely-heterogenous nature of the class.

\subsection{Strong {\nv} Absorption Systems}

The first class of intrinsic absorption-line systems that we note are those which exhibit strong {\nv} absorption equivalent width relative to {\hi} Ly\,$\alpha$. Three quasars (and a possible fourth) in this sample exhibit absorption-line systems consistent with the ``strong {\nv}'' family of M07: PG\,$1206+459$, RX\,J$1230.8+0115$, and 3C\,351 (and possibly PKS\,$0312-77$). RX\,J$1230.8+0115$\ and 3C\,351 have been examined extensively by \citet{gan03b} and \citet{yuan02}, respectively. The absorption in these two quasars consists of complexes of closely spaced lines, separated by unabsorbed segments of continuum \citep{gan01b}. In all quasars, the absorption is highly saturated as evidenced by the relative depths of the doublet lines (i.e., similar normalized flux levels for both members of the doublet). The absorption profiles are not black, however, because the coverage fraction is less than unity. Saturation combined with partial covering of the background source is also partly responsible for the large observed strength of {\nv} relative to Ly\,$\alpha$. Like the ``strong {\nv}'' absorbers in the M07 sample, these also tend to have high ionization (i.e., typically there are no accompanying low-ionization species detected).

The ``strong {\nv}'' absorbers represented in this sample place interesting constraints on models of outflows (with regard to the drivers and geometry of the flow) which we discuss in the following sections. It is interesting to note a recent study that used a background QSO to probe the extended gaseous halo of a foreground AGN shows strong {\nv} absorption (as well as {\neviii} ) at the foreground AGN redshift but at least 68 kpc in projection from the AGN host galaxy \citep{ding03,tripp11}. Such strong {\nv} absorption is rarely observed in the gaseous halos of ordinary galaxies and is more similar to the strong N V lines that we see in intrinsic systems.  This may suggest that nitrogen-rich AGN outflows are able to propagate to large distances away from the nucleus.

\subsubsection{Line-locking}

The complex of absorption systems intrinsic to RX\,J$1230.8+0115$ appears to be line-locked in the {\nv} and {\ovi} doublets, and {\hi} Ly\,$\beta$. Likewise, the associated absorbers toward PG\,$0117+213$\ appear to be line-locked in {\ovi} and {\hi} Ly\,$\beta$, as are systems in the 3C\,351 intrinsic absorbers. There have also been other reports of line-locking between the stronger member of the {\nv} doublet and {\hi} Ly\,$\alpha$\ \citep[i.e., the ``ghost of Ly\,$\alpha$,'' e.g., ][]{korista93,arav96}.

If the outflow picture is indeed correct for these intrinsic systems, and the apparent line-locking is real, it would have the following implications. First, the part of the flow associated with line-locked pairs of systems is likely directed toward the observer. Otherwise, projection effects would preclude the detection of blended lines (the relative velocities in the direction of the observer would not be those required to detect line-locked pairs). Second, driving via radiation pressure must be important (and dominant) in these systems. Moreover, the dominant component of radiation pressure must be line-driving by the UV resonance lines ({\nv}$\lambda\lambda$1238.821, 1242.804, \ovi$\lambda\lambda$1031.926, 1037.617, and {\hi} Ly\,$\alpha, \beta$). Consequently, this may imply a narrow range in ionization conditions so that the species from which the lines originate are sufficiently abundant. Numerical treatments of line-driven winds still need to account for this effect, and the conditions in which it occurs.

\subsubsection{Constraints on the Wind-Driving Mechanism from PG\,$1206+459$}

Of the four quasars in the strong {\nv} family presented here, only PG\,$1206+459$\ has intrinsic absorption that does not appear within 5000\,\kms\ of the quasar emission redshift\footnote{The redshift is that reported by the Sloan Digital Sky Survey using a template that measures the redshift of, in this case, the {\mgii}. The template includes an average shift of 160\,\kms\ from the narrow {\oiii} $\lambda$5007 line \citep{vdb01} to more accurately reproduce systemic redshifts in the ensemble average.}. The maximum velocity of absorption is 19,400\,\kms.

PG\,$1206+459$\ also happens to be the most luminous quasar in this subsample with a luminosity of $\lambda L_\lambda(3000\,\mathrm{\AA}) = 5.7 \times 10^{46}$\ erg~s$^{-1}$. With this UV luminosity, the quasar is capable of radiatively driving an outflow with a terminal velocity of $\sim 130,000$\,\kms\ \citep{gan07b}. The ejection velocities of the strong {\nv} systems in this sample are consistent with a scenario in which the outflows are radiatively driven. Nevertheless, there are a number of possible reasons that we measure a maximum velocity of absorption that is much less than the largest possible terminal velocity for a quasar of this luminosity. One possibility is that the direction of the outflow is not toward the observer, but at an angle. If this is the case, we would infer an angle of 82$^\circ$. If the outflow is equatorial \citep[an assumption not supported by recent observations of some BAL quasars, e.g.,][]{brotherton06,gp07,dipompeo10}, then our line of sight should lie within within 10 degrees of the rotational axis of the accretion disk.

An alternative possibility is that the gas is too ionized (or insufficiently ionized) as to allow for the most efficient use of ultraviolet photons to radiatively drive the flow. \citet{gallsc06} have shown that the maximum velocity of absorption is also correlated with the parameter $\Delta \alpha_{\mathrm{ox}}$\footnote{$\alpha_{\mathrm{ox}}$\ is a two-point spectral index computed from the fluxes at 2500\,\AA\ and 2\,keV \citep{tea}. This quantity is correlated\ with the 2500\,\AA\ luminosity \citep[e.g.,][]{steffen06,strateva05}. $\Delta \alpha_{\mathrm{ox}}$\ is the difference between the measured $\alpha_{\mathrm{ox}}$\ and the expected value of $\alpha_{\mathrm{ox}}$\ based on the 2500\,\AA\ luminosity.}. $\Delta \alpha_{\mathrm{ox}}$\ is a measure of the apparent absorption optical depth at 2\,keV, and the reported correlation is such that objects with the largest amount of soft X-ray absorption also have the largest measured maximum velocity of UV absorption. (In the radiative-driving scenario, the X-ray-absorbing gas is typically not the same as the UV-absorbing gas, but rather a separate phase of the outflow that shields the UV-absorbing gas.) For the strong {\nv} outflow observed in the spectrum of PG\,$1206+459$, this is an entirely plausible, and testable hypothesis.

The rest-frame 2500\,\AA\ flux is covered by the Sloan spectrum of PG\,$1206+459$\ and we estimate a power-law continuum of the form $f_\lambda = 8 \times 10^{-16} (\lambda/8000)^{-1.7}$\ (from a ``by-eye'' fit visually accounting for emission from the {\feii} UV multiplets). The luminosity density at a rest-frame wavelength of 2500\,\AA\ is then $5.4 \times 10^{31}$\ erg s$^{-1}$\ Hz$^{-1}$, implying a benchmark $\alpha_\mathrm{ox} = -1.7 \pm 0.2$\ from \citet[][see their equation 2]{steffen06}. [Note that the uncertainty reflects only the quoted rms scatter in the $\alpha_\mathrm{ox}-\log L_\nu(2500\,\mathrm{\AA})$\ relationship.] \citet{piconcelli05} report an XMM 12.2 kilosecond observation of PG\,$1206+459$\ on 11 May 2002 that is described well by a $\Gamma=2$\ power-law in the 2-12\,keV band, and a slightly flatter $\Gamma=1.74$\ power-law over the range 0.3-12\,keV, with a normalization of $2.4 \times 10^{-4}$\ photons cm$^{-2}$\ s$^{-1}$\ keV$^{-1}$\ at 1\,kev. This corresponds to an observed flux at rest-frame 2\,keV of $f_\nu = 1.68 \times 10^{-30}$\ erg cm$^{-2}$\ s$^{-1}$\ Hz$^{-1}$, which implies an observed $\alpha_\mathrm{ox} = -1.5$. Thus, $\Delta \alpha_\mathrm{ox} = +0.2 \pm 0.2$, consistent with minimal soft X-ray absorption, as expected from the radiative-driving hypothesis.

Of course, there could be a combination of reasons why the observed velocity of the {\nv} absorbers is lower than the maximum possible value. More complicated scenarios involving a combination of, say, the preceding two ideas -- orientation and shielding of the outflow -- may reflect the truth more closely. Furthermore, other factors that we have not discussed may be important, such as the column density of the UV-absorbing gas, or the location of the gas in the velocity-stratified outflow. Here, we only conclude that we cannot rule out either of the two simplest explanations.

\subsection{Associated \ovi}

The importance of studying associated {\ovi} absorption lies in the fact that the photon energy needed to produce the {\ovi} ion is intermediate between the energies needed to produce the ions observed in the near UV (e.g., {\civ} and {\nv}) and those detected in the X-ray band (e.g., {\ovii} and {\oviii}). If there is a relation between the medium detected through UV absorption lines and the medium detected through X-ray absorption edges \citep[as argued, for example, by ][]{scott04b,cren03,ngc3783ii,ngc3783iii,mon01,brandt,mew95}, then this relation can be explored with the help of {\ovi} lines, which probe gas with an intermediate ionization state between that of UV and X-ray absorbers. The overall fractions of associated {\ovi} and X-ray warm-absorbers appear similar \citep[e.g.,][and this work]{gb08,dunn07,myr07}.

We detect many associated {\ovi} (27) absorption systems, but it is not clear where these arise. So far, 2 systems have been examined in detail, the ones toward PG\,$1116+215$ and HE\,$0226-4110$. From a redshift coincidence, \citet{sembach04b} attribute the associated {\ovi} absorption in the PG\,$1116+215$\ sight-line to a field galaxy. The HE\,$0226-4110$\ sight-line shows absorption in a wide range of ions: {\hi}, {{\ciii}, {O\,{\sc iii-vi}}, {\neviii}  \citep{gan06a}. Analysis of the detailed kinematics and ionization conditions shows that it most likely arises in the quasar host galaxy and probably no closer to the central engine than the narrow-line region.

By our estimates, $\approx60$\% of associated {\ovi} absorbers are directly related to either the quasar central engine, quasar host galaxy, or nearby environment (see \S\ref{sec:assocovi}), with only 14-21\% of those systems showing evidence (from partial coverage or line-locking) of an origin in the outflow. The significance of this potentially low fraction is not clear. If we assume that intrinsic absorbers fully occult the compact UV continuum \citep[at least in general, but see ][for a counter-example]{gan99}, then a coverage fraction less than unity must arise from a partial occultation of the broad emission-line region (BELR). The BELR is known to have an ionization stratification \citep[e.g., ][ and references therein]{op02}. Thus, the {\ovi} BELR (that is occulted by the {\ovi} associated absorbers) is more compact than the {\civ} BELR that is occulted by {\civ} associated absorbers. Furthermore, the effective coverage fraction which we have measured is the mean of the partial coverage fraction of each of the emitting regions (continuum, BELR), weighted by their relative fluxes \citep[e.g.,][]{gan99}. On average, the {\ovi} emission line is about a third as strong as the {\civ} emission line while the local continuum level is similar \citep{vdb01}, so partial coverage of the {\ovi} BELR would be weighted less than partial coverage of the {\civ} BELR. It is probable that the combination of these two factors - a relatively more compact, and less luminous {\ovi} BELR - explain the lack of partial coverage among associated {\ovi} that are directly related to the quasar. Further efforts to identify the truly intrinsic absorbers from this sample in an efficient manner, then, requires spectral monitoring to look for time variability.

One more property of associated {\ovi} systems is worth noting. The absorption profiles, in general, are not smooth as expected from a smooth outflow. They are clumpy, featuring discrete kinematic components, and are essentially indistinguishable from intervening systems. Thus, they could arise in interstellar gas from either the host galaxy or satellite galaxies (as in the case of HE\,0226-4110). However, this is difficult to reconcile with our estimate that 60\% of these systems are related in some fashion with the presence quasar (see \S\ref{sec:assocovi}). In a couple cases, namely 3C\,351 \citep{yuan02} and RX\,J$1230.8+0115$\ \citep{gan03b}, the evidence for an association with the quasar outflow is compelling. However, in most other cases, it is not. These other absorbers could arise from the interaction between the quasar outflow and the host galaxy ISM \citep[e.g.,][]{kp09}, further out in a superwind, or in the local environment of the quasar.

One issue that remains to be addressed is the nature of absorption by the host galaxy and whether a redshift path density estimate of the number of intervening absorbers correctly accounts for absorbers in the host galaxy that are unrelated to the quasar outflow. If absorption by the quasar host galaxy is not taken into account, then it is possible that the excess absorption is the interstellar medium of the host galaxy, not the quasar outflow. We have made the assumption, however, that such an accounting is proper, inasmuch as absorption by galaxies is part of the redshift path density (in addition to other sources like the warm-hot phase of the intergalactic medium) and that the host galaxy is not different in those respects. This assumption may or may not be valid. In general, sight-lines do not pass through the disks of intervening galaxies, whereas they would certainly pass through both the disk and halo of the host galaxy. The same is true for the Milky Way -- all sight-lines pass through the Galactic halo and disk, and Galactic {\ovi} absorption is always observed \citep[e.g.,][]{savage03}. On the other hand, as mentioned above, our spectra cover the wavelength range for associated {\ovi} absorption in 29 quasars, but only detect it 55\% of the time. Hence, unlike the Milky Way, the quasar host galaxies do not always produce an {\ovi} absorption feature. This may be due to ionization effects (e.g., perhaps the ISM of the quasar host galaxy is more highly ionized than that of the Milky Way as a consequence of the quasar radiation field) or blowout of gas. If the host galaxy indeed does not generally produce absorption in the same sense as intervening galaxies, then our estimate of the fraction of associated {\ovi} that are due to the quasar outflow may be a lower limit. Further information on the location of these systems (e.g., from variability and absorption-line diagnostics) is needed to improve the assessment.

If these systems are part of the quasar outflow, then the general accretion-disk/wind model needs to be altered to show more clumpiness, such as instabilities in the wind itself, or in the interface region between the wind and the lower-density plasma above the wind \citep[e.g.,][for associated {\civ} absorption]{gan01a}. More observational constraints on the location and physical conditions is needed to rule out possibilities. Higher S/N, high resolution spectra would help to diagnose physical conditions. Detection of excited states or time variability would help to constrain the density and location of the gas relative to the compact continuum. In addition, deep imaging of the quasar fields to consider the distribution of galaxies near the quasar redshift would be useful.

\section{Summary}
\label{sec:summary}

We have conducted a search for intrinsic absorption in the spectra of quasars available in the archive of high spectral resolution HST/STIS echelle (E140M, E230M) observations. We have carried out partial coverage tests of all detected moderate- to high- ionization absorption-line doublets along the 36 sight-lines independent of the putative ejection velocity of the system. Our findings are as follows.
\begin{itemize}
\item[1.] Overall, we find that 10--19\% of absorption-line systems detected in a moderate- to high- ionization doublet show evidence of an intrinsic origin as indicated by  dilution of the absorption trough.

\item[2.] In this sample, there is a greater fraction of intrinsic systems in radio-loud quasars (18--41\%) than in radio-quiet quasars (7-12\%), but this result may be unduly influenced by a single object (3C\,351).

\item[3.] We find little evidence for evolution in the frequency of intrinsic absorption down to the equivalent width limits of the survey, dividing our sample into a low and intermediate redshift bins, and adding the M07 sample as a high-redshift bin. However, we find marginal evidence for the evolution in the fraction of quasars exhibiting intrinsic NALs from 9--18\% at low redshift, to 14--29\% at intermediate redshift, to $>$50\% at high-redshift (M07).

\item[4.] Contrary to \citet{rich01a}, we find no significant differences in the velocity distribution of absorption-line systems (either \ovi-selected, {\civ}-selected, or overall) between radio-loud and radio-quiet objects. This is likely a result of the luminosities of the samples used. As in other surveys, we find an excess of absorption-line systems near the quasars redshift (so-called associated absorbers). In spite of its statistical insignificance, we do find high-velocity intrinsic absorbers at velocities up to 19,000\,\kms.

\item[5.] We find that the {\nv} $\lambda\lambda$1238.821, 1242.804 doublet offers the most efficient means at finding intrinsic absorbers. 30--40\% of {\nv}-selected absorbers show signs of an intrinsic origin. This is probably not surprising given the overabundance of nitrogen in AGN reported by some authors \citep[e.g., ][]{kg02,hf99}.

\item[6.] We confirm the existence of a ``strong {\nv}'' family of intrinsic absorbers claimed by M07, occurring in 11-22\% of quasars in this sample. Overall, these systems are consistent with the radiative-driving scenario of quasar outflows, with some showing line-locking, and others showing observed maximum velocities that depend on the amount of putative X-ray shielding gas. If we take the properties of individual strong {\nv} collectively as representative of the whole sample of strong {\nv} systems, we find that the gas must lie outside (or co-spatially with) the broad emission-line region.

\item[7.] We do not find ``strong {\civ}'' absorbers in this sample. However, we present another family of intrinsic absorbers - associated {\ovi} - that were not found by M07 since their spectra did not cover the relevant (rest-frame) wavelength region and because, at the redshift of their sample, contamination by the Ly\,$\alpha$\ forest was severe \citep[see also][for associated {\ovi} in lower-redshift/luminosity AGN]{dunn07}. These occur in $\sim$55\% of quasars in this sample, with 61\% statistically likely to have an intrinsic origin, generally have kinematics of discrete/clumpy components, and peak in putative ejection velocity at the velocity of the broad emission-line region. Statistically discounting the 39\%\ of systems that may not be intrinsic drops the fraction of quasars hosting an {\ovi} absorber to 24--44\%. Furthermore, of the 61\% of systems statistically expected to be related to the quasars, only 14--21\% actually show evidence of an intrinsic origin in the form of partial coverage. Better quality data (higher resolution and signal-to-noise ratio) or synoptic observations are required to further isolate truly intrinsic systems. The kinematic and dynamical properties of the associated {\ovi} absorbers favor a location similar to that of associated {\civ} absorbers as noted by \citet{gan01a}.
\end{itemize}

~\\
The authors wish to thank Charles Danforth for useful discussions and the anonymous referee for a thoughtful review. RG acknowledges support provided under Program Number HST-AR-10296.06-A by NASA through a grant from the Space Telescope Science Institute, which is operated by the Associated of Universities for Research in Astronomy, Incorporated, under NASA contract NAS5-26555. This work was also supported by NASA grant NAG5-10817 and NSF grant AST-0807993.

This research has made use of the NASA/IPAC Extragalactic Database (NED) which is operated by the Jet Propulsion Laboratory, California Institute of Technology, under contract with the National Aeronautics and Space Administration. Funding for the Sloan Digital Sky Survey (SDSS) and SDSS-II has been provided by the Alfred P. Sloan Foundation, the Participating Institutions, the National Science Foundation, the U.S. Department of Energy, the National Aeronautics and Space Administration, the Japanese Monbukagakusho, and the Max Planck Society, and the Higher Education Funding Council for England. The SDSS Web site is http://www.sdss.org/.

The SDSS is managed by the Astrophysical Research Consortium (ARC) for the Participating Institutions. The Participating Institutions are the American Museum of Natural History, Astrophysical Institute Potsdam, University of Basel, University of Cambridge, Case Western Reserve University, The University of Chicago, Drexel University, Fermilab, the Institute for Advanced Study, the Japan Participation Group, The Johns Hopkins University, the Joint Institute for Nuclear Astrophysics, the Kavli Institute for Particle Astrophysics and Cosmology, the Korean Scientist Group, the Chinese Academy of Sciences (LAMOST), Los Alamos National Laboratory, the Max-Planck-Institute for Astronomy (MPIA), the Max-Planck-Institute for Astrophysics (MPA), New Mexico State University, Ohio State University, University of Pittsburgh, University of Portsmouth, Princeton University, the United States Naval Observatory, and the University of Washington.


\begin{thebibliography}{141}
\expandafter\ifx\csname natexlab\endcsname\relax\def\natexlab#1{#1}\fi

\bibitem[{{Anderson} {et~al}\mbox{.}(1987){Anderson}, {Weymann}, {Foltz}, \&
  {Chaffee}}]{and87}
{Anderson} S.~F., {Weymann} R.~J., {Foltz} C.~B., {Chaffee} F.~H., 1987, \aj,
  94, 278

\bibitem[{{Arav}(1996)}]{arav96}
{Arav} N., 1996, \apj, 465, 617

\bibitem[{{Arav} {et~al}\mbox{.}(1999){Arav}, {Becker}, {Laurent-Muehleisen},
  {Gregg}, {White}, {Brotherton}, \& {de Kool}}]{arav99b}
{Arav} N., {Becker} R.~H., {Laurent-Muehleisen} S.~A., {Gregg} M.~D., {White}
  R.~L., {Brotherton} M.~S., {de Kool} M., 1999, \apj, 524, 566

\bibitem[{{Arav} {et~al}\mbox{.}(1994){Arav}, {Li}, \& {Begelman}}]{alb94}
{Arav} N., {Li} Z.-Y., {Begelman} M.~C., 1994, \apj, 432, 62

\bibitem[\protect\citeauthoryear{Arav et al.}{2007}]{arav07}
Arav N., et al., 2007, ApJ, 658, 829

\bibitem[{{Asplund} {et~al}\mbox{.}(2005){Asplund}, {Grevesse}, \&
  {Sauval}}]{ags05}
{Asplund} M., {Grevesse} N., {Sauval} A.~J., 2005, in ASP Conf. Ser.: Cosmic
  abundances as records of stellar evolution and nucleosynthesis, {Bash} F.~N.,
  {Barnes} T.~G., eds., ASP, San Francisco, p. {in press}

\bibitem[{{Barlow} {et~al}\mbox{.}(1997){Barlow}, {Hamann}, \&
  {Sargent}}]{bhs97}
{Barlow} T.~A., {Hamann} F., {Sargent} W.~L.~W., 1997, in ASP Conf. Ser. 128:
  Mass Ejection from Active Galactic Nuclei, {Arav} N., {Shlosman} I.,
  {Weymann} R.~J., eds., ASP, San Francisco, p.~13

\bibitem[{{Barthel} {et~al}\mbox{.}(1997){Barthel}, {Tytler}, \&
  {Vestergaard}}]{btv97}
{Barthel} P.~D., {Tytler} D.~R., {Vestergaard} M., 1997, in ASP Conf. Ser. 128:
  Mass Ejection from Active Galactic Nuclei, p.~48

\bibitem[{{Becker} {et~al}\mbox{.}(2000){Becker}, {White}, {Gregg},
  {Brotherton}, {Laurent-Muehleisen}, \& {Arav}}]{becker00}
{Becker} R.~H., {White} R.~L., {Gregg} M.~D., {Brotherton} M.~S.,
  {Laurent-Muehleisen} S.~A., {Arav} N., 2000, \apj, 538, 72

\bibitem[{{Becker} {et~al}\mbox{.}(2001){Becker}, {White}, {Gregg},
  {Laurent-Muehleisen}, {Brotherton}, {Impey}, {Chaffee}, {Richards},
  {Helfand}, {Lacy}, {Courbin}, \& {Proctor}}]{fbqs3}
{Becker} R.~H. {et~al.}, 2001, \apjs, 135, 227

\bibitem[{{Becker} {et~al}\mbox{.}(1995){Becker}, {White}, \&
  {Helfand}}]{first}
{Becker} R.~H., {White} R.~L., {Helfand} D.~J., 1995, \apj, 450, 559

\bibitem[{{Begelman}(2006)}]{begelman06}
{Begelman} M.~C., 2006, \apj, 643, 1065

\bibitem[{{Blaes} {et~al}\mbox{.}(2011){Blaes}, {Krolik}, {Hirose}, \&
  {Shabaltas}}]{blaes2011}
{Blaes} O., {Krolik} J.~H., {Hirose} S., {Shabaltas} N., 2011, \apj, 733, 110

\bibitem[\protect\citeauthoryear{Borguet et al.}{2012}]{bor12} Borguet B.~C.~J., Edmonds D., Arav N.,
Dunn J., Kriss G.~A., 2012, ApJ, 751, 107

\bibitem[\protect\citeauthoryear{Borguet et al.}{2013}]{bor13} Borguet B.~C.~J., Arav N., Edmonds D.,
Chamberlain C., Benn C., 2013, ApJ, 762, 49

\bibitem[{{Bowen} {et~al}\mbox{.}(1994){Bowen}, {Osmer}, {Blades}, {Tytler},
  {Cottrell}, {Fan}, \& {Lanzetta}}]{bowen94}
{Bowen} D.~V., {Osmer} S.~J., {Blades} J.~C., {Tytler} D., {Cottrell} L., {Fan}
  X.-M., {Lanzetta} K.~M., 1994, \aj, 107, 461

\bibitem[{{Brandt} {et~al}\mbox{.}(2000){Brandt}, {Laor}, \& {Wills}}]{brandt}
{Brandt} W.~N., {Laor} A., {Wills} B.~J., 2000, \apj, 528, 637

\bibitem[{{Braun} \& {Milgrom}(1989)}]{bm89}
{Braun} E., {Milgrom} M., 1989, \apj, 342, 100

\bibitem[{{Brotherton} {et~al}\mbox{.}(2006){Brotherton}, {de Breuck}, \&
  {Schaefer}}]{brotherton06}
{Brotherton} M.~S., {de Breuck} C., {Schaefer} J.~J., 2006, \mnras, 372, L58

\bibitem[{{Brotherton} {et~al}\mbox{.}(1997){Brotherton}, {Tran}, {van
  Breugel}, {Dey}, \& {Antonucci}}]{broth97}
{Brotherton} M.~S., {Tran} H.~D., {van Breugel} W., {Dey} A., {Antonucci} R.,
  1997, \apjl, 487, L113

\bibitem[{{Brotherton} {et~al}\mbox{.}(1998){Brotherton}, {van Breugel},
  {Smith}, {Boyle}, {Shanks}, {Croom}, {Miller}, \& {Becker}}]{bro98}
{Brotherton} M.~S., {van Breugel} W., {Smith} R.~J., {Boyle} B.~J., {Shanks}
  T., {Croom} S.~M., {Miller} L., {Becker} R.~H., 1998, \apjl, 505, L7

\bibitem[{{Brotherton} {et~al}\mbox{.}(1994){Brotherton}, {Wills}, {Steidel},
  \& {Sargent}}]{bro94}
{Brotherton} M.~S., {Wills} B.~J., {Steidel} C.~C., {Sargent} W.~L.~W., 1994,
  \apj, 423, 131

\bibitem[{{Churchill} {et~al}\mbox{.}(1999{\natexlab{a}}){Churchill}, {Rigby},
  {Charlton}, \& {Vogt}}]{crcv99}
{Churchill} C.~W., {Rigby} J.~R., {Charlton} J.~C., {Vogt} S.~S.,
  1999{\natexlab{a}}, \apjs, 120, 51

\bibitem[{{Churchill} {et~al}\mbox{.}(1999{\natexlab{b}}){Churchill},
  {Schneider}, {Schmidt}, \& {Gunn}}]{cssg99}
{Churchill} C.~W., {Schneider} D.~P., {Schmidt} M., {Gunn} J.~E.,
  1999{\natexlab{b}}, \aj, 117, 2573

\bibitem[{{Condon} {et~al}\mbox{.}(1998){Condon}, {Cotton}, {Greisen}, {Yin},
  {Perley}, {Taylor}, \& {Broderick}}]{nvss}
{Condon} J.~J., {Cotton} W.~D., {Greisen} E.~W., {Yin} Q.~F., {Perley} R.~A.,
  {Taylor} G.~B., {Broderick} J.~J., 1998, \aj, 115, 1693

\bibitem[{{Crenshaw} {et~al}\mbox{.}(2003){Crenshaw}, {Kraemer}, {Gabel},
  {Kaastra}, {Steenbrugge}, {Brinkman}, {Dunn}, {George}, {Liedahl}, {Paerels},
  {Turner}, \& {Yaqoob}}]{cren03}
{Crenshaw} D.~M. {et~al.}, 2003, \apj, 594, 116

\bibitem[Crenshaw \& Kraemer(2012)]{ck12} Crenshaw, D.~M., \& Kraemer, S.~B.\ 2012, \apj, 753, 75

\bibitem[{{de Grijp} {et~al}\mbox{.}(1992){de Grijp}, {Keel}, {Miley},
  {Goudfrooij}, \& {Lub}}]{degrijp92}
{de Grijp} M.~H.~K., {Keel} W.~C., {Miley} G.~K., {Goudfrooij} P., {Lub} J.,
  1992, \aaps, 96, 389

\bibitem[{{Di Matteo} {et~al}\mbox{.}(2008){Di Matteo}, {Colberg}, {Springel},
  {Hernquist}, \& {Sijacki}}]{tdm08}
{Di Matteo} T., {Colberg} J., {Springel} V., {Hernquist} L., {Sijacki} D.,
  2008, \apj, 676, 33

\bibitem[{{Di Matteo} {et~al}\mbox{.}(2005){Di Matteo}, {Springel}, \&
  {Hernquist}}]{tdm05}
{Di Matteo} T., {Springel} V., {Hernquist} L., 2005, \nat, 433, 604

\bibitem[{{Ding} {et~al}\mbox{.}(2003){Ding}, {Charlton}, {Churchill}, \&
  {Palma}}]{ding03}
{Ding} J., {Charlton} J.~C., {Churchill} C.~W., {Palma} C., 2003, \apj, 590,
  746

\bibitem[{{DiPompeo} {et~al}\mbox{.}(2010){DiPompeo}, {Brotherton}, {Becker},
  {Tran}, {Gregg}, {White}, \& {Laurent-Muehleisen}}]{dipompeo10}
{DiPompeo} M.~A., {Brotherton} M.~S., {Becker} R.~H., {Tran} H.~D., {Gregg}
  M.~D., {White} R.~L., {Laurent-Muehleisen} S.~A., 2010, \apjs, 189, 83

\bibitem[{{Dunn} {et~al}\mbox{.}(2007){Dunn}, {Crenshaw}, {Kraemer}, \&
  {Gabel}}]{dunn07}
{Dunn} J.~P., {Crenshaw} D.~M., {Kraemer} S.~B., {Gabel} J.~R., 2007, \aj, 134,
  1061

\bibitem[\protect\citeauthoryear{Dunn et al.}{2010}]{dunn10}
Dunn J.~P., et al., 2010, ApJ, 709, 611

\bibitem[{{Elvis}(2000)}]{elvis00}
{Elvis} M., 2000, \apj, 545, 63

\bibitem[{{Espey}(1993)}]{esp93}
{Espey} B.~R., 1993, \apjl, 411, L59

\bibitem[{{Everett} {et~al}\mbox{.}(2009){Everett}, {Gallagher}, \&
  {Keating}}]{egk09}
{Everett} J., {Gallagher} S., {Keating} S., 2009, in American Institute of
  Physics Conference Series, Vol. 1201, American Institute of Physics
  Conference Series, {S.~Heinz \& E.~Wilcots}, ed., pp. 56--59

\bibitem[{{Everett}(2005)}]{everett05}
{Everett} J.~E., 2005, \apj, 631, 689

\bibitem[{{Falomo} {et~al}\mbox{.}(1993){Falomo}, {Pesce}, \& {Treves}}]{fpt93}
{Falomo} R., {Pesce} J.~E., {Treves} A., 1993, \apjl, 411, L63

\bibitem[Fields et al.(2007)]{fields07} Fields, D.~L., Mathur,
S., Krongold, Y., Williams, R., \& Nicastro, F.\ 2007, \apj, 666, 828

\bibitem[{{Foltz} {et~al}\mbox{.}(1988){Foltz}, {Chaffee}, {Weymann}, \&
  {Anderson}}]{foltz88}
{Foltz} C.~B., {Chaffee} F.~H., {Weymann} R.~J., {Anderson} S.~F., 1988, in QSO
  Absorption Lines; Probing the Universe; Proceedings of the QSO Absorption
  Line Meeting, {Blades} J.~C., {Turnshek} D.~A., {Norman} C.~A., eds.,
  Cambridge University Press, Cambridge, pp. 53--65

\bibitem[{{Foltz} {et~al}\mbox{.}(1986){Foltz}, {Weymann}, {Peterson}, {Sun},
  {Malkan}, \& {Chaffee}}]{foltz86}
{Foltz} C.~B., {Weymann} R.~J., {Peterson} B.~M., {Sun} L., {Malkan} M.~A.,
  {Chaffee} F.~H., 1986, \apj, 307, 504

\bibitem[{{Fukugita} {et~al}\mbox{.}(1996){Fukugita}, {Ichikawa}, {Gunn},
  {Doi}, {Shimasaku}, \& {Schneider}}]{ugriz1}
{Fukugita} M., {Ichikawa} T., {Gunn} J.~E., {Doi} M., {Shimasaku} K.,
  {Schneider} D.~P., 1996, \aj, 111, 1748

\bibitem[{{Gabel} {et~al}\mbox{.}(2003{\natexlab{a}}){Gabel}, {Crenshaw},
  {Kraemer}, {Brandt}, {George}, {Hamann}, {Kaiser}, {Kaspi}, {Kriss},
  {Mathur}, {Mushotzky}, {Nandra}, {Netzer}, {Peterson}, {Shields}, {Turner},
  \& {Zheng}}]{ngc3783ii}
{Gabel} J.~R. {et~al.}, 2003{\natexlab{a}}, \apj, 583, 178

\bibitem[{{Gabel} {et~al}\mbox{.}(2003{\natexlab{b}}){Gabel}, {Crenshaw},
  {Kraemer}, {Brandt}, {George}, {Hamann}, {Kaiser}, {Kaspi}, {Kriss},
  {Mathur}, {Mushotzky}, {Nandra}, {Netzer}, {Peterson}, {Shields}, {Turner},
  \& {Zheng}}]{ngc3783iii}
{Gabel} J.~R. {et~al.}, 2003{\natexlab{b}}, \apj, 595, 120

\bibitem[{{Gallagher} {et~al}\mbox{.}(2006){Gallagher}, {Brandt}, {Chartas},
  {Priddey}, {Garmire}, \& {Sambruna}}]{gallsc06}
{Gallagher} S.~C., {Brandt} W.~N., {Chartas} G., {Priddey} R., {Garmire} G.~P.,
  {Sambruna} R.~M., 2006, \apj, 644, 709

\bibitem[{{Ganguly} {et~al}\mbox{.}(2001{\natexlab{a}}){Ganguly}, {Bond},
  {Charlton}, {Eracleous}, {Brandt}, \& {Churchill}}]{gan01a}
{Ganguly} R., {Bond} N.~A., {Charlton} J.~C., {Eracleous} M., {Brandt} W.~N.,
  {Churchill} C.~W., 2001{\natexlab{a}}, \apj, 549, 133

\bibitem[{{Ganguly} \& {Brotherton}(2008)}]{gb08}
{Ganguly} R., {Brotherton} M.~S., 2008, \apj, 672, 102

\bibitem[{{Ganguly} {et~al}\mbox{.}(2007){Ganguly}, {Brotherton}, {Cales},
  {Scoggins}, {Shang}, \& {Vestergaard}}]{gan07b}
{Ganguly} R., {Brotherton} M.~S., {Cales} S., {Scoggins} B., {Shang} Z.,
  {Vestergaard} M., 2007, \apj, 665, 990

\bibitem[{{Ganguly} {et~al}\mbox{.}(2001{\natexlab{b}}){Ganguly}, {Charlton},
  \& {Bond}}]{gan01b}
{Ganguly} R., {Charlton} J.~C., {Bond} N.~A., 2001{\natexlab{b}}, \apjl, 553,
  L101

\bibitem[{{Ganguly} {et~al}\mbox{.}(1999){Ganguly}, {Eracleous}, {Charlton}, \&
  {Churchill}}]{gan99}
{Ganguly} R., {Eracleous} M., {Charlton} J.~C., {Churchill} C.~W., 1999, \aj,
  117, 2594

\bibitem[{{Ganguly} {et~al}\mbox{.}(2003){Ganguly}, {Masiero}, {Charlton}, \&
  {Sembach}}]{gan03b}
{Ganguly} R., {Masiero} J., {Charlton} J.~C., {Sembach} K.~R., 2003, \apj, 598,
  922

\bibitem[{{Ganguly} {et~al}\mbox{.}(2004){Ganguly}, {Sembach}, {Charlton},
  {Eracleous}, {Palma}, \& {Tripp}}]{gan04}
{Ganguly} R., {Sembach} K., {Charlton} J., {Eracleous} M., {Palma} C., {Tripp}
  T., 2004, in Astronomical Society of the Pacific Conference Series, p. 243

\bibitem[{{Ganguly} {et~al}\mbox{.}(2006){Ganguly}, {Sembach}, {Tripp},
  {Savage}, \& {Wakker}}]{gan06a}
{Ganguly} R., {Sembach} K.~R., {Tripp} T.~M., {Savage} B.~D., {Wakker} B.~P.,
  2006, \apj, 645, 868

\bibitem[{{Gaskell}(1983)}]{gaskell83}
{Gaskell} C.~M., 1983, \apjl, 267, L1

\bibitem[{{Ghosh} \& {Punsly}(2007)}]{gp07}
{Ghosh} K.~K., {Punsly} B., 2007, ArXiv e-prints, 704

\bibitem[{{Giandoni} {et~al}\mbox{.}(2003){Giandoni}, {Kobulnicky},
  {Prochaska}, {Hwang}, \& {Kiminki}}]{gia03}
{Giandoni} S.~S., {Kobulnicky} H.~A., {Prochaska} J., {Hwang} S., {Kiminki}
  D.~C., 2003, American Astronomical Society Meeting Abstracts, 203

\bibitem[{{Goodrich}(1997)}]{goodrich97}
{Goodrich} R.~W., 1997, \apj, 474, 606

\bibitem[{{Goodrich} \& {Miller}(1995)}]{good95}
{Goodrich} R.~W., {Miller} J.~S., 1995, \apjl, 448, L73

\bibitem[{{Gregg} {et~al}\mbox{.}(2000){Gregg}, {Becker}, {Brotherton},
  {Laurent-Muehleisen}, {Lacy}, \& {White}}]{gregg00}
{Gregg} M.~D., {Becker} R.~H., {Brotherton} M.~S., {Laurent-Muehleisen} S.~A.,
  {Lacy} M., {White} R.~L., 2000, \apj, 544, 142

\bibitem[{{Gregg} {et~al}\mbox{.}(2006){Gregg}, {Becker}, \& {de
  Vries}}]{gbd06}
{Gregg} M.~D., {Becker} R.~H., {de Vries} W., 2006, \apj, 641, 210

\bibitem[{{Gregg} {et~al}\mbox{.}(1996){Gregg}, {Becker}, {White}, {Helfand},
  {McMahon}, \& {Hook}}]{gregg96}
{Gregg} M.~D., {Becker} R.~H., {White} R.~L., {Helfand} D.~J., {McMahon} R.~G.,
  {Hook} I.~M., 1996, \aj, 112, 407

\bibitem[{{Griffith} \& {Wright}(1993)}]{gw93}
{Griffith} M.~R., {Wright} A.~E., 1993, \aj, 105, 1666

\bibitem[{{Hall} {et~al}\mbox{.}(2002){Hall}, {Anderson}, {Strauss}, {York},
  {Richards}, {Fan}, {Knapp}, {Schneider}, {Vanden Berk}, {Geballe}, {Bauer},
  {Becker}, {Davis}, {Rix}, {Nichol}, {Bahcall}, {Brinkmann}, {Brunner},
  {Connolly}, {Csabai}, {Doi}, {Fukugita}, {Gunn}, {Haiman}, {Harvanek},
  {Heckman}, {Hennessy}, {Inada}, {Ivezi{\'c}}, {Johnston}, {Kleinman},
  {Krolik}, {Krzesinski}, {Kunszt}, {Lamb}, {Long}, {Lupton}, {Miknaitis},
  {Munn}, {Narayanan}, {Neilsen}, {Newman}, {Nitta}, {Okamura}, {Pentericci},
  {Pier}, {Schlegel}, {Snedden}, {Szalay}, {Thakar}, {Tsvetanov}, {White}, \&
  {Zheng}}]{hallai}
{Hall} P.~B. {et~al.}, 2002, \apjs, 141, 267

\bibitem[{{Hamann} {et~al}\mbox{.}(1997{\natexlab{a}}){Hamann}, {Barlow},
  {Cohen}, {Junkkarinen}, \& {Burbidge}}]{ham97c}
{Hamann} F., {Barlow} T., {Cohen} R.~D., {Junkkarinen} V., {Burbidge} E.~M.,
  1997{\natexlab{a}}, in ASP Conf. Ser. 128: Mass Ejection from Active Galactic
  Nuclei, {Arav} N., {Shlosman} I., {Weymann} R.~J., eds., ASP, San Francisco,
  p.~19

\bibitem[{{Hamann} {et~al}\mbox{.}(1997{\natexlab{b}}){Hamann}, {Barlow},
  {Cohen}, {Junkkarinen}, \& {Burbidge}}]{ham97d}
{Hamann} F., {Barlow} T.~A., {Cohen} R.~D., {Junkkarinen} V., {Burbidge} E.~M.,
  1997{\natexlab{b}}, in ASP Conf. Ser. 128: Mass Ejection from Active Galactic
  Nuclei, {Arav} N., {Shlosman} I., {Weymann} R.~J., eds., ASP, San Francisco,
  p. 187

\bibitem[{{Hamann} \& {Ferland}(1999)}]{hf99}
{Hamann} F., {Ferland} G., 1999, \araa, 37, 487

\bibitem[{{Hamann} {et~al}\mbox{.}(2011){Hamann}, {Kanekar}, {Prochaska},
  {Murphy}, {Ellison}, {Malec}, {Milutinovic}, \& {Ubachs}}]{ham11}
{Hamann} F., {Kanekar} N., {Prochaska} J.~X., {Murphy} M.~T., {Ellison} S.,
  {Malec} A.~L., {Milutinovic} N., {Ubachs} W., 2011, \mnras, 410, 1957

\bibitem[{{Hewitt} \& {Burbidge}(1989)}]{hb89}
{Hewitt} A., {Burbidge} G., 1989, \apjs, 69, 1

\bibitem[{{Hopkins} \& {Elvis}(2010)}]{he10}
{Hopkins} P.~F., {Elvis} M., 2010, \mnras, 401, 7

\bibitem[{{Hopkins} {et~al}\mbox{.}(2007){Hopkins}, {Hernquist}, {Cox},
  {Robertson}, \& {Krause}}]{hopkins07e}
{Hopkins} P.~F., {Hernquist} L., {Cox} T.~J., {Robertson} B., {Krause} E.,
  2007, \apj, 669, 67

\bibitem[{{Huchra} {et~al}\mbox{.}(1990){Huchra}, {Geller}, {Henry}, \&
  {Postman}}]{huchra90}
{Huchra} J.~P., {Geller} M.~J., {Henry} J.~P., {Postman} M., 1990, \apj, 365,
  66

\bibitem[{{Jannuzi} {et~al}\mbox{.}(1998){Jannuzi}, {Bahcall}, {Bergeron},
  {Boksenberg}, {Hartig}, {Kirhakos}, {Sargent}, {Savage}, {Schneider},
  {Turnshek}, {Weymann}, \& {Wolfe}}]{kpxiii}
{Jannuzi} B.~T. {et~al.}, 1998, \apjs, 118, 1

\bibitem[{{Jannuzi} {et~al}\mbox{.}(1996){Jannuzi}, {Hartig}, {Kirhakos},
  {Sargent}, {Turnshek}, {Weymann}, {Bahcall}, {Bergeron}, {Boksenberg},
  {Savage}, {Schneider}, \& {Wolfe}}]{kp96}
{Jannuzi} B.~T. {et~al.}, 1996, \apjl, 470, L11

\bibitem[{{Jauncey} {et~al}\mbox{.}(1978){Jauncey}, {Wright}, {Peterson}, \&
  {Condon}}]{jauncey78}
{Jauncey} D.~L., {Wright} A.~E., {Peterson} B.~A., {Condon} J.~J., 1978, \apjl,
  219, L1

\bibitem[{{Jenkins} {et~al}\mbox{.}(2003){Jenkins}, {Bowen}, {Tripp},
  {Sembach}, {Leighly}, {Halpern}, \& {Lauroesch}}]{jenkins03}
{Jenkins} E.~B., {Bowen} D.~V., {Tripp} T.~M., {Sembach} K.~R., {Leighly}
  K.~M., {Halpern} J.~P., {Lauroesch} J.~T., 2003, \aj, 125, 2824

\bibitem[{{Jones} {et~al}\mbox{.}(2005){Jones}, {Saunders}, {Read}, \&
  {Colless}}]{6dfdr2}
{Jones} D.~H., {Saunders} W., {Read} M.~A., {Colless} M., 2005, PASA, {in press
  (astro-ph/0505068)}

\bibitem[{{Kellermann} {et~al}\mbox{.}(1989){Kellermann}, {Sramek}, {Schmidt},
  {Shaffer}, \& {Green}}]{kel89}
{Kellermann} K.~I., {Sramek} R., {Schmidt} M., {Shaffer} D.~B., {Green} R.,
  1989, \aj, 98, 1195

\bibitem[{{Kim-Quijano} {et~al}\mbox{.}(2003){Kim-Quijano}, {Baum},
  {Leitherer}, {Voit}, {Keyes}, {Mobasher}, {Corbin}, {Hsu}, \& {Brown}}]{stis}
{Kim-Quijano} J. {et~al.}, 2003, {STIS Intrument Handbook, Version 7.0}. STScI,
  Baltimore

\bibitem[{{Konigl} \& {Kartje}(1994)}]{kk94}
{Konigl} A., {Kartje} J.~F., 1994, \apj, 434, 446

\bibitem[{{Korista} {et~al}\mbox{.}(1993){Korista}, {Voit}, {Morris}, \&
  {Weymann}}]{korista93}
{Korista} K.~T., {Voit} G.~M., {Morris} S.~L., {Weymann} R.~J., 1993, \apjs,
  88, 357

\bibitem[{{Kriss}(2002)}]{kriss02}
{Kriss} G.~A., 2002, in ASP Conf. Ser. 255: Mass Outflow in Active Galactic
  Nuclei: New Perspectives, {Crenshaw} D.~M., {Kraemer} S.~B., {George} I.~M.,
  eds., ASP, San Francisco, p.~69

\bibitem[{{Krolik} \& {Kriss}(2001)}]{kk01}
{Krolik} J.~H., {Kriss} G.~A., 2001, \apj, 561, 684

\bibitem[{{Kuraszkiewicz} \& {Green}(2002)}]{kg02}
{Kuraszkiewicz} J.~K., {Green} P.~J., 2002, \apjl, 581, L77

\bibitem[{{Kurosawa} \& {Proga}(2009)}]{kp09}
{Kurosawa} R., {Proga} D., 2009, \apj, 693, 1929

\bibitem[{{Lehner} {et~al}\mbox{.}(2009){Lehner}, {Prochaska}, {Kobulnicky},
  {Cooksey}, {Howk}, {Williger}, \& {Cales}}]{lehner09}
{Lehner} N., {Prochaska} J.~X., {Kobulnicky} H.~A., {Cooksey} K.~L., {Howk}
  J.~C., {Williger} G.~M., {Cales} S.~L., 2009, \apj, 694, 734

\bibitem[{{Lupton} {et~al}\mbox{.}(1999){Lupton}, {Gunn}, \& {Szalay}}]{ugriz2}
{Lupton} R.~H., {Gunn} J.~E., {Szalay} A.~S., 1999, \aj, 118, 1406

\bibitem[{{Marziani} {et~al}\mbox{.}(1996){Marziani}, {Sulentic},
  {Dultzin-Hacyan}, {Calvani}, \& {Moles}}]{mar96}
{Marziani} P., {Sulentic} J.~W., {Dultzin-Hacyan} D., {Calvani} M., {Moles} M.,
  1996, \apjs, 104, 37

\bibitem[{{Mathur} {et~al}\mbox{.}(1995){Mathur}, {Elvis}, \& {Wilkes}}]{mew95}
{Mathur} S., {Elvis} M., {Wilkes} B., 1995, \apj, 452, 230

\bibitem[{{McKernan} {et~al}\mbox{.}(2007){McKernan}, {Yaqoob}, \&
  {Reynolds}}]{myr07}
{McKernan} B., {Yaqoob} T., {Reynolds} C.~S., 2007, \mnras, 379, 1359

\bibitem[{{Miller} {et~al}\mbox{.}(1990){Miller}, {Peacock}, \& {Mead}}]{mpm90}
{Miller} L., {Peacock} J.~A., {Mead} A.~R.~G., 1990, \mnras, 244, 207

\bibitem[{{Milne}(1926)}]{milne26}
{Milne} E.~A., 1926, \mnras, 86, 459

\bibitem[{{Misawa} {et~al}\mbox{.}(2007){Misawa}, {Charlton}, {Eracleous},
  {Ganguly}, {Tytler}, {Kirkman}, {Suzuki}, \& {Lubin}}]{misawa07}
{Misawa} T., {Charlton} J.~C., {Eracleous} M., {Ganguly} R., {Tytler} D.,
  {Kirkman} D., {Suzuki} N., {Lubin} D., 2007, \apjs, 171, 1

\bibitem[{{Misawa} {et~al}\mbox{.}(2005){Misawa}, {Eracleous}, {Charlton}, \&
  {Tajitsu}}]{misawa05}
{Misawa} T., {Eracleous} M., {Charlton} J.~C., {Tajitsu} A., 2005, \apj, 629,
  115

\bibitem[\protect\citeauthoryear{Moe et al.}{2009}]{moe09}
Moe M., Arav N., Bautista M.~A., Korista K.~T., 2009, ApJ, 706, 525

\bibitem[{{M{\o}ller} \& {Jakobsen}(1987)}]{mj87}
{M{\o}ller} P., {Jakobsen} P., 1987, \apjl, 320, L75

\bibitem[{{Monier} {et~al}\mbox{.}(2001){Monier}, {Mathur}, {Wilkes}, \&
  {Elvis}}]{mon01}
{Monier} E.~M., {Mathur} S., {Wilkes} B., {Elvis} M., 2001, \apj, 559, 675

\bibitem[{{Morton}(2003)}]{morton03}
{Morton} D.~C., 2003, \apjs, 149, 205

\bibitem[\protect\citeauthoryear{Morris et al.}{1986}]{mor86}
Morris S.~L., Weymann R.~J., Foltz C.~B., Turnshek D.~A., Shectman S.,
Price C., Boroson T.~A., 1986, ApJ, 310, 40

\bibitem[{{Murray} \& {Chiang}(1997)}]{mur97}
{Murray} N., {Chiang} J., 1997, \apj, 474, 91

\bibitem[{{Murray} {et~al}\mbox{.}(1995){Murray}, {Chiang}, {Grossman}, \&
  {Voit}}]{mur95}
{Murray} N., {Chiang} J., {Grossman} S.~A., {Voit} G.~M., 1995, \apj, 451, 498

\bibitem[{{Oke} \& {Schild}(1970)}]{os70}
{Oke} J.~B., {Schild} R.~E., 1970, \apj, 161, 1015

\bibitem[{{Onken} \& {Peterson}(2002)}]{op02}
{Onken} C.~A., {Peterson} B.~M., 2002, \apj, 572, 746

\bibitem[{{Piconcelli} {et~al}\mbox{.}(2005){Piconcelli}, {Jimenez-Bail{\'o}n},
  {Guainazzi}, {Schartel}, {Rodr{\'{\i}}guez-Pascual}, \&
  {Santos-Lle{\'o}}}]{piconcelli05}
{Piconcelli} E., {Jimenez-Bail{\'o}n} E., {Guainazzi} M., {Schartel} N.,
  {Rodr{\'{\i}}guez-Pascual} P.~M., {Santos-Lle{\'o}} M., 2005, \aap, 432, 15

\bibitem[{{Proga} \& {Kallman}(2004)}]{pk04}
{Proga} D., {Kallman} T.~R., 2004, \apj, 616, 688

\bibitem[{{Proga} {et~al}\mbox{.}(2000){Proga}, {Stone}, \& {Kallman}}]{psk00}
{Proga} D., {Stone} J.~M., {Kallman} T.~R., 2000, \apj, 543, 686

\bibitem[{{Reimers} {et~al}\mbox{.}(1998){Reimers}, {Hagen},
  {Rodriguez-Pascual}, \& {Wisotzki}}]{rei98}
{Reimers} D., {Hagen} H.-J., {Rodriguez-Pascual} P., {Wisotzki} L., 1998, \aap,
  334, 96

\bibitem[{{Richards}(2001)}]{rich01b}
{Richards} G.~T., 2001, \apjs, 133, 53

\bibitem[{{Richards} {et~al}\mbox{.}(2002{\natexlab{a}}){Richards}, {Gregg},
  {Becker}, \& {White}}]{rich02}
{Richards} G.~T., {Gregg} M.~D., {Becker} R.~H., {White} R.~L.,
  2002{\natexlab{a}}, \apjl, 567, L13

\bibitem[{{Richards} {et~al}\mbox{.}(2001){Richards}, {Laurent-Muehleisen},
  {Becker}, \& {York}}]{rich01a}
{Richards} G.~T., {Laurent-Muehleisen} S.~A., {Becker} R.~H., {York} D.~G.,
  2001, \apj, 547, 635

\bibitem[{{Richards} {et~al}\mbox{.}(2002{\natexlab{b}}){Richards}, {Vanden
  Berk}, {Reichard}, {Hall}, {Schneider}, {SubbaRao}, {Thakar}, \&
  {York}}]{rich02b}
{Richards} G.~T., {Vanden Berk} D.~E., {Reichard} T.~A., {Hall} P.~B.,
  {Schneider} D.~P., {SubbaRao} M., {Thakar} A.~R., {York} D.~G.,
  2002{\natexlab{b}}, \aj, 124, 1

\bibitem[{{Richards} {et~al}\mbox{.}(1999){Richards}, {York}, {Yanny},
  {Kollgaard}, {Laurent--Muehleisen}, \& {vanden Berk}}]{rich99}
{Richards} G.~T., {York} D.~G., {Yanny} B., {Kollgaard} R.~I.,
  {Laurent--Muehleisen} S.~A., {vanden Berk} D.~E., 1999, \apj, 513, 576

\bibitem[{{Rodr{\'{\i}}guez Hidalgo} {et~al}\mbox{.}(2011){Rodr{\'{\i}}guez
  Hidalgo}, {Hamann}, \& {Hall}}]{rhh11}
{Rodr{\'{\i}}guez Hidalgo} P., {Hamann} F., {Hall} P., 2011, \mnras, 411, 247

\bibitem[{{Rosenberg} {et~al}\mbox{.}(2003){Rosenberg}, {Ganguly}, {Giroux}, \&
  {Stocke}}]{ros03}
{Rosenberg} J.~L., {Ganguly} R., {Giroux} M.~L., {Stocke} J.~T., 2003, \apj,
  591, 677

\bibitem[{{Sargent} {et~al}\mbox{.}(1988){Sargent}, {Steidel}, \&
  {Boksenberg}}]{sbs88}
{Sargent} W.~L.~W., {Steidel} C.~C., {Boksenberg} A., 1988, \apjs, 68, 539

\bibitem[{{Savage} {et~al}\mbox{.}(2003){Savage}, {Sembach}, {Wakker},
  {Richter}, {Meade}, {Jenkins}, {Shull}, {Moos}, \& {Sonneborn}}]{savage03}
{Savage} B.~D. {et~al.}, 2003, \apjs, 146, 125

\bibitem[{{Scannapieco} \& {Oh}(2004)}]{so04}
{Scannapieco} E., {Oh} S.~P., 2004, \apj, 608, 62

\bibitem[{{Scargle}(1973)}]{scargle73}
{Scargle} J.~D., 1973, \apj, 179, 705

\bibitem[{{Schmidt} \& {Green}(1983)}]{sg83}
{Schmidt} M., {Green} R.~F., 1983, \apj, 269, 352

\bibitem[{{Schneider} {et~al}\mbox{.}(2007){Schneider}, {Hall}, {Richards},
  {Strauss}, {Vanden Berk}, {Anderson}, {Brandt}, {Fan}, {Jester}, {Gray},
  {Gunn}, {SubbaRao}, {Thakar}, {Stoughton}, {Szalay}, {Yanny}, {York},
  {Bahcall}, {Barentine}, {Blanton}, {Brewington}, {Brinkmann}, {Brunner},
  {Castander}, {Csabai}, {Frieman}, {Fukugita}, {Harvanek}, {Hogg},
  {Ivezi{\'c}}, {Kent}, {Kleinman}, {Knapp}, {Kron}, {Krzesi{\'n}ski}, {Long},
  {Lupton}, {Nitta}, {Pier}, {Saxe}, {Shen}, {Snedden}, {Weinberg}, \&
  {Wu}}]{sdssqso5}
{Schneider} D.~P. {et~al.}, 2007, \aj, 134, 102

\bibitem[{{Schneider} {et~al}\mbox{.}(1993){Schneider}, {Hartig}, {Jannuzi},
  {Kirhakos}, {Saxe}, {Weymann}, {Bahcall}, {Bergeron}, {Boksenberg},
  {Sargent}, {Savage}, {Turnshek}, \& {Wolfe}}]{kpii}
{Schneider} D.~P. {et~al.}, 1993, \apjs, 87, 45

\bibitem[{{Scott} {et~al}\mbox{.}(2004){Scott}, {Kriss}, {Brotherton}, {Green},
  {Hutchings}, {Shull}, \& {Zheng}}]{scott04b}
{Scott} J.~E., {Kriss} G.~A., {Brotherton} M., {Green} R.~F., {Hutchings} J.,
  {Shull} J.~M., {Zheng} W., 2004, \apj, 615, 135

\bibitem[{{Sembach} \& {Savage}(1992)}]{ss92}
{Sembach} K.~R., {Savage} B.~D., 1992, \apjs, 83, 147

\bibitem[{{Sembach} {et~al}\mbox{.}(2004){Sembach}, {Tripp}, {Savage}, \&
  {Richter}}]{sembach04b}
{Sembach} K.~R., {Tripp} T.~M., {Savage} B.~D., {Richter} P., 2004, \apjs, 155,
  351

\bibitem[{{Shin} {et~al}\mbox{.}(2010){Shin}, {Ostriker}, \& {Ciotti}}]{soc10}
{Shin} M.-S., {Ostriker} J.~P., {Ciotti} L., 2010, \apj, 711, 268

\bibitem[{{Sramek} \& {Weedman}(1980)}]{sw80}
{Sramek} R.~A., {Weedman} D.~W., 1980, \apj, 238, 435

\bibitem[{{Steffen} {et~al}\mbox{.}(2006){Steffen}, {Strateva}, {Brandt},
  {Alexander}, {Koekemoer}, {Lehmer}, {Schneider}, \& {Vignali}}]{steffen06}
{Steffen} A.~T., {Strateva} I., {Brandt} W.~N., {Alexander} D.~M., {Koekemoer}
  A.~M., {Lehmer} B.~D., {Schneider} D.~P., {Vignali} C., 2006, \aj, 131, 2826

\bibitem[{{Stocke} {et~al}\mbox{.}(1992){Stocke}, {Morris}, {Weymann}, \&
  {Foltz}}]{stocke92}
{Stocke} J.~T., {Morris} S.~L., {Weymann} R.~J., {Foltz} C.~B., 1992, \apj,
  396, 487

\bibitem[{{Strateva} {et~al}\mbox{.}(2005){Strateva}, {Brandt}, {Schneider},
  {Vanden Berk}, \& {Vignali}}]{strateva05}
{Strateva} I.~V., {Brandt} W.~N., {Schneider} D.~P., {Vanden Berk} D.~G.,
  {Vignali} C., 2005, \aj, 130, 387

\bibitem[{{Strauss} {et~al}\mbox{.}(1992){Strauss}, {Huchra}, {Davis}, {Yahil},
  {Fisher}, \& {Tonry}}]{strauss92}
{Strauss} M.~A., {Huchra} J.~P., {Davis} M., {Yahil} A., {Fisher} K.~B.,
  {Tonry} J., 1992, \apjs, 83, 29

\bibitem[{{Tananbaum} {et~al}\mbox{.}(1986){Tananbaum}, {Avni}, {Green},
  {Schmidt}, \& {Zamorani}}]{tea}
{Tananbaum} H., {Avni} Y., {Green} R.~F., {Schmidt} M., {Zamorani} G., 1986,
  \apj, 305, 57

\bibitem[{{Tripp} {et~al}\mbox{.}(2005){Tripp}, {Jenkins}, {Bowen},
  {Prochaska}, {Aracil}, \& {Ganguly}}]{tripp05}
{Tripp} T.~M., {Jenkins} E.~B., {Bowen} D.~V., {Prochaska} J.~X., {Aracil} B.,
  {Ganguly} R., 2005, \apj, 619, 714

\bibitem[{{Tripp} {et~al}\mbox{.}(1998){Tripp}, {Lu}, \& {Savage}}]{tls98}
{Tripp} T.~M., {Lu} L., {Savage} B.~D., 1998, \apj, 508, 200

\bibitem[\protect\citeauthoryear{Tripp, Lu, \& Savage}{1996}]{tls96}
Tripp T.~M., Lu L., Savage B.~D., 1996, ApJS, 102, 239

\bibitem[{{Tripp} {et~al}\mbox{.}(2011){Tripp}, {Meiring}, {Prochaska},
  {Willmer}, {Howk}, {Werk}, {Jenkins}, {Bowen}, {Lehner}, {Sembach}, {Thom},
  \& {Tumlinson}}]{tripp11}
{Tripp} T.~M. {et~al.}, 2011, Science, 334, 952

\bibitem[{{Tripp} {et~al}\mbox{.}(2008){Tripp}, {Sembach}, {Bowen}, {Savage},
  {Jenkins}, {Lehner}, \& {Richter}}]{tripp08}
{Tripp} T.~M., {Sembach} K.~R., {Bowen} D.~V., {Savage} B.~D., {Jenkins} E.~B.,
  {Lehner} N., {Richter} P., 2008, \apjs, 177, 39

\bibitem[{{Turnshek}(1988)}]{turn88b}
{Turnshek} D.~A., 1988, in QSO Absorption Lines; Probing the Universe;
  Proceedings of the QSO Absorption Line Meeting, {Blades} J.~C., {Turnshek}
  D.~A., {Norman} C.~A., eds., Cambridge University Press, Cambridge, pp.
  17--46

\bibitem[{{Turnshek} {et~al}\mbox{.}(1988){Turnshek}, {Grillmair}, {Foltz}, \&
  {Weymann}}]{turn88}
{Turnshek} D.~A., {Grillmair} C.~J., {Foltz} C.~B., {Weymann} R.~J., 1988,
  \apj, 325, 651

\bibitem[{{Tytler} \& {Fan}(1992)}]{tf92}
{Tytler} D., {Fan} X.-M., 1992, \apjs, 79, 1

\bibitem[{{Vanden Berk} {et~al}\mbox{.}(2001){Vanden Berk}, {Richards},
  {Bauer}, {Strauss}, {Schneider}, {Heckman}, {York}, {Hall}, {Fan}, {Knapp},
  {Anderson}, {Annis}, {Bahcall}, {Bernardi}, {Briggs}, {Brinkmann}, {Brunner},
  {Burles}, {Carey}, {Castander}, {Connolly}, {Crocker}, {Csabai}, {Doi},
  {Finkbeiner}, {Friedman}, {Frieman}, {Fukugita}, {Gunn}, {Hennessy},
  {Ivezi{\'c}}, {Kent}, {Kunszt}, {Lamb}, {Leger}, {Long}, {Loveday}, {Lupton},
  {Meiksin}, {Merelli}, {Munn}, {Newberg}, {Newcomb}, {Nichol}, {Owen}, {Pier},
  {Pope}, {Rockosi}, {Schlegel}, {Siegmund}, {Smee}, {Snir}, {Stoughton},
  {Stubbs}, {SubbaRao}, {Szalay}, {Szokoly}, {Tremonti}, {Uomoto}, {Waddell},
  {Yanny}, \& {Zheng}}]{vdb01}
{Vanden Berk} D.~E. {et~al.}, 2001, \aj, 122, 549

\bibitem[{{Vestergaard}(2003)}]{ves03}
{Vestergaard} M., 2003, \apj, 599, 116

\bibitem[{{Weymann} {et~al}\mbox{.}(1991){Weymann}, {Morris}, {Foltz}, \&
  {Hewett}}]{weymann91}
{Weymann} R.~J., {Morris} S.~L., {Foltz} C.~B., {Hewett} P.~C., 1991, \apj,
  373, 23

\bibitem[{{Weymann} {et~al}\mbox{.}(1985){Weymann}, {Turnshek}, \&
  {Christiansen}}]{weymann85}
{Weymann} R.~J., {Turnshek} D.~A., {Christiansen} W.~A., 1985, in Astrophysics
  of Active Galaxies and Quasi-Stellar Objects, {Miller} J.~S., ed., pp.
  333--365

\bibitem[{{Weymann} {et~al}\mbox{.}(1979){Weymann}, {Williams}, {Peterson}, \&
  {Turnshek}}]{wwpt}
{Weymann} R.~J., {Williams} R.~E., {Peterson} B.~M., {Turnshek} D.~A., 1979,
  \apj, 234, 33

\bibitem[{{White} {et~al}\mbox{.}(2000){White}, {Becker}, {Gregg},
  {Laurent-Muehleisen}, {Brotherton}, {Impey}, {Petry}, {Foltz}, {Chaffee},
  {Richards}, {Oegerle}, {Helfand}, {McMahon}, \& {Cabanela}}]{second}
{White} R.~L. {et~al.}, 2000, \apjs, 126, 133

\bibitem[{{Wilkes}(1986)}]{wilkes86}
{Wilkes} B.~J., 1986, \mnras, 218, 331

\bibitem[{{Wills} {et~al}\mbox{.}(1999){Wills}, {Brandt}, \& {Laor}}]{wbl99}
{Wills} B.~J., {Brandt} W.~N., {Laor} A., 1999, \apjl, 520, L91

\bibitem[{{Wisotzki} {et~al}\mbox{.}(2000){Wisotzki}, {Christlieb}, {Bade},
  {Beckmann}, {Koehler}, {Vanelle}, \& {Reimers}}]{wis00}
{Wisotzki} L., {Christlieb} N., {Bade} N., {Beckmann} V., {Koehler} T.,
  {Vanelle} C., {Reimers} D., 2000, VizieR Online Data Catalog, 335, 80077

\bibitem[{{Wright} \& {Otrupcek}(1990)}]{pkscat}
{Wright} A., {Otrupcek} R., 1990, in PKS Catalog (1990), p.~0

\bibitem[{{Yuan} {et~al}\mbox{.}(2002){Yuan}, {Green}, {Brotherton}, {Tripp},
  {Kaiser}, \& {Kriss}}]{yuan02}
{Yuan} Q., {Green} R.~F., {Brotherton} M., {Tripp} T.~M., {Kaiser} M.~E.,
  {Kriss} G.~A., 2002, \apj, 575, 687

\end{thebibliography}


\appendix

\section{Notes on the Survey}

In Table~\ref{tab:parcov}, we list absorption-line systems that were detected along with our assessments for whether the \ovi$\lambda\lambda$1031.926, 1037.617, {\nv} $\lambda\lambda$1239.821, 1242.804, {\siliv} $\lambda\lambda$1393.760, 1402.773, and {\civ} $\lambda\lambda$1548.204, 1550.781 doublets show evidence for partial coverage. An `a' is appended to the redshift if it appears within 5000\,\kms\ of the quasar redshift listed in Table~\ref{tab:sample}. See \S\ref{sec:pcsys} for a description of our classification scheme.

\begin{table*}
\vspace{-4ex}
\begin{minipage}{170mm}
\caption{Partial Coverage Tests For Absorption-Line Systems}
\label{tab:parcov}
\begin{tabular}{llccccllcccc}
\hline\hline
{Target} & {Redshift$^\mathrm{a}$} & {O\,{\sc vi}} & {N\,{\sc v}} & {Si\,{\sc iv}} & {C\,{\sc iv}} & {Target} & {Redshift$^\mathrm{a}$} & {O\,{\sc vi}} & {N\,{\sc v}} & {Si\,{\sc iv}} & {C\,{\sc iv}} \\ \hline
PG 0117+213     & 0.5763  & \nodata & \nodata & \nodata & CON                & PG 1211+143     & 0.0511  & \nodata & ND      & CON     & CON     \\
                & 1.2541  & CNE     & ND      & \nodata & \nodata            &                 & 0.0623  & \nodata & ND      & ND      & CNE     \\
                & 1.3250  & CNE     & CNE     & \nodata & \nodata            & PG 1216+069     & \multicolumn{5}{c}{No High-Ionization Metal-line Doublets} \\
                & 1.3385  & CNE     & ND      & \nodata & \nodata            & Mark 205        & 0.0047  & \nodata & ND      & CNE     & CNE \\
                & 1.3427  & CON     & CNE     & \nodata & \nodata            & 3C273           & \multicolumn{5}{c}{No High-Ionization Metal-line Doublets} \\
                & 1.4241  & CON     & CNE     & \nodata & \nodata            & RX J1230.8-0115 & 0.0057  & \nodata & CON     & CON     & CON     \\
                & 1.4464  & POS     & ND      & \nodata & \nodata            &                 & 0.1000a & \nodata & PC      & ND      & PC      \\
                & 1.4477  & CNE     & CNE     & \nodata & \nodata            &                 & 0.1058a & \nodata & LL      & ND      & PC      \\
                & 1.4949a & LL      & ND      & \nodata & \nodata            &                 & 0.1093a & \nodata & LL      & ND      & CON     \\
                & 1.5090a & LL      & ND      & \nodata & \nodata            &                 & 0.1172a & \nodata & CON     & ND      & \nodata \\
Ton S210        & \multicolumn{5}{c}{No High-Ionization Metal-line Doublets} & PG 1241+176     & 0.5507  & \nodata & \nodata & \nodata & CON     \\
HE 0226-4110    & 0.2070  & CON     & \nodata & \nodata & \nodata            &                 & 0.5584  & \nodata & \nodata & \nodata & CNE     \\
                & 0.4925a & CNE     & \nodata & \nodata & \nodata            &                 & 1.2717a & CON     & ND      & \nodata & \nodata \\
PKS 0232-04     & 0.7390  & \nodata & \nodata & CNE     & CNE                & PG 1248+401     & 0.5648  & \nodata & \nodata & \nodata & CON     \\
                & 0.8078  & \nodata & \nodata & CNE     & CNE                &                 & 0.7011  & \nodata & \nodata & ND      & CNE     \\
                & 0.8674  & \nodata & ND      & ND      & CNE                &                 & 0.7730  & \nodata & \nodata & CON     & CON     \\
                & 1.3561  & CON     & ND      & \nodata & \nodata            &                 & 0.8548  & \nodata & CNE     & ND      & CON     \\
PKS 0312-77     & 0.2017  & CNE     & CNE     & CNE     & \nodata            & PG 1259+593     & 0.3197  & ND      & CNE     & \nodata & \nodata \\  
                & 0.2025  & CON     & POS     & CNE     & \nodata            & PKS 1302-102    & 0.2256  & CNE     & CNE     & ND      & \nodata \\  
PKS 0405-123    & 0.1671  & CNE     & CNE     & POS     & \nodata            & CSO 873         & 0.6611  & \nodata & \nodata & ND      & CON     \\
                & 0.4951  & CON     & \nodata & \nodata & \nodata            & PG 1444+407     & 0.2673a & CON     & ND      & \nodata & \nodata \\
PKS 0454-22     & 0.3318  & \nodata & \nodata & CON     & ND                 & PG 1630+377     & 0.9143  & \nodata & CNE     & CNE     & CNE     \\
                & 0.3815  & \nodata & \nodata & ND      & CON                &                 & 0.9528  & \nodata & ND      & CNE     & CNE     \\
                & 0.4023  & \nodata & \nodata & ND      & CNE                &                 & 1.0961  & \nodata & ND      & CNE     & \nodata \\
                & 0.4744  & \nodata & \nodata & CON     & CON                &                 & 1.1749  & \nodata & ND      & CNE     & \nodata \\
                & 0.4833  & \nodata & \nodata & CNE     & CON                &                 & 1.3243  & CON     & ND      & \nodata & \nodata \\
HE 0515-4414    & 0.9406  & \nodata & ND      & CNE     & CON                &                 & 1.4333a & CNE     & ND      & \nodata & \nodata \\
                & 1.1470  & \nodata & ND      & CNE     & \nodata            & PG 1634+706     & 0.2122  & \nodata & \nodata & \nodata & CNE     \\
                & 1.3858  & CNE     & ND      & \nodata & \nodata            &                 & 0.6536  & \nodata & CNE     & CON     & CON     \\
                & 1.6020  & CON     & \nodata & \nodata & \nodata            &                 & 0.8181  & ND      & ND      & CNE     & CON     \\
                & 1.6736a & CNE     & \nodata & \nodata & \nodata            &                 & 0.9056  & CNE     & CNE     & CNE     & CON     \\
                & 1.6971a & CNE     & \nodata & \nodata & \nodata            &                 & 0.9904  & ND      & ND      & CON     & CON     \\
HS 0624+6907    & 0.0635  & \nodata & ND      & CNE     & CON                &                 & 1.0414  & CON     & ND      & CON     & \nodata \\
                & 0.0757  & \nodata & ND      & CNE     & CNE                &                 & 1.0876  & CNE     & CNE     & ND      & \nodata \\
HS 0747+4259    & 0.5154  & \nodata & \nodata & \nodata & CON                &                 & 1.3415a & POS     & ND      & \nodata & \nodata \\
                & 0.6368  & \nodata & \nodata & ND      & CON                & 3C351.0$^{b}$   & 0.2211  & CNE     & ND      & CNE     & \nodata \\
                & 0.7150  & \nodata & \nodata & ND      & CNE                &                 & 0.2857  & CNE     & ND      & \nodata & \nodata \\
                & 1.5952  & CON     & \nodata & \nodata & \nodata            &                 & 0.3166  & POS     & ND      & \nodata & \nodata \\
                & 1.6131  & CNE     & \nodata & \nodata & \nodata            &                 & 0.3619a & PC      & ND      & \nodata & \nodata \\
HS 0810+2554    & 0.8313  & \nodata & ND      & CNE     & CON                &                 & 0.3631a & CON     & LL      & \nodata & \nodata \\
                & 1.2580  & CNE     & ND      & \nodata & \nodata            &                 & 0.3645a & LL      & PC      & \nodata & \nodata \\
                & 1.3524  & CNE     & ND      & \nodata & \nodata            &                 & 0.3673a & PC      & LL      & \nodata & \nodata \\
                & 1.4862a & CON     & \nodata & \nodata & \nodata            &                 & 0.3719a & LL      & CON     & \nodata & \nodata \\
                & 1.4948a & CON     & \nodata & \nodata & \nodata            & PG 1718+481     & 0.3448  & \nodata & \nodata & ND      & CNE     \\
                & 1.5025a & CON     & \nodata & \nodata & \nodata            &                 & 0.8665  & CNE     & ND      & ND      & \nodata \\
PG 0953+415     & 0.06806 & \nodata & CNE     & ND      & CNE                &                 & 0.8928  & CNE     & ND      & ND      & \nodata \\
                & 0.2335a & CON     & ND      & ND      & \nodata            &                 & 1.0065  & POS     & ND      & \nodata & \nodata \\
Mrk 132         & 1.7322  & CNE     & \nodata & \nodata & \nodata            &                 & 1.0318  & CON     & CON     & \nodata & \nodata \\
Ton 28          & 0.3302a & CON     & ND      & \nodata & \nodata            &                 & 1.0548a & CON     & ND      & \nodata & \nodata \\
3C 249.1        & 0.3080a & CNE     & ND      & \nodata & \nodata            &                 & 1.0874a & POS     & ND      & \nodata & \nodata \\
                & 0.3136a & CNE     & CNE     & \nodata & \nodata            & H 1821+643      & 0.2133  & CNE     & ND      & ND      & \nodata \\
PG 1116+215     & 0.1385  & CNE     & ND      & CNE     & \nodata            &                 & 0.2249  & CON     & ND      & ND      & \nodata \\
                & 0.1655a & CON     & ND      & ND      & \nodata            &                 & 0.2453  & CON     & CNE     & \nodata & \nodata \\
PKS 1127-145    & \multicolumn{5}{c}{No High-Ionization Metal-line Doublets} &                 & 0.2665  & CON     & ND      & \nodata & \nodata \\
PG 1206+459     & 0.7338  & \nodata & \nodata & ND      & CNE                &                 & 0.2967a & POS     & ND      & \nodata & \nodata \\
                & 0.9254  & \nodata & CON     & CNE     & CNE                & PHL 1811        & 0.0809  & \nodata & ND      & CON     & CON     \\
                & 0.9276  & \nodata & CON     & CON     & CON                &                 & 0.1765  & CNE     & ND      & \nodata & \nodata \\
                & 0.9343  & \nodata & ND      & CON     & CON                &                 & 0.1919a & CON     & CNE     & ND      & \nodata \\
                & 1.0280  & \nodata & PC      & ND?     & \nodata            & PKS 2155-304    & \multicolumn{5}{c}{No High-Ionization Metal-line Doublets} \\ \hline
\end{tabular}
Comments: {CNE: Cannot Evaluate; CON: Consistent With 1:2 Doublet Ratio; LL: Line-Locked Pair; ND: Doublet Not Detected; POS: Possible Partial Coverage; PC: Partial Coverage Likely}\\
$^\mathrm{a}$: {An 'a' next to the absorption-line redshift marks systems that appears within the 5000\,\kms\ of the emission redshift listed in Table~\ref{tab:sample}.}\\
$^\mathrm{b}$: {For the associated absorbers in the spectrum of 3C\,351.0, we use the measurements from \citet[][Table 3]{yuan02}. Specifically, we adopt their values for velocities and coverage fractions for components A, D, F, I, and M.}
\end{minipage}
\end{table*}

\clearpage

\section{The Sample}
\label{sec:sample}

In this section, we describe the systems in our systematic search that either lie within 5000\,\kms\ of the quasar redshift listed in Table~\ref{tab:sample} or exhibit the signature of partial coverage. In some cases, these systems have been studied in other works. In such cases, we point the reader to the appropriate references.

{\noindent \bf PG\,0117+213 ($\zabs$=1.4949, 1.5090) --} There are two associated absorption line systems in the sight line toward this quasar at $\zabs=1.4949$\ and $\zabs=1.5090$. In Figure~\ref{fig:pg0117}, we present plots of detected transitions for these two systems.
In each of these systems, we detect transitions from the {\hi} Lyman series, {{\ciii} $\lambda$977, and the {\ovi} $\lambda\lambda$1031.926,1037.617 doublet. The velocity separation of the absorption line systems is such that the {\ovi} doublets are apparently line-locked. That is, the 1031.926\,\AA\ and 1037.617\,\AA\ transitions of the $\zabs=1.4949$\ system are respectively blended with the Ly\,$\beta$\ and 1031.926\,\AA\ transitions of the $\zabs=1.5090$\ system. The various blends (both between transitions in these two systems as well as with transitions from systems at other redshifts) makes it difficult to assess
whether the {\ovi} doublet exhibits partial coverage. We note, however, that the strong Lyman lines of {\hi}\ reach zero flux levels indicating that no dilution exists.


{\noindent \bf HE 0226-4110 ($\zabs$=0.4925) --} As presented and discussed by \citet{gan06a}, this absorption-line system is detected in four adjacent ionization stages of oxygen - {\oiii} $\lambda$832.927, {\oiv} $\lambda$787.711, {\ov} $\lambda$629.720, and {\ovi} $\lambda\lambda$1031.926, 1037.617 - as well as the {\hi} Lyman series, {{\ciii} $\lambda$977.020, {\niv} $\lambda$765.148, {\suliv} $\lambda$748.400, and {\neviii}  $\lambda$770.409. The {\hi} Ly\,$\beta$, {{\ciii} $\lambda$977.020, and {\ov} $\lambda$629.730 transitions reach zero flux levels, indicating no unocculted flux. Likewise, modulo a blend in the {\ovi} $\lambda$1037.617 with Galactic {\civ} $\lambda$1548.204, the doublet is consistent with a 1:2 true optical depth ratio.


{\noindent \bf PKS 0312-77 ($\zabs$=0.2029) --} As shown in Figure~\ref{fig:pks0312sys},
an absorption-line system is detected in the E140M spectrum of this quasar over a very wide range of ionization species, from {\hi}\ to \ovi. Both {\nv}\ and \ovi\ show smooth profiles possibly indicative of an outflow origin. The results of the coverage fraction test on the \ovi, {\nv}, and {\siliv} doublets is presented in Figure~\ref{fig:pks0312pc}. Neither \ovi, nor {\siliv} doublet shows signs of trough-dilution to a high confidence. That is, the derived coverage fractions are consistent with unity at the $3\sigma$\ level. The {\siliv} doublet resides in a low signal-to-noise region of the spectrum and it is difficult to evaluate the presence/absence of partial coverage. Hence, we assign a ``Cannot Evaluate'' classification to the {\siliv} doublet, but  a ``Consistent'' classification to the {\ovi} doublet. However, the {\nv} doublet, though weak, appears to be inconsistent with full coverage in most velocity bins to greater than $3\sigma$\ confidence, even accounting for continuum placement uncertainties \citep{ss92}. \citet{gan04}\ speculated that this absorber may only occult the compact continuum source and that the dilution of the {\nv} profiles arises due to Ly$\alpha$\ flux from the broad emission line region. However, \citet{gia03} and \citet{lehner09} report on galaxy redshifts in the field and find several galaxies at the same redshift of the absorber. In addition, the lower-ionization and neutral species show saturated troughs that reach zero flux levels implying no partial coverage in these species. Together, these imply that indeed the absorber is unlikely to be related to the quasar, or quasar host.


{\noindent \bf HE\,0515-4414 ($\zabs$=1.6736, 1.6971) --} There is one securely-identified associated absorption line system toward HE\,0515-4414, at $\zabs=1.69736$, that is detected in the {\hi} Lyman series, {\niii} $\lambda$989.799, and the {\ovi} $\lambda\lambda$1031.926,1037.617 doublet. These are shown in left panels of Figure~\ref{fig:he0515}. The {\hi} Ly\,$\beta$\ and Ly\,$\gamma$\ profiles reach zero flux
levels, indicating that the neutral species fully occult the background source. ({\hi} Ly\,$\gamma$\ appears to be blended with absorption unrelated to the absorption-line system.) The {\ovi} $\lambda$1037.617 profile is blended, making it difficult to examine potential partial coverage. Another possible associated absorption-line system may reside at $\zabs=1.6971$. In this system, we detect {\hi} Ly\,$\beta$, {{\ciii} $\lambda$977.020, and the {\ovi} $\lambda\lambda$1031.926, 1037.617 doublet. The Ly\,$\beta$\ and {\ovi} $\lambda$1031.926 profiles are blended with intervening Ly\,$\alpha$\ forest lines, which makes the secure identification of this system difficult (and the evaluation of possible partial
coverage impossible). However, the alignment between the {{\ciii} $\lambda$977.020 and {\ovi} $\lambda$1037.617 profiles leads us to conclude that there is probably a real absorption-lines system at this redshift.


{\noindent \bf HS 0810+2554 ($\zabs$=1.4862, 1.4948, 1.5025):} There are three associated systems toward this quasar. The system at $\zabs=1.5025$\ (Figure~\ref{fig:hs0810a}) is detected in {\hi} Ly\,$\beta$, Ly\,$\gamma$, {{\ciii} $\lambda$977.020, and the {\ovi} $\lambda\lambda$1031.926, 1037.617 doublet. The {\ovi} doublet is saturated and reaches
a zero flux level, indicating that the high-ionization gas fully occults the quasar central engine. In Figure~\ref{fig:hs0810b}, we show the detected {\ovi} $\lambda\lambda$1031.926, 1037.617 doublet in the $\zabs=1.4948$\ system, velocity-aligned flux profiles on the left, and evaluation of the velocity-dependent partial coverage on the right. The system at $\zabs=1.4948$\ is detected only in the {\ovi} $\lambda\lambda$1031.926, 1037.617 doublet. No other lines, most notably notably {\hi} Ly\,$\beta$, are detected. This system does not show evidence for partial coverage. The system at $\zabs=1.4862$\ (Figure~\ref{fig:hs0810c})
shows the same features at the $\zabs=1.4948$, detection of the {\ovi} $\lambda\lambda$1031.926, 1037.617 doublet with no evidence for partial coverage, and no detection of {\hi} Ly\,$\beta$.


{\noindent \bf PG\,0953+415 ($\zabs$=0.2335) --} This associated absorption-line system, previously reported by \citet{kpxiii} and \citet{gan01a} in spectra from the HST Faint Object Spectrograph (HST/FOS), is detected only in the {\hi} Lyman series, and the {\ovi} $\lambda\lambda$1031.926, 1037.617 doublet. [\citet{kpxiii} only reported detection of {\hi} Ly\,$\alpha$, while \citet{gan01a} reported detection of {\hi} Ly\,$\alpha$\
and the {\civ} $\lambda\lambda$1548.204, 1550.781 doublet.] \citet{tripp08} also provide measurement of the absorption lines in this system. Figure~\ref{fig:pg0953} shows velocity-aligned flux profiles of the transitions (left) and an evaluation of velocity-dependent partial coverage (right) in the {\ovi} doublet. The {\ovi} doublet is consistent with fully occulting the quasar central engine.

{\noindent \bf Ton\,28 ($\zabs$=0.3302) --} This associated absorption-line system is detected only in the {\ovi} $\lambda\lambda$1031.926, 1037.617 doublet. The {\nv} $\lambda\lambda$1238.821, 1242.804 doublet is covered but not detected. Figure~\ref{fig:ton28} shows velocity-aligned flux profiles of the transitions (left) and an evaluation of velocity-dependent partial coverage (right) in the {\ovi} doublet. The {\ovi} doublet is consistent with fully occulting the quasar central engine.


{\noindent \bf Mrk 132 ($\zabs$=1.7322) --} This associated absorption-line system is detected in all Lyman series lines and includes a sharp Lyman limit break. The system is also detected in {\cii} $\lambda$1036.337, {{\ciii} $\lambda$977.020, and the {\ovi} $\lambda\lambda$1031.926, 1037.617 doublet. Figure~\ref{fig:mrk132sys} shows velocity-aligned flux profiles of selected Lyman series and metal-line transitions. The {\hi} Lyman series lines are saturated with flat-bottomed trough that reach zero flux indicating full occultation of the background by neutral species. Likewise, the {\ovi} shows no evidence for partial coverage via our test (Figure~\ref{fig:mrk132pc}).


{\noindent \bf 3C 249.1 ($\zabs$=0.3080, 0.3136) --} [AKA PG\,1100+772] The quasar 3C 249.1 harbors two associated absorption-line systems at $\zabs=0.3080$\ and $\zabs=0.3136$. The $\zabs=0.3080$\ system, previously reported by \citet{kpxiii}, \citet{gan01a}, and \citet{tripp08} is detected in the {\hi} Lyman series, and the {\ovi} $\lambda\lambda$1031.926, 1037.617 and {\nv} $\lambda\lambda$1238.821, 1242.804 doublets. In Figure~\ref{fig:3c249a}, we present velocity-aligned flux profiles of these detected transitions and the velocity-dependent coverage fraction analysis. Neither doublet shows evidence for partial coverage. In addition,
comparison to \citet{gan01a} shows no variation in the equivalent width of the {\ovi} $\lambda$1031.926 transition. The associated absorption-line system at $\zabs=0.3631$, which appears reward of the quoted emission redshift, is detected as a narrow component in the {\hi} Lyman series, {{\ciii} $\lambda$977.020, and the {\ovi} $\lambda\lambda$1031.926, 1037.617 doublet. Velocity-aligned flux profiles of these transitions and a partial coverage analysis of the {\ovi} doublet is shown in Figure~\ref{fig:3c249b}. While most of the {\ovi} is inconsistent with full coverage, the profile is too narrow and the profile weak. The apparent trough-dilution could result from unresolved saturation. The saturated {\hi} Ly\,$\alpha$\ profile reaches zero flux levels indicating that the absorber does fully occult the quasar central engine.


{\noindent \bf PG\,1116+215 ($\zabs$=0.1655) --} This associated absorption-line system is detected in the {\ovi} $\lambda\lambda$1031.926, 1037.617 doublet and {\hi} Ly$\alpha$-$\beta$. The system was first reported by \citet{tls98} in {\hi} Ly$\alpha$\ absorption, and then by \citet[][see their \S5.16,7.1]{sembach04b} in {\ovi} absorption. Both papers claim, based on velocity alignments with galaxies in the field of the quasar \citep{tls98}, and the lack of evidence for diluted troughs in the {\ovi} doublet, that the system is unlikely to arise from gas related to the quasar. We confirm that assessment with our partial coverage analysis. As also pointed out by both papers, this system appears to be a satellite of a stronger, lower-ionization {\hi} Ly$\alpha$\ system at $\zabs=0.1661$. \citet{tls98} also reported an {\hi} Ly\,$\alpha$\ absorber at $\zabs=0.17366$. \citet{sembach04b} corroborate this find with associated {\ovi} absorption at $\zabs=0.1734$. The doublet lies on the damping wings of the Galactic Ly\,$\alpha$\ line and does not meet our detection criterion. Thus, in the interest of uniformity, we do not include it in our analysis.


{\noindent \bf PG\,1206+459 ($\zabs$=1.0280) --} In the E230M spectrum of PG\,1206+459, we detect an intrinsic absorption-line system at $\zabs=1.0280$\ in {\hi} Ly$\alpha$\ and the {\nv} $\lambda\lambda$1238.821,1242.804 doublet. The spectrum covers the rest-frame wavelength range 1120-1538,\AA. This system has not been identified in other works and we present a velocity-stacked plot of the detected transitions in Figure~\ref{fig:pg1206}. The absorption profiles are very strong and smooth, indicative of a possible origin in the accretion-disk outflow. This type of system has often been termed a ``mini-BAL'' \citep[e.g., ][]{cssg99,bhs97}.


{\noindent \bf RX\,J1230.8+0115 ($\zabs$=0.1000, 0.1058, 0.1093, 0.1172) --} There are several intrinsic systems within 4600\,\kms\ of  the quasar redshift in this sight line passing through the Virgo cluster. These have been analyzed in detail by \citet[][see their Table 1, and Figures 3 and 4]{gan03b}, who present evidence for partial coverage in the {\civ} $\lambda\lambda$1548.204, 1550.781 and {\nv} $\lambda\lambda$1238.821, 1242.804 doublets for the systems at $\zabs=0.1000$, 0.1058, and 0.1093. The system at $\zabs=0.1172$, which is at the quasar redshift, shows no evidence for partial coverage. In addition, the three systems showing partial coverage are also apparently line-locked, $\zabs=0.1000$\ and 0.1058 in {\ovi} and {\hi} Ly$\beta$\ (transitions detected in spectra from the {\it Far Ultraviolet Spectroscopic Explorer}), $\zabs=0.1058$\ and 0.1093 in {\nv}.


{\noindent \bf PG\,1241+176 ($\zabs$=1.2717) --} In the E230M spectrum of this quasar, we detect an associated absorption-line system in {\hi} Ly$\alpha$\ and the {\ovi} $\lambda\lambda$1031.926, 1037.617 doublet (Figure~\ref{fig:pg1241}). The core of the {\ovi} $\lambda$1031.926 line in blended with the Galactic {\feii} $\lambda$2344.214 line. The data for the doublet are fairly noisy and the fluxes are consistent with full coverage in spite of the blend.


{\noindent \bf PG\,1444+407 ($\zabs$=0.2673) --} This associated absorption-line system, previously reported by \citet{tripp08}, is detected in {\hi} Ly\,$\alpha$, and the {\ovi} $\lambda\lambda$1031.926, 1037.617 doublet (Figure~\ref{fig:pg1444}). The system is peculiar in the sense that the {\hi} and {\ovi} profiles do not trace each other. The {\hi} appears in a single component that is offset from the \ovi, which appears in two discrete components. {\hi} Ly\,$\beta$\ is covered by the spectra, but is not detected as expected based on the strength of the Ly\,$\alpha$\ line, assuming there are no partial coverage effects. If partial coverage of the {\hi} were present, then the Ly\,$\beta$\ transition would be stronger (potentially as strong as Ly\,$\alpha$). In any case, this absorption feature is kinematically distinct from the \ovi-bearing gas. Hence it is unclear if there is and relationship between them. The {\ovi} doublet shows no evidence for partial coverage.


{\noindent \bf PG\,1630+377 ($\zabs$=1.4333) --} This associated
absorption-line system is detected in the {\hi} Lyman series ($\alpha$ - $\epsilon$), and the {\ovi} $\lambda\lambda$1031.926, 1037.617 doublet (Figure~\ref{fig:pg1630}). The {\nv} $\lambda\lambda$1238.821, 1242.804 doublet is covered, but not detected. The {\ovi} is very weak and narrow and does not allow for a proper/definitive determination of partial coverage, as unresolved saturation could yield non-unity coverage fractions. However, the {\hi} Lyman series down to Ly\,$\delta$\ reach zero flux levels, indicating full coverage in the {\hi} ion. Furthermore, like PG\,$1444+407$, the kinematics of the {\hi} appear distinct from the \ovi, with significant {\hi} at $v\lesssim0$\,\kms, but no detected {\ovi} absorption at those velocities.


{\noindent \bf PG\,1634+706 ($\zabs$=1.3415) --} This associated absorption-line system is detected in {\hi} Ly\,$\alpha$, and the {\ovi} $\lambda\lambda$1031.926, 1037.617 doublet (Figure~\ref{fig:pg1634}). Higher order Lyman series lines are not detected. The {\nv} $\lambda\lambda$1238.821, 1242.804 doublet is covered, but not
detected. The {\ovi} lines are broader than the instrumental profile, but a partial coverage analysis is inconclusive. The coverage fraction is marginally inconsistent with unity at the 1--2$\sigma$\ level, but this may be a result of noise or unresolved saturation. Furthermore, like PG\,$1444+407$\ and PG\,$1630+377$, the kinematics of the {\hi} appear distinct from the \ovi, with {\hi} kinematically offset from the {\ovi} by $v\sim20$\,\kms.


{\noindent \bf 3C\,351 ($\zabs$=0.3166, 0.3617, 0.3631, 0.3648, 0.3673, 0.3719) --} The associated absorption-line systems ($\zabs = 0.3617-0.3719$) have been studied in detail by \citet{yuan02}. In the E140M spectrum, the absorption is detected in the {\hi}
Lyman series ($\alpha-\epsilon$), the {\ovi} $\lambda\lambda$1031.926, 1037.617 doublet, and the {\nv} $\lambda\lambda$1238.821, 1242.804 doublet. \citet{yuan02} report 15 individual components (labeled A-O) in the ejection velocity range $v=37 - 2750$\,\kms. The absorption is complex and, unlike RX\,J1230.8+0115, it is not clear how to group components into isolated
systems in order to compare with other absorption-line systems. Thus, we use the \citet{yuan02} components A, D, F, I, and M as representative of the complex. One additional system at
$\zabs=0.3166$\ (corresponding to a putative ejection velocity 12370\,\kms) is also present in the spectrum of 3C\,351. This absorption-line system is detected in {\hi} Ly\,$\alpha$ --
Ly\,$\delta$, and the {\ovi} $\lambda\lambda$1031.926, 1037.617 doublet (Figure~\ref{fig:3c351}). The {\nv} $\lambda\lambda$1238.821, 1242.804 doublet is covered, but not detected. The {\ovi} doublet in this system is marginally inconsistent with full coverage, but this may be a result of noise or unresolved saturation.


{\noindent \bf PG\,1718+481 ($\zabs$=1.0548, 1.0874) --} We detect two associated absorption-line systems in the E230M spectrum of this quasar. The $\zabs=1.0548$\ is detected in {\hi} Ly\,$\alpha,\beta$, and the {\ovi} $\lambda\lambda$1031.926, 1037.617 doublet (Figure~\ref{fig:pg1718a}). The {\nv} $\lambda\lambda$1238.821, 1242.804 doublet is covered, but not detected. The {\ovi} doublet is consistent with full coverage. The $\zabs=1.0874$\ is detected in {\hi} Ly\,$\alpha,\beta$, and the {\ovi} $\lambda\lambda$1031.926, 1037.617 doublet. The {\nv} $\lambda\lambda$1238.821, 1242.804 doublet is covered, but not detected. The system is detected with two kinematic components, with one component ($v\sim100$\,\kms\ relative to $\zabs = 1.0874$) bearing the bulk of the {\hi} and the other ($v\sim0$\,\kms\ relative to $\zabs = 1.0874$) bearing the \ovi. The {\ovi} doublet in this system is marginally inconsistent with full coverage, but this may be a result of noise or unresolved saturation.


{\noindent \bf H 1821+643 ($\zabs$=0.2967) --} This associated absorption-line system, previously reported by \citet{tripp08}, is detected in the {\hi} Lyman series ($\alpha-\delta$), {{\ciii} $\lambda$977.020, and the {\ovi} $\lambda\lambda$1031.926, 1037.617 doublet (Figure~\ref{fig:h1821}). The {\nv} $\lambda\lambda$1238.821, 1242.804 doublet is covered, but not detected. The {\hi} Ly\,$\alpha$\ and Ly\,$\beta$\ lines reach zero flux indicating full coverage in that ion. Similarly, the {\ovi} doublet shows complete consistency with full coverage (i.e., a coverage fraction of unity).


{\noindent \bf PHL 1811 ($\zabs$=0.1765,0.1919) --} We detect two associated absorption-line systems in the STIS spectrum of this quasar, which were previously reported by \citet{tripp08}. The $\zabs=0.1765$\ absorption-line system, previously reported by \citet{jenkins03} from Lyman series, {{\ciii}, and {\oi} absorption from a FUSE spectrum, is detected in {\hi} Ly\,$\alpha$\ and Ly\,$\beta$, and the weaker member of the {\ovi} $\lambda\lambda$1031.926, 1037.617 doublet (Figure~\ref{fig:phl1811a}). The stronger member is wiped out by the Galactic {\hi} Ly\,$\alpha$\ line, while the weaker member sits on the
red damping wing. The blending precludes testing for partial coverage. However, we note that {\hi} Ly\,$\alpha$\ reaches zero flux, indicating full coverage in that ion. \citet{jenkins03} so report on a galaxy about 340\,kpc from the sightline and offset from the absorber redshift by $\sim100$\,\kms, well within the stellar velocity dispersion. The $\zabs=0.1919$\ absorption-line system, previously reported by \citet{jenkins03} in Ly\,$\alpha$\ absorption in a STIS-G140L spectrum, is detected in the {\ovi} $\lambda\lambda$1031.926, 1037.617 doublet (Figure~\ref{fig:phl1811b}). The stronger member of the {\nv} $\lambda\lambda$1238.821, 1242.804 doublet is also marginally detected. Lines from the {\hi} Lyman series and the {\siliv} $\lambda\lambda$1393.760, 1402.773 doublet are also covered by the spectrum, but none are detected. Both members of the  {\ovi} $\lambda\lambda$1031.926, 1037.617 doublet reach zero flux indicating full coverage in that ion. \citet{jenkins03} report two galaxies near the redshift of this absorber (G142 at $z=0.1917$, $\rho = 22h_{70}^{-1}$\,kpc, and G151 at $z=0.1927$, $\rho = 90h_{70}^{-1}$\,kpc), either of which could be responsible for the absorber.


\clearpage
\begin{figure*}
\includegraphics[width=1.8\columnwidth]{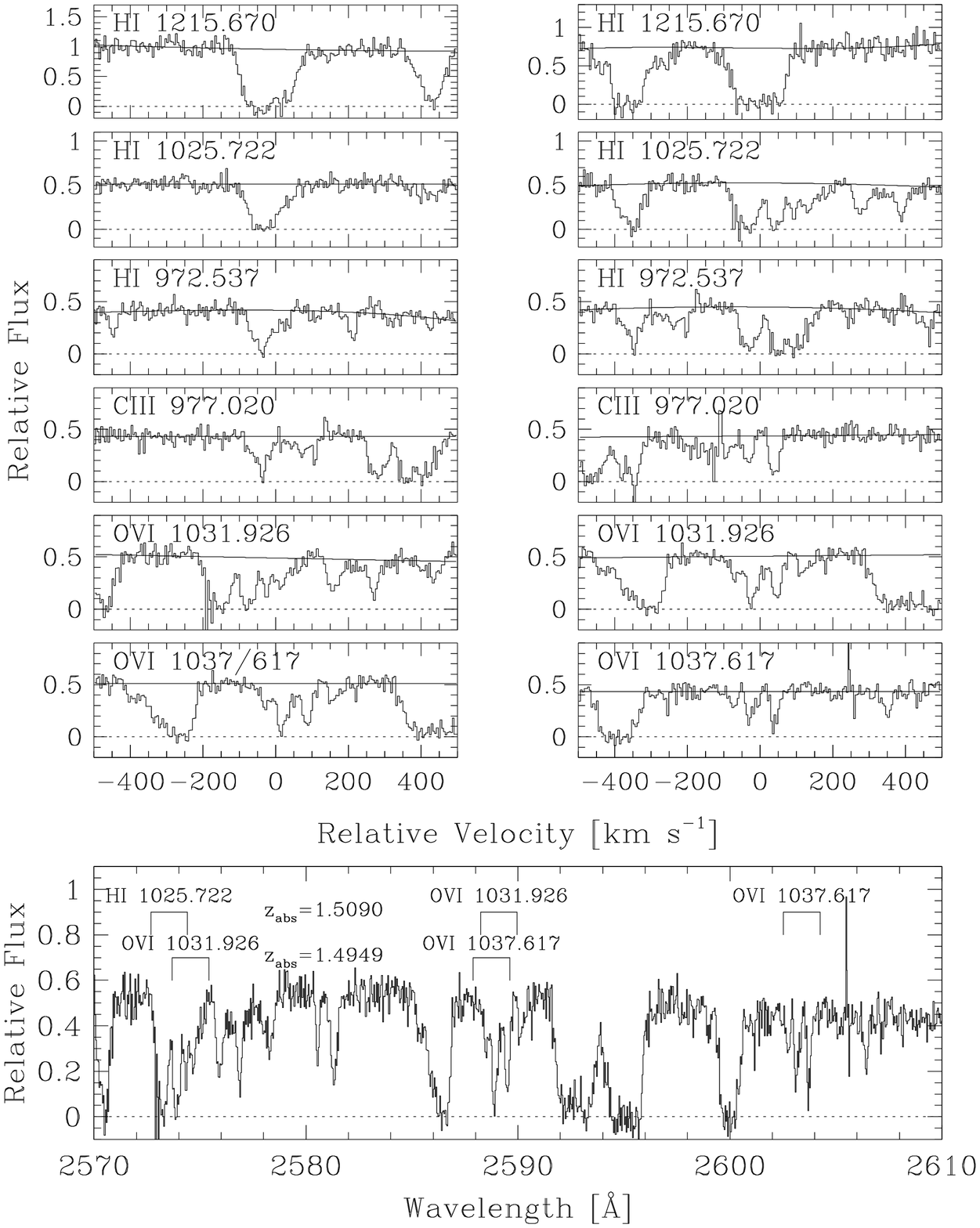}
\protect\caption[pg0117 system plot]{Velocity-aligned absorption profiles of detected transitions for the associated absorption line systems toward PG\,0117+213 at $\zabs=1.4949$\ (top left) and $\zabs=1.5090$\ (top right). Absorption profiles are plotted in the rest-frame of the system (i.e., $v=0$\,\kms\ at $\zabs=1.4949$\ for the left panel, and $v=0$\,\kms\ at $\zabs=1.5090$\ for the right panel). These two systems are apparently line-locked in \ovi $\lambda\lambda$1031.926,1037.617 and \hi Ly\,$\beta$, as illustrated in the lower panel.} \label{fig:pg0117}
\end{figure*}
\begin{figure*}
\includegraphics[width=0.99\columnwidth]{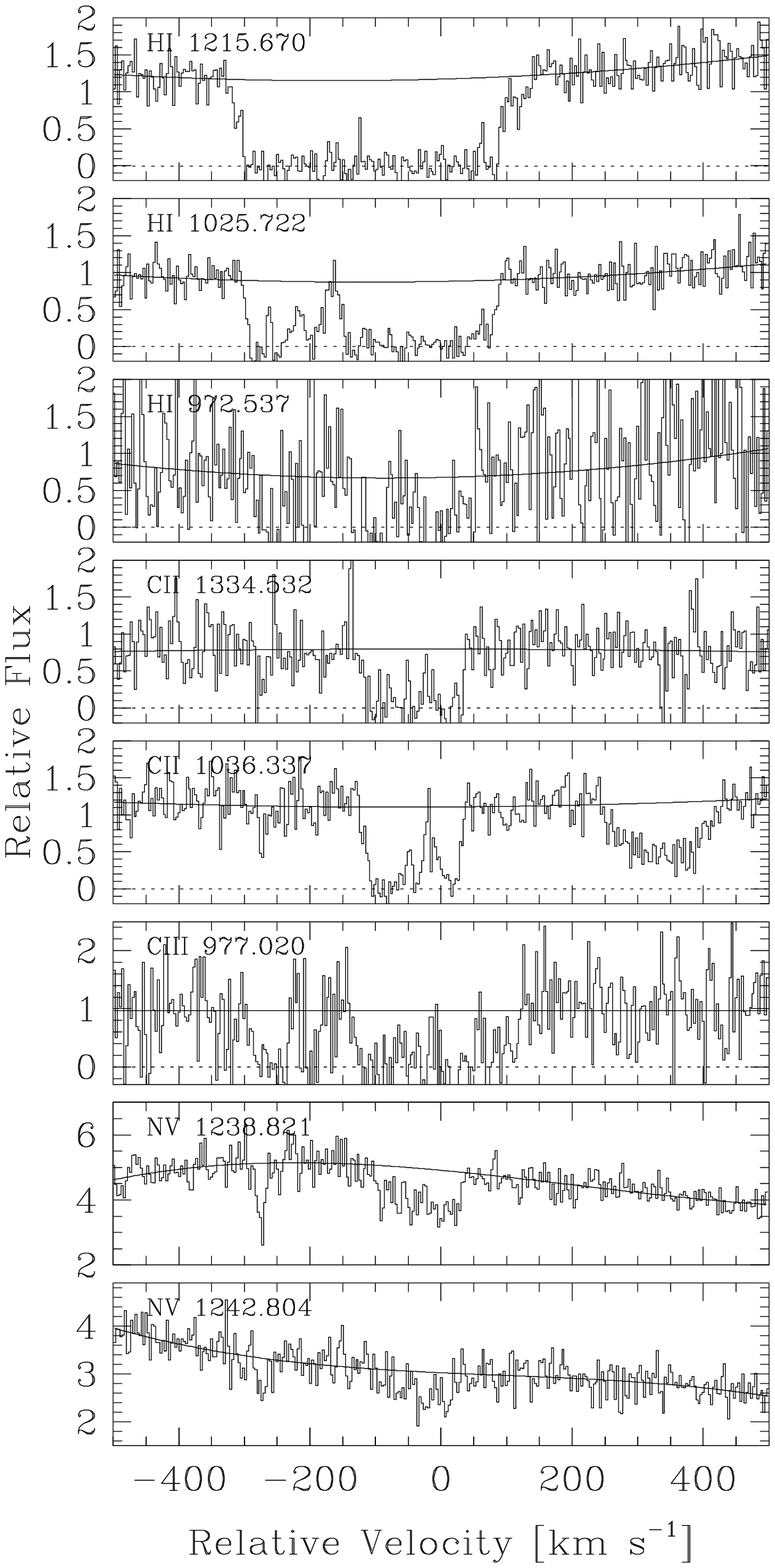}
\includegraphics[width=0.99\columnwidth]{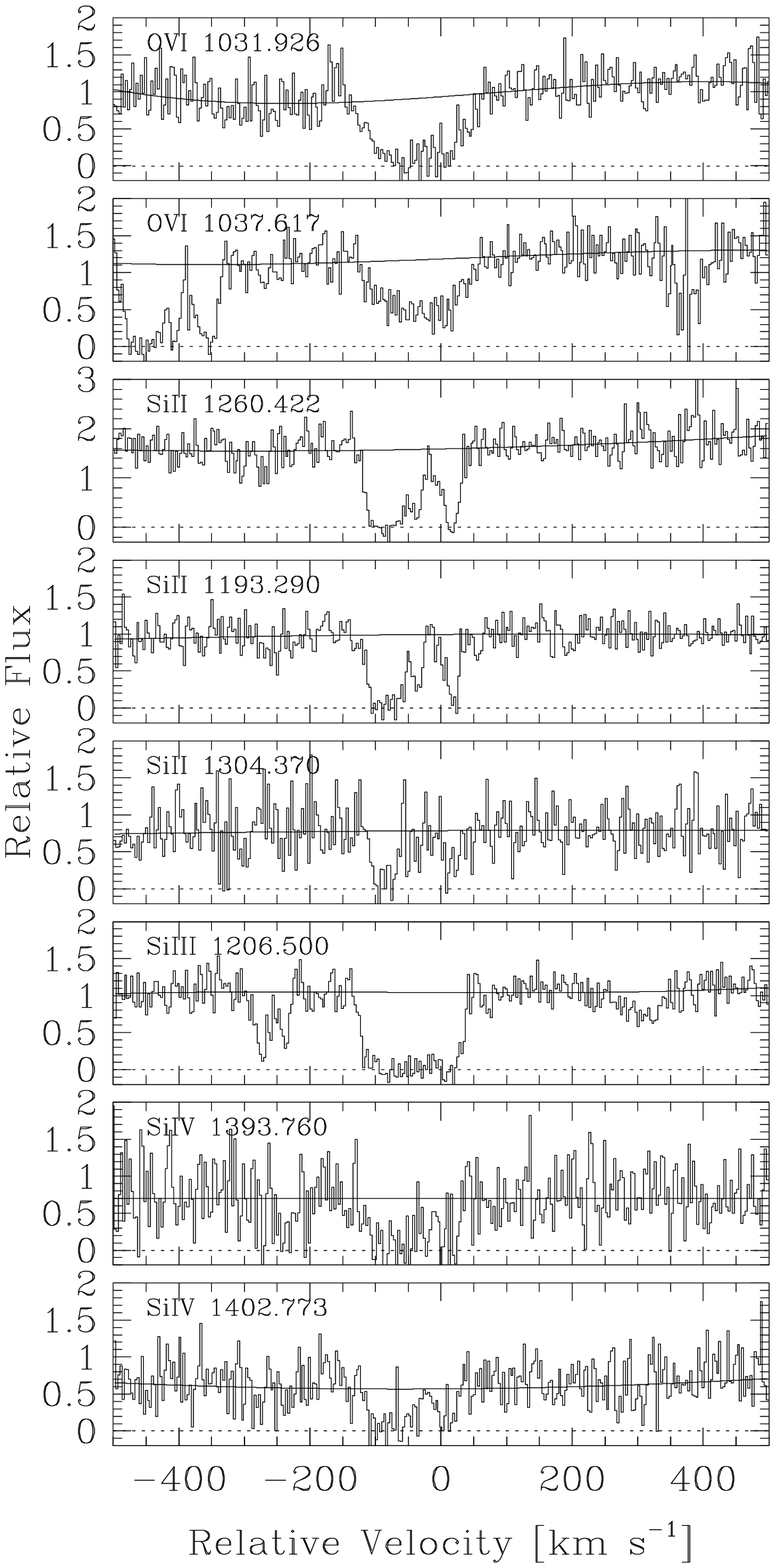}
\protect\caption[pks0312 system plot]{Velocity-aligned absorption profiles of detected transitions for the associated absorption-line system toward PKS\,0312-77 at zabs = 0.2029. (The quoted redshift refers to the velocity zero-point in each panel.)} \label{fig:pks0312sys}
\end{figure*}

\begin{figure*}
\includegraphics[width=0.99\columnwidth]{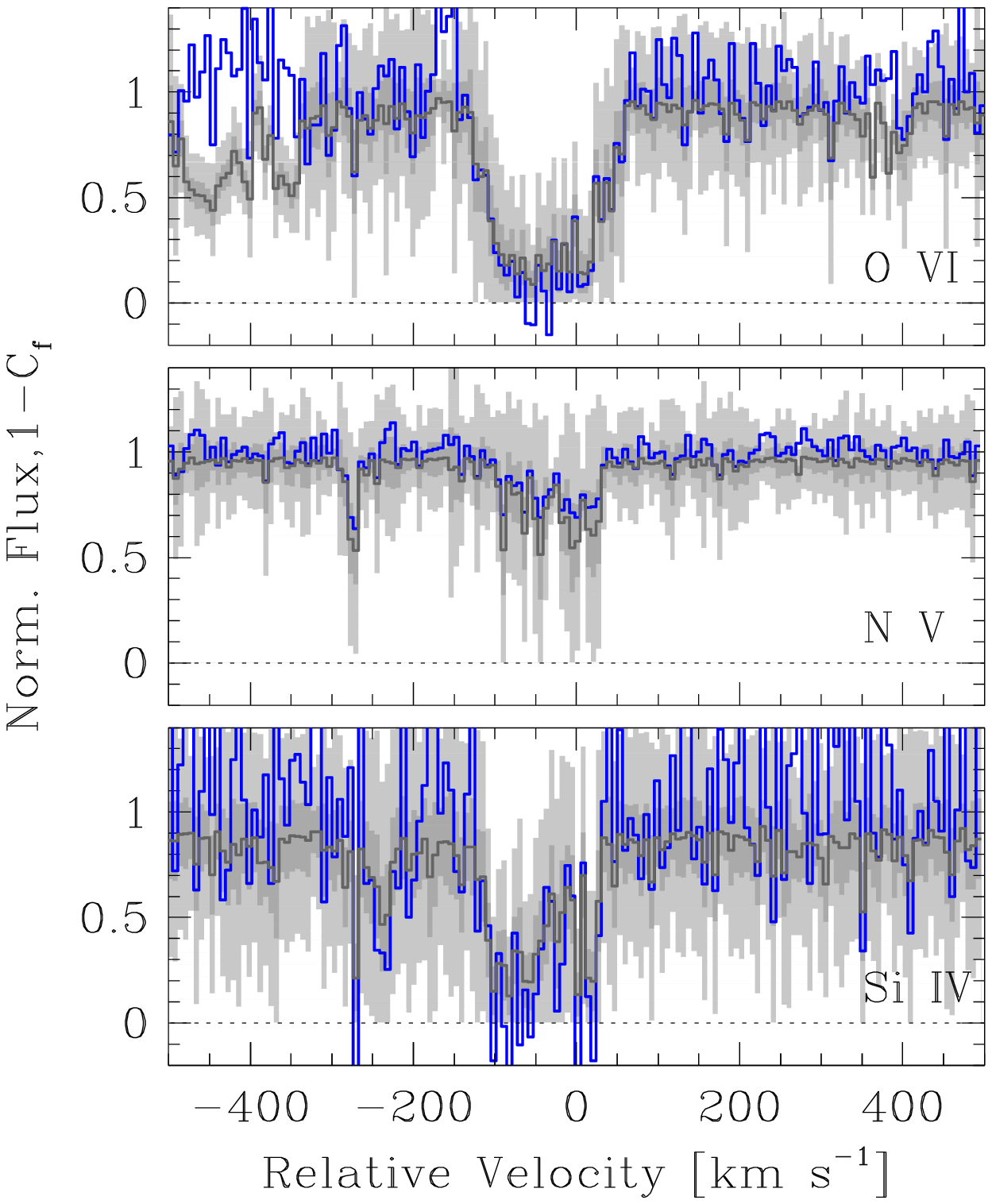}
\protect\caption[pks0312 partial coverage plot]{Evaluation of coverage fractions for the \ovi\ (top), \nv\ (middle), and \siliv\ (bottom) doublets for the $\zabs=0.2029$\ absorption-line system toward PKS\,0312-77. In each panel, the normalized flux profile of the stronger transition is shown as a black histogram (blue histogram in color version). For clarity, each bin represents one resolution element. The nominal velocity-dependent coverage fraction is shown as black points (black histogram in color version) with the grey shadings indicating the $1\sigma$\ (dark grey) and $3\sigma$\ (lighter grey) confidence limits} \label{fig:pks0312pc}
\end{figure*}
\begin{figure*}
\includegraphics[width=0.99\columnwidth]{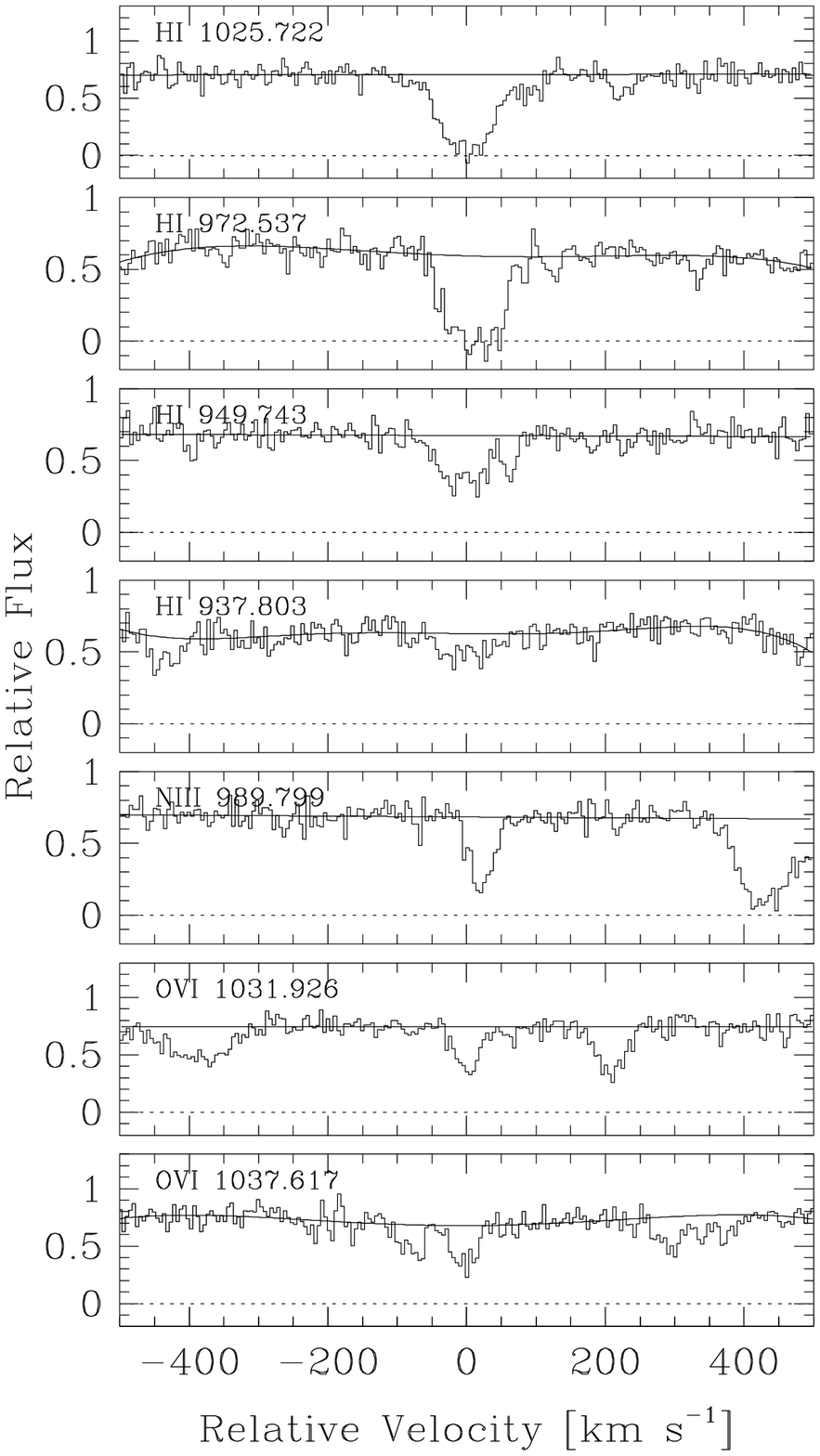}
\includegraphics[width=0.99\columnwidth]{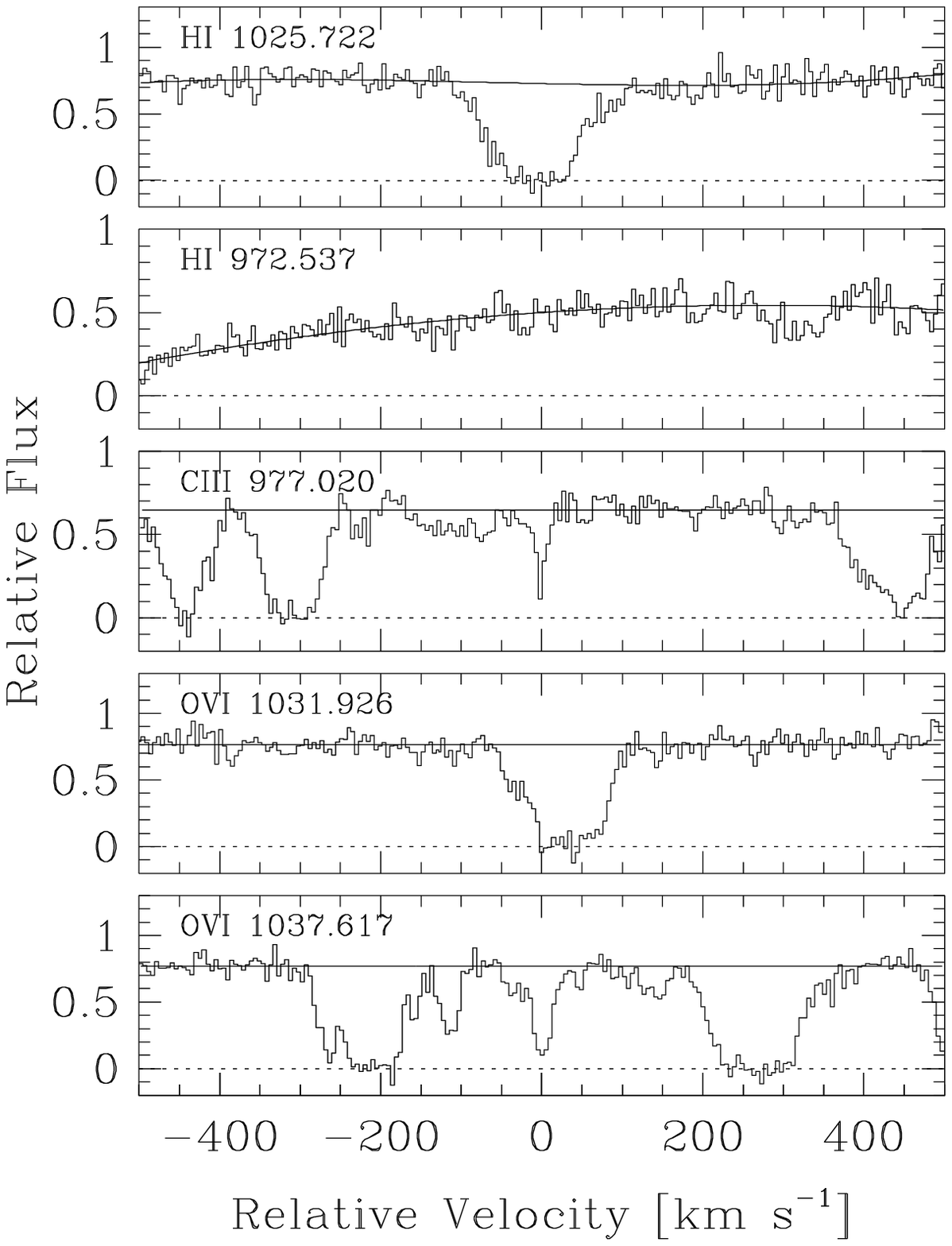}
\protect\caption[HE 0515-4414 system plot]{Velocity-aligned absorption profiles of transitions for the associated absorption line systems at $\zabs=1.6736$\ (left) and $\zabs=1.6971$\ (right) toward HE 0515-4414. All transitions are detected by our criteria except \hi\ Ly\,$\gamma$\ in the $\zabs=1.6971$\ system. It is only shown for comparison with \hi\ Ly\,$\beta$. (The quoted redshifts refer to the velocity zero-points.)}
\label{fig:he0515}
\end{figure*}
\begin{figure*}
\includegraphics[width=0.99\columnwidth]{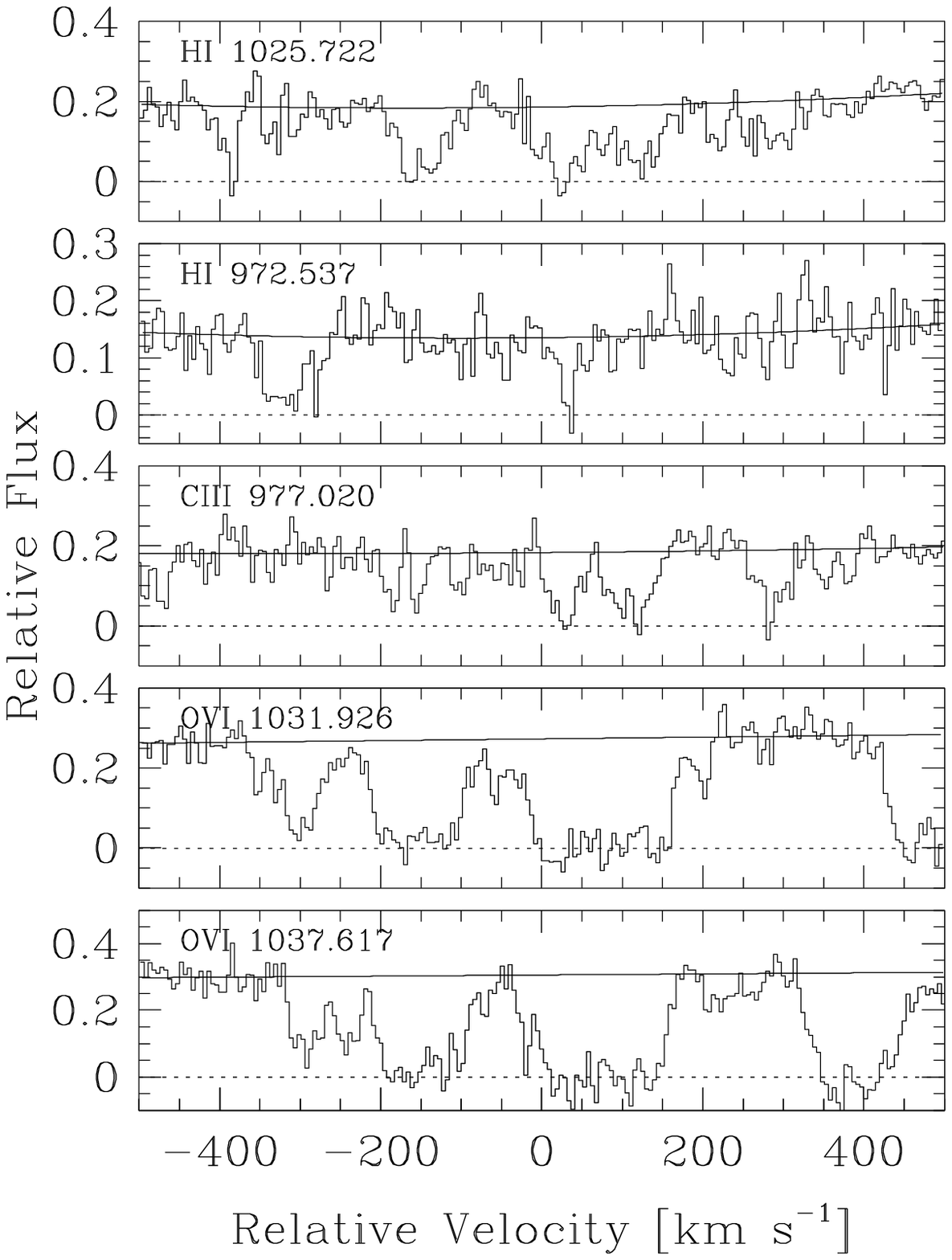}
\includegraphics[width=0.99\columnwidth]{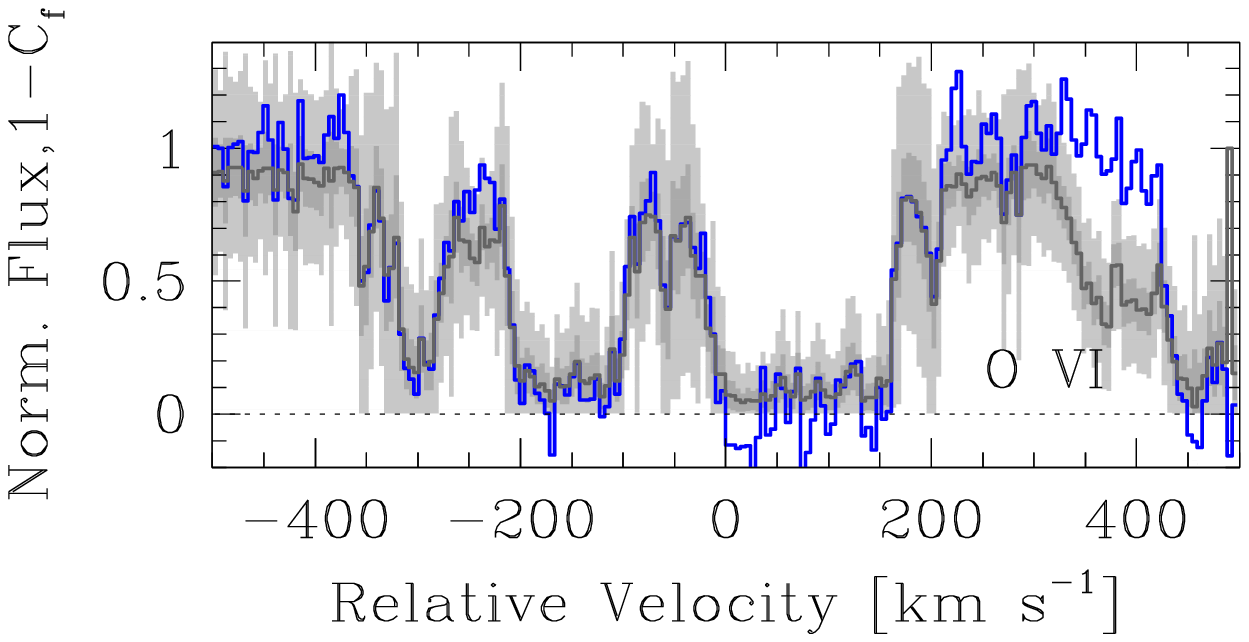}
\protect\caption[HS 0810+2554 system plot]{On the left, we present velocity-aligned absorption profiles of detected transitions for the associated absorption line system at $\zabs=1.5205$\ toward the quasar HS\,0810+2554. (The quoted redshift refers to the velocity zero-point.) On the right, we show the results of the partial coverage test on the \ovi\ doublet. The normalized flux profile of the stronger transition is shown as a black histogram (blue histogram in color version). The nominal velocity-dependent coverage fraction is shown as black points (black histogram in color version) with the grey shadings indicating the $1\sigma$\ (dark grey) and $3\sigma$\ (lighter grey) confidence limits.} \label{fig:hs0810a}
\end{figure*}

\begin{figure*}
\includegraphics[width=0.99\columnwidth]{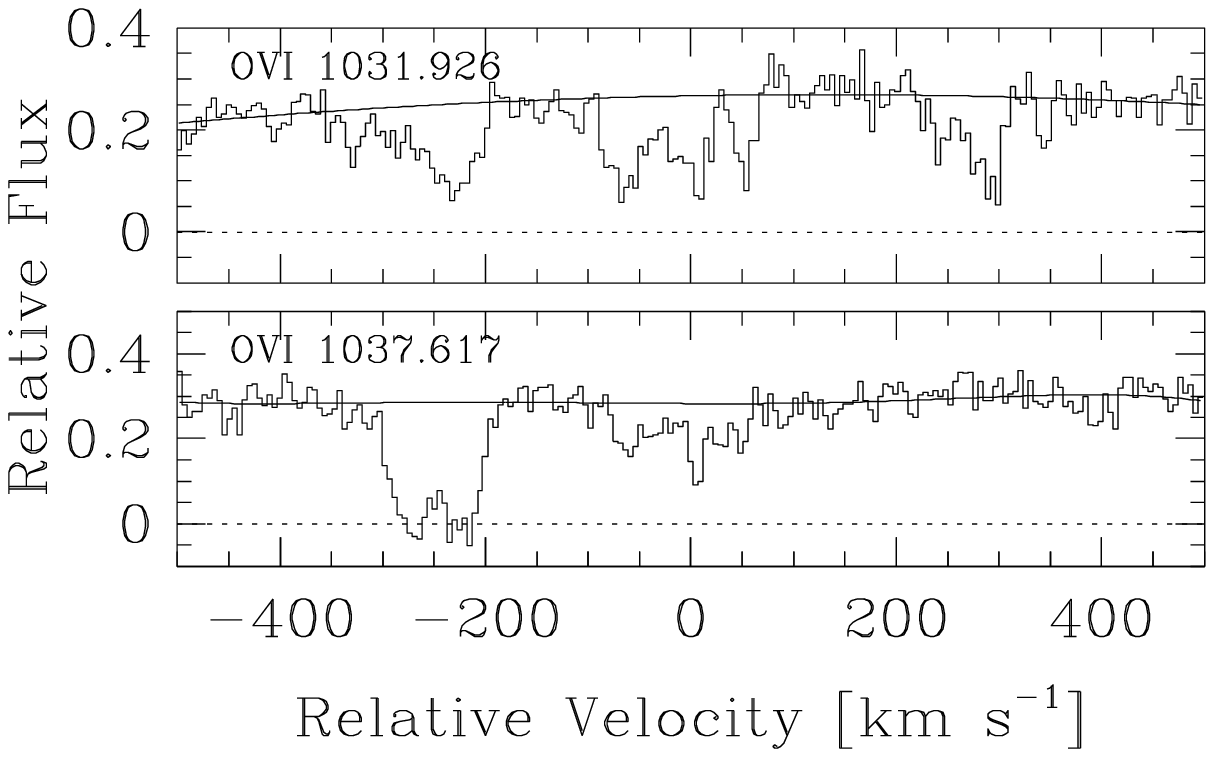}
\includegraphics[width=0.99\columnwidth]{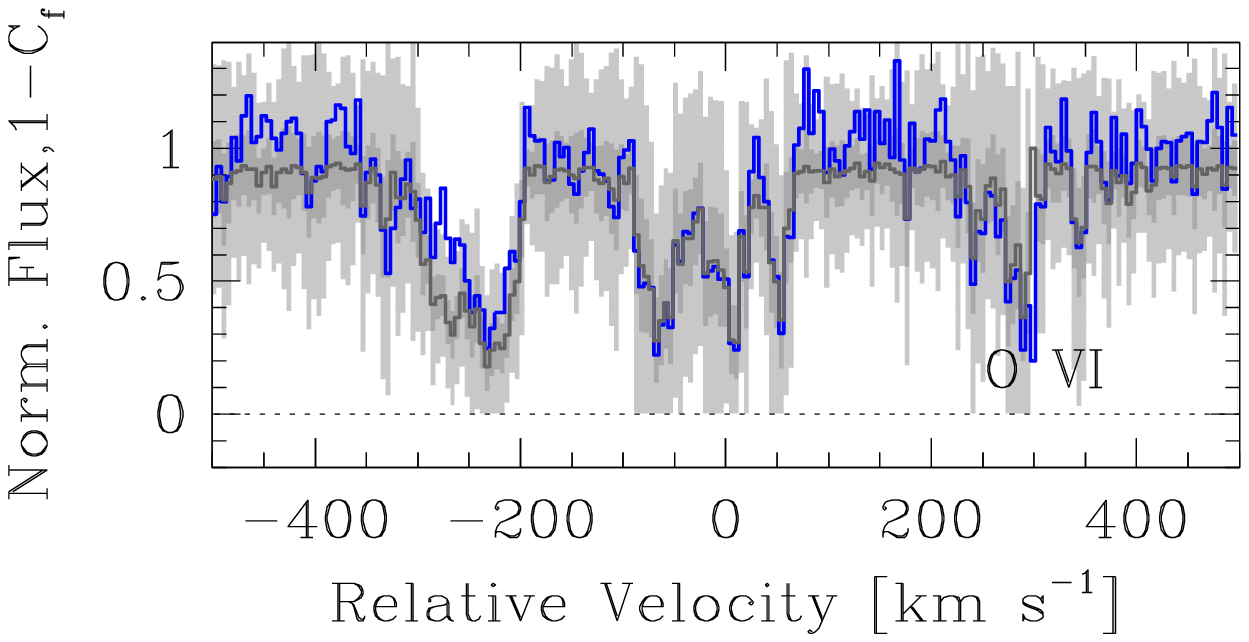}
\protect\caption[HS 0810+2554 system plot]{On the left, we present velocity-aligned absorption profiles of detected transitions for the associated absorption line system at $\zabs=1.4948$\ toward the quasar HS\,0810+2554. (The quoted redshift refers to the velocity zero-point.) On the right, we show the results of the partial coverage test on the \ovi\ doublet. The normalized flux profile of the stronger transition is shown as a black histogram (blue histogram in color version). The nominal velocity-dependent coverage fraction is shown as black points (black histogram in color version) with the grey shadings indicating the $1\sigma$\ (dark grey) and $3\sigma$\ (lighter grey) confidence limits.} \label{fig:hs0810b}
\end{figure*}

\begin{figure*}
\includegraphics[width=0.99\columnwidth]{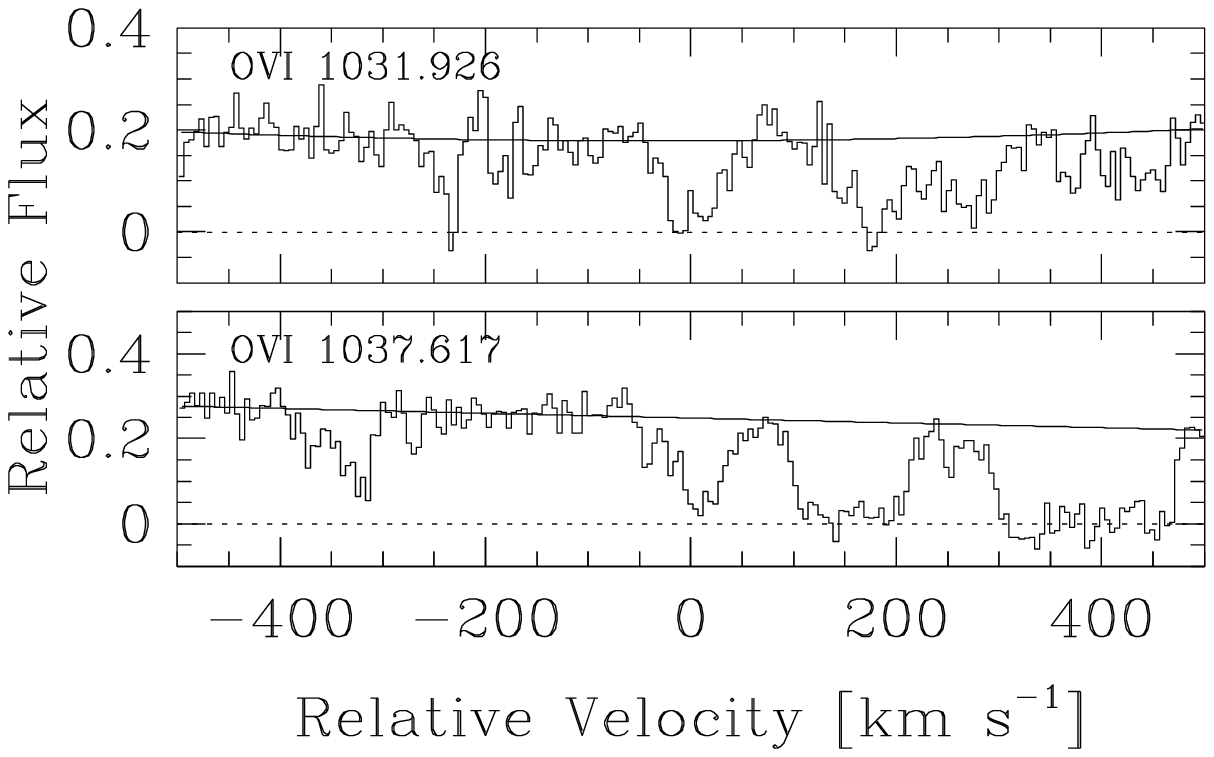}
\includegraphics[width=0.99\columnwidth]{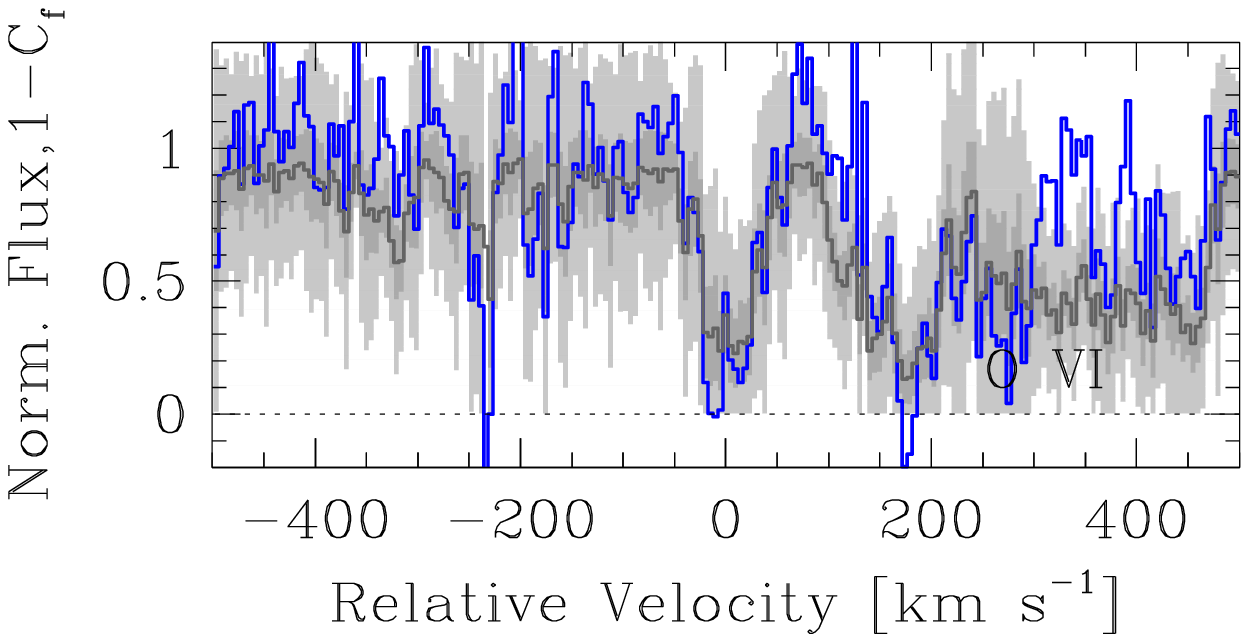}
\protect\caption[HS 0810+2554 system plot]{On the left, we present velocity-aligned absorption profiles of detected transitions for the associated absorption line system at $\zabs=1.4862$\ toward the quasar HS\,0810+2554. (The quoted redshift is the velocity zero-point.) On the right, we show the results of the partial coverage test on the \ovi\ doublet. The normalized flux profile of the stronger transition is shown as a black histogram (blue histogram in color version). The nominal velocity-dependent coverage fraction is shown as black points (black histogram in color version) with the grey shadings indicating the $1\sigma$\ (dark grey) and $3\sigma$\ (lighter grey) confidence limits.} \label{fig:hs0810c}
\end{figure*}
\begin{figure*}
\includegraphics[width=0.99\columnwidth]{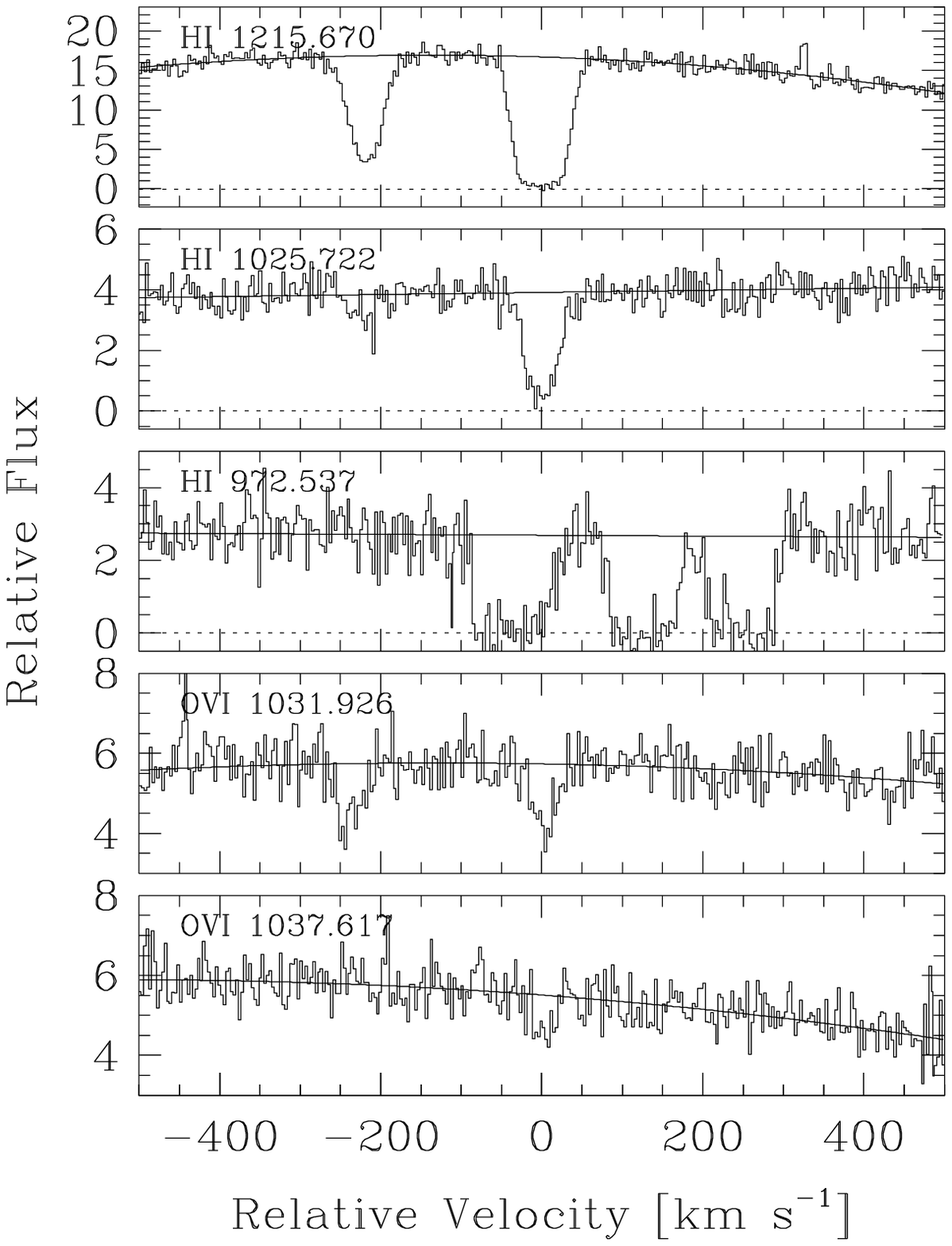}
\includegraphics[width=0.99\columnwidth]{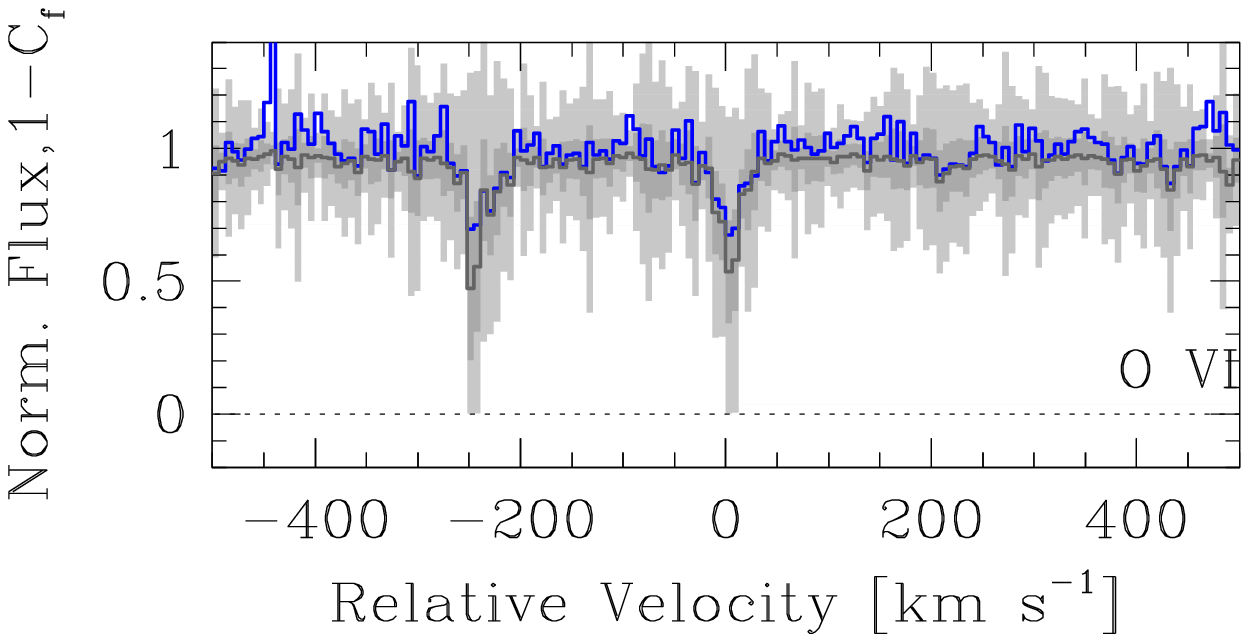}
\protect\caption[PG\,0953+415 system plot]{On the left, we present velocity-aligned absorption profiles of detected transitions for the associated absorption line system at $\zabs=0.2335$\ observed toward PG\,0953+415. (The quoted redshift is the velocity zero-point.) On the right, we show the results of the partial coverage test on the \ovi\ doublet. For clarity, each bin represents one resolution element. The normalized flux profile of the stronger transition is shown as a black histogram (blue histogram in color version). The nominal velocity-dependent coverage fraction is shown as black points (black histogram in color version) with the grey shadings indicating the $1\sigma$\ (dark grey) and $3\sigma$\ (lighter grey) confidence limits.} \label{fig:pg0953}
\end{figure*}
\begin{figure*}
\includegraphics[width=0.99\columnwidth]{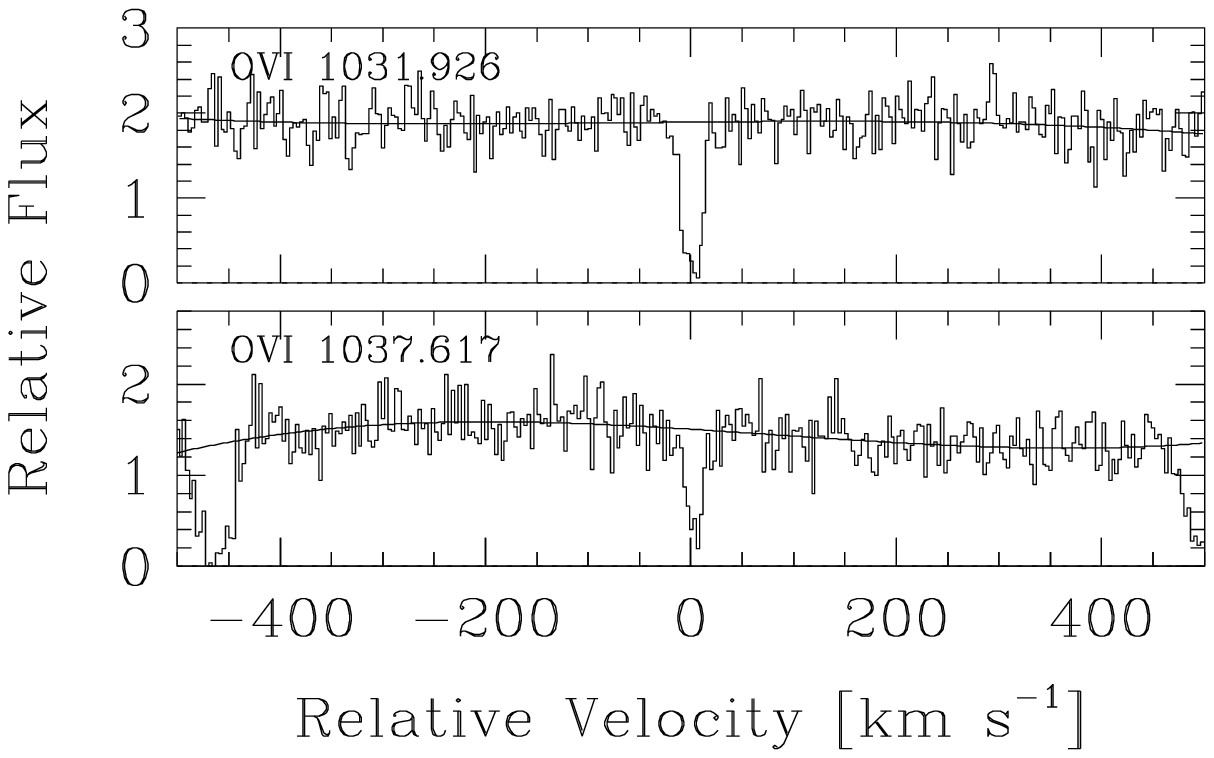}
\includegraphics[width=0.99\columnwidth]{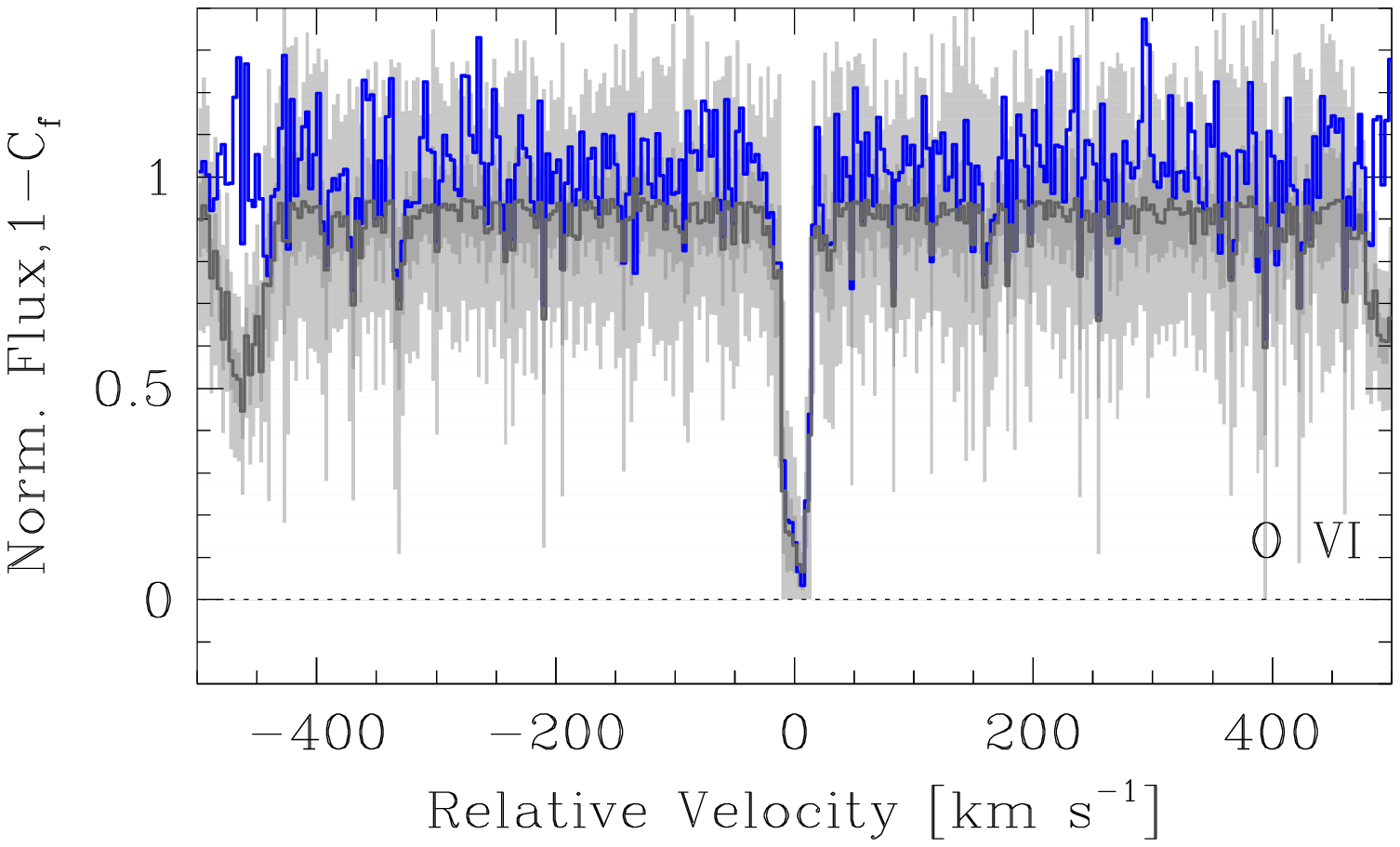}
\protect\caption[Ton\,28 system plot]{On the left, we present velocity-aligned absorption profiles of detected transitions for the associated absorption line system at $\zabs=0.3302$\ observed toward Ton\,28. (The quoted redshift is the velocity zero-point.) On the right, we show the results of the partial coverage test on the \ovi\ doublet. The normalized flux profile of the stronger transition is shown as a black histogram (blue histogram in color version). The nominal velocity-dependent coverage fraction is shown as black points (black histogram in color version) with the grey shadings indicating the $1\sigma$\ (dark grey) and $3\sigma$\ (lighter grey) confidence limits.} \label{fig:ton28}
\end{figure*}
\begin{figure*}
\includegraphics[width=0.99\columnwidth]{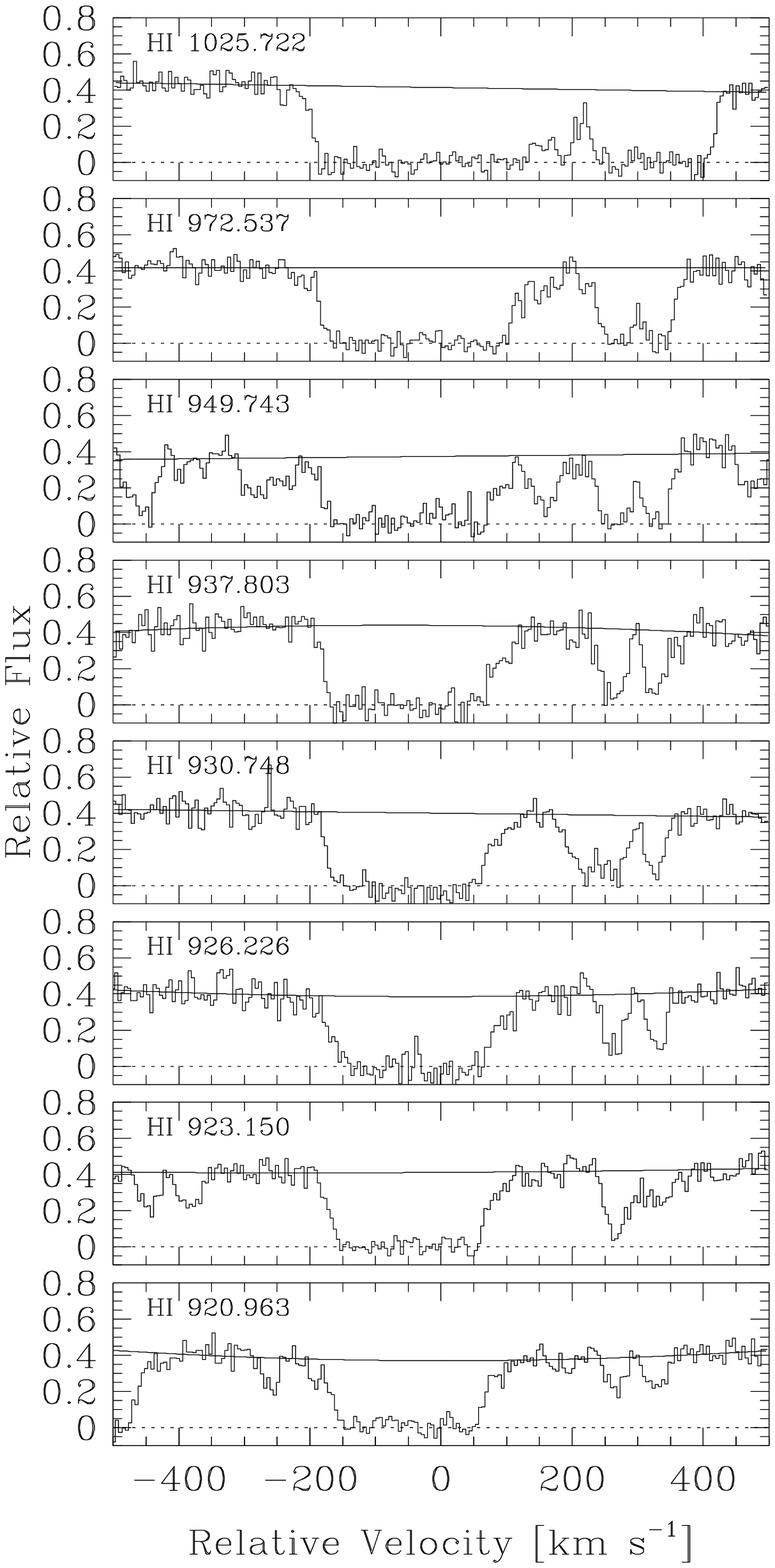}
\includegraphics[width=0.99\columnwidth]{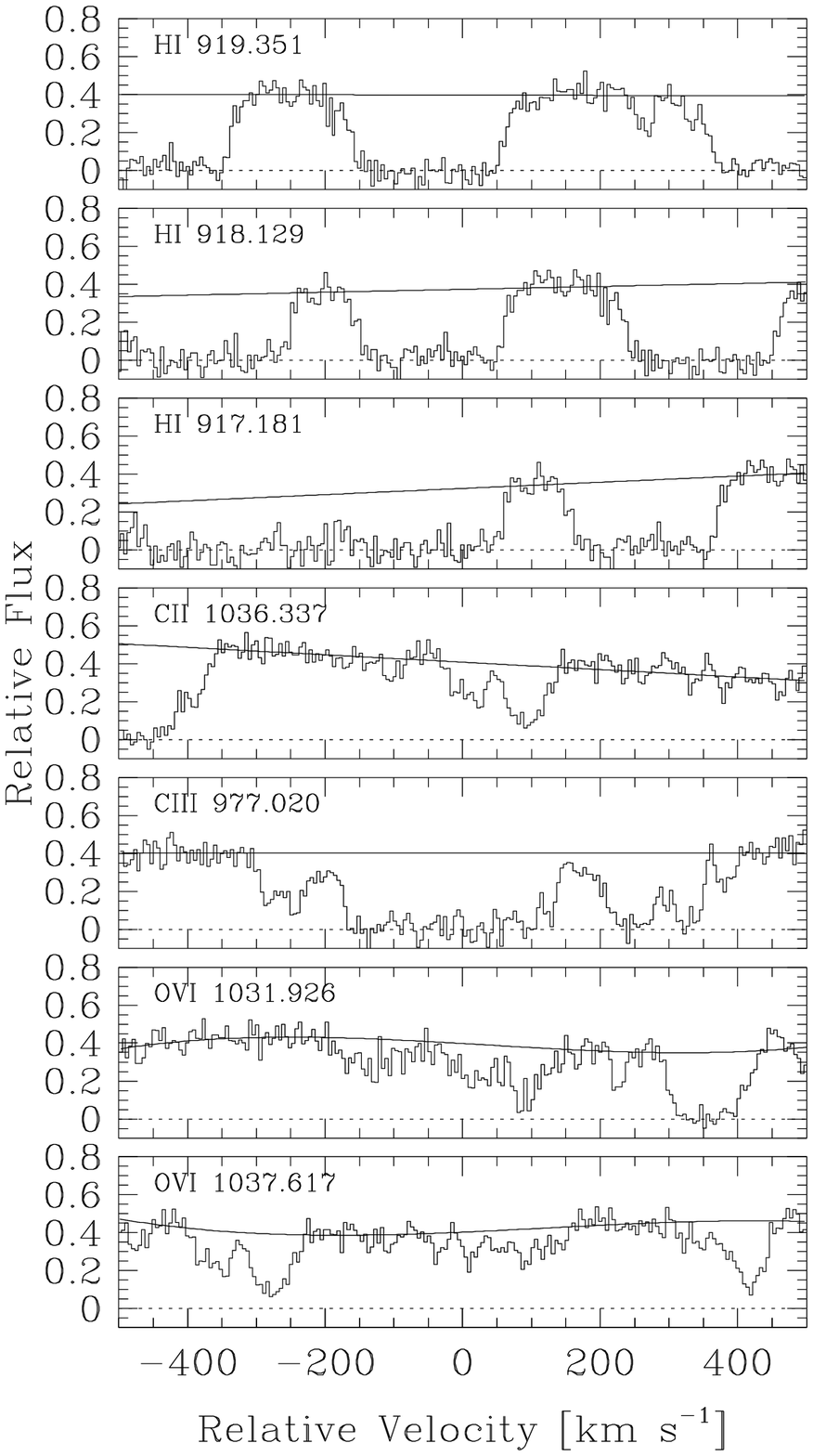}
\protect\caption[Mrk 132 system plot]{Velocity-aligned absorption profiles of detected transitions for the associated absorption line system at $\zabs=1.7322$\ observed toward Mrk 132. (The quoted redshift is the velocity zero-point.)}
\label{fig:mrk132sys}
\end{figure*}

\begin{figure*}
\includegraphics[width=0.99\columnwidth]{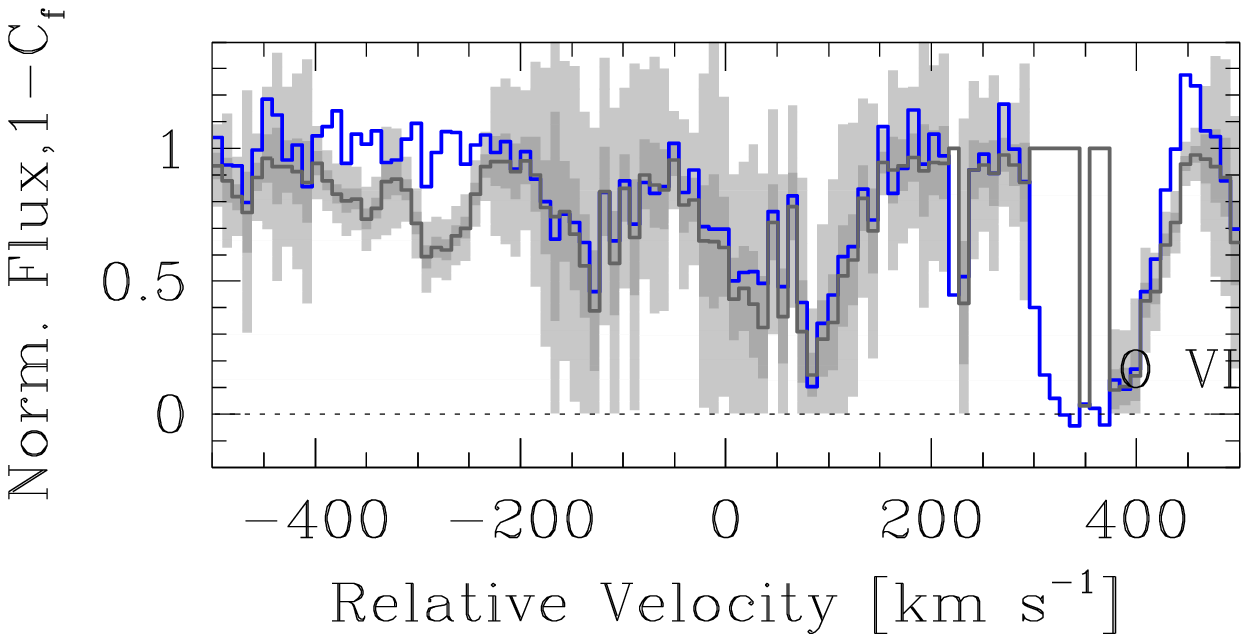}
\protect\caption[Mrk 132 partial coverage plot]{Evaluation of velocity-dependent coverage fraction for the \ovi\ doublet for the $\zabs=1.7322$\ absorption-line system toward Mrk 132. For clarity, profiles are shown sampled a one bin per resolution element. The normalized flux profile of the stronger transition is shown as a black histogram (blue histogram in color version). To improve signal-to-noise, the \ovi\ profile is shown with a sampling of one bin per resolution element. The nominal velocity-dependent coverage fraction is shown as black points (black histogram in color version) with the grey shadings indicating the $1\sigma$\ (dark grey) and $3\sigma$\ (lighter grey) confidence limits} \label{fig:mrk132pc}
\end{figure*}
\begin{figure*}
\includegraphics[width=0.99\columnwidth]{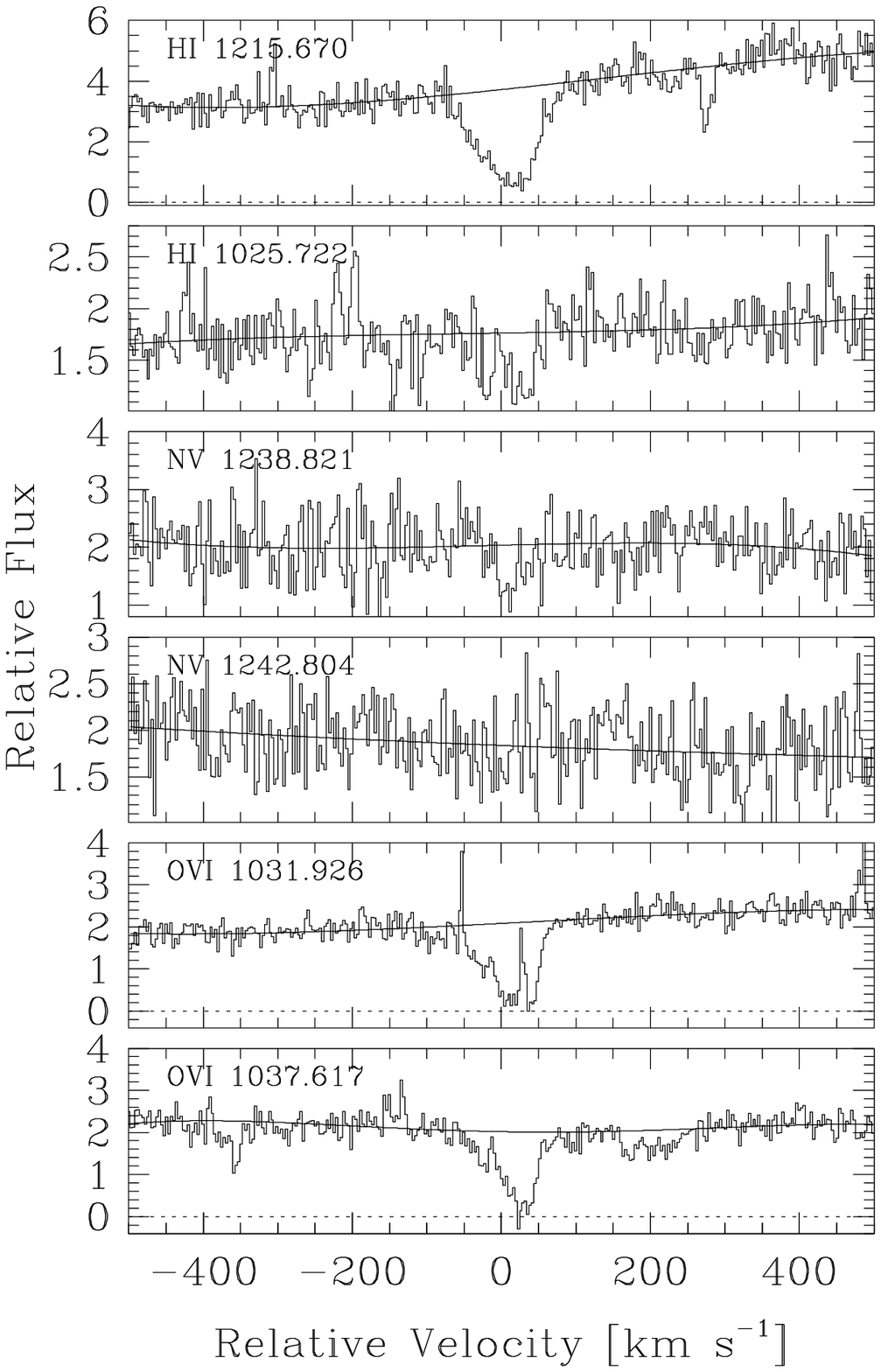}
\includegraphics[width=0.99\columnwidth]{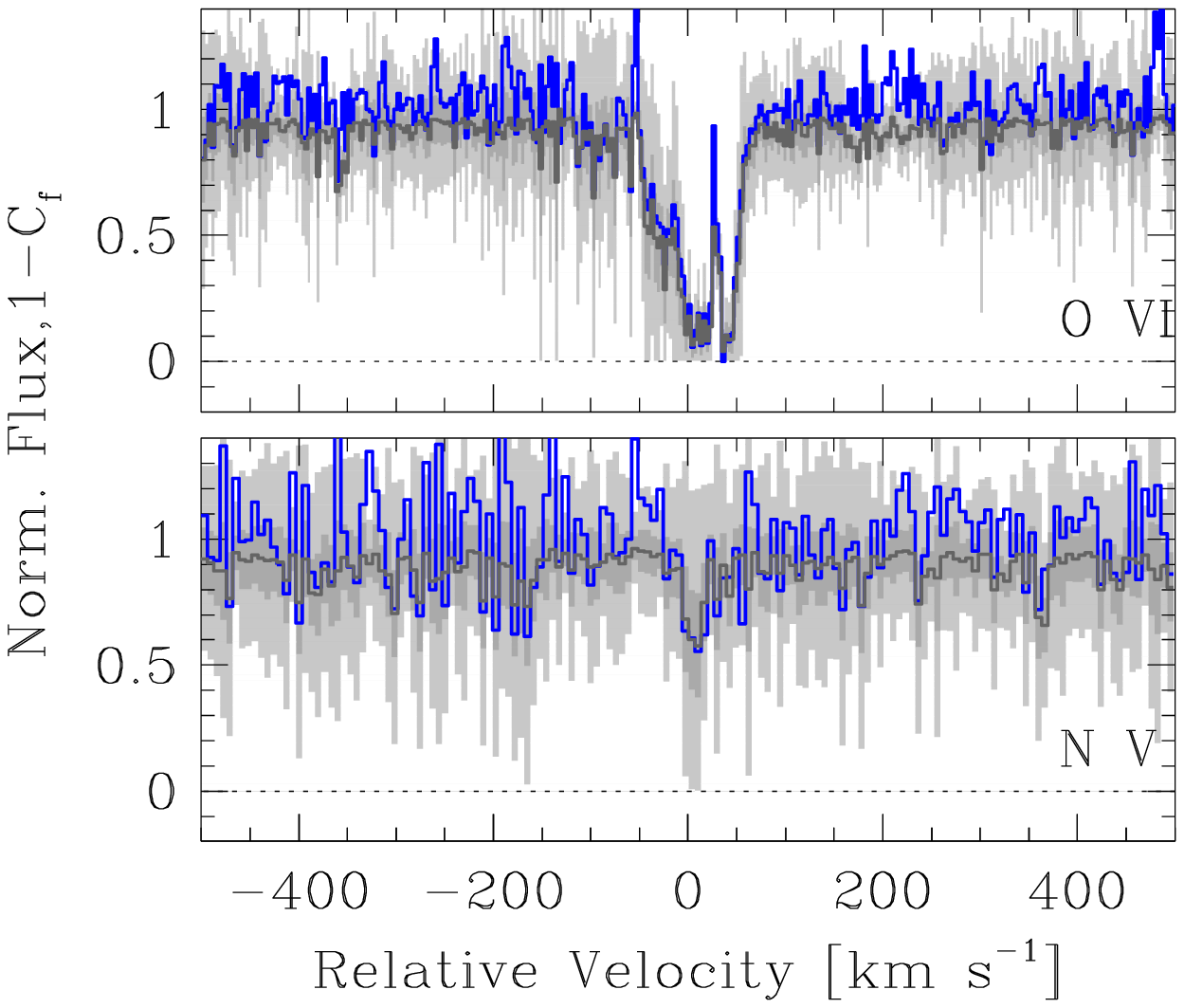}
\protect\caption[3C 249.1 system plot 2]{On the left, we present velocity-aligned absorption profiles of detected transitions for the associated absorption line system at $\zabs=0.3080$\ observed toward 3C 249.1. (The quoted redshift is the velocity zero-points.) On the right, we show the results of the partial coverage test on the \ovi\ and \nv\ doublets. In each panel on the right, the normalized flux profile of the stronger transition is shown as a black histogram (blue histogram in color version). To improve signal-to-noise, the \nv\ profile is shown with a sampling of one bin per resolution element. The nominal velocity-dependent coverage fraction is shown as black points (black histogram in color version) with the grey shadings indicating the $1\sigma$\ (dark grey) and $3\sigma$\ (lighter grey) confidence limits. Note that there is a hot pixel at $v\sim+25$\,\kms in the \ovi\ 1031.926\,\AA\ absorption profile.} \label{fig:3c249a}
\end{figure*}

\begin{figure*}
\includegraphics[width=0.99\columnwidth]{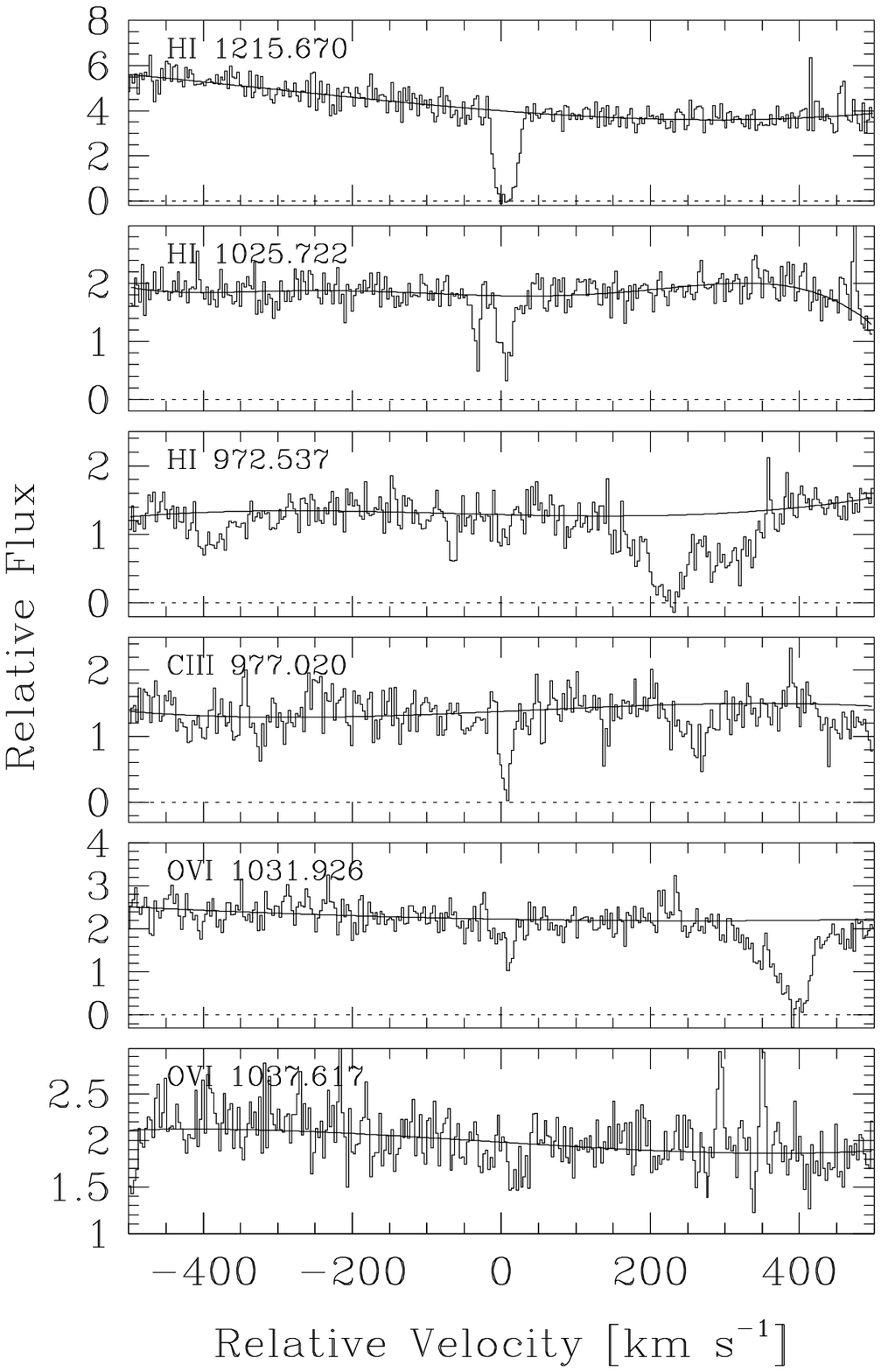}
\includegraphics[width=0.99\columnwidth]{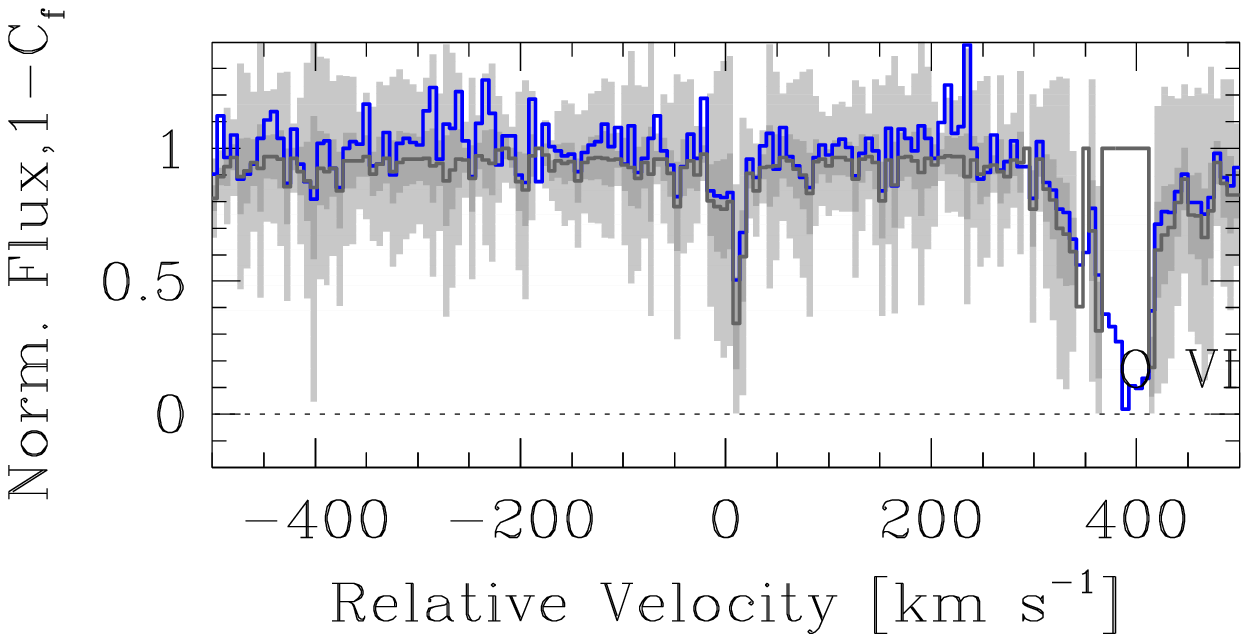}
\protect\caption[3C 249.1 system plot 1]{On the left, we present velocity-aligned absorption profiles of detected transitions for the associated absorption line system at $\zabs=0.3136$\ observed toward 3C 249.1.(The quoted redshift is the velocity zero-point.) On the right, we show the results of the partial coverage test on the \ovi\ doublet. The normalized flux profile of the stronger transition is shown as a black histogram (blue histogram in color version). To improve signal-to-noise, the \ovi profile is shown with a sampling of one bin per resolution element. The nominal velocity-dependent coverage fraction is shown as black points (black histogram in color version) with the grey shadings indicating the $1\sigma$\ (dark grey) and $3\sigma$\ (lighter grey) confidence limits.}
\label{fig:3c249b}
\end{figure*}
\begin{figure*}
\includegraphics[width=0.99\columnwidth]{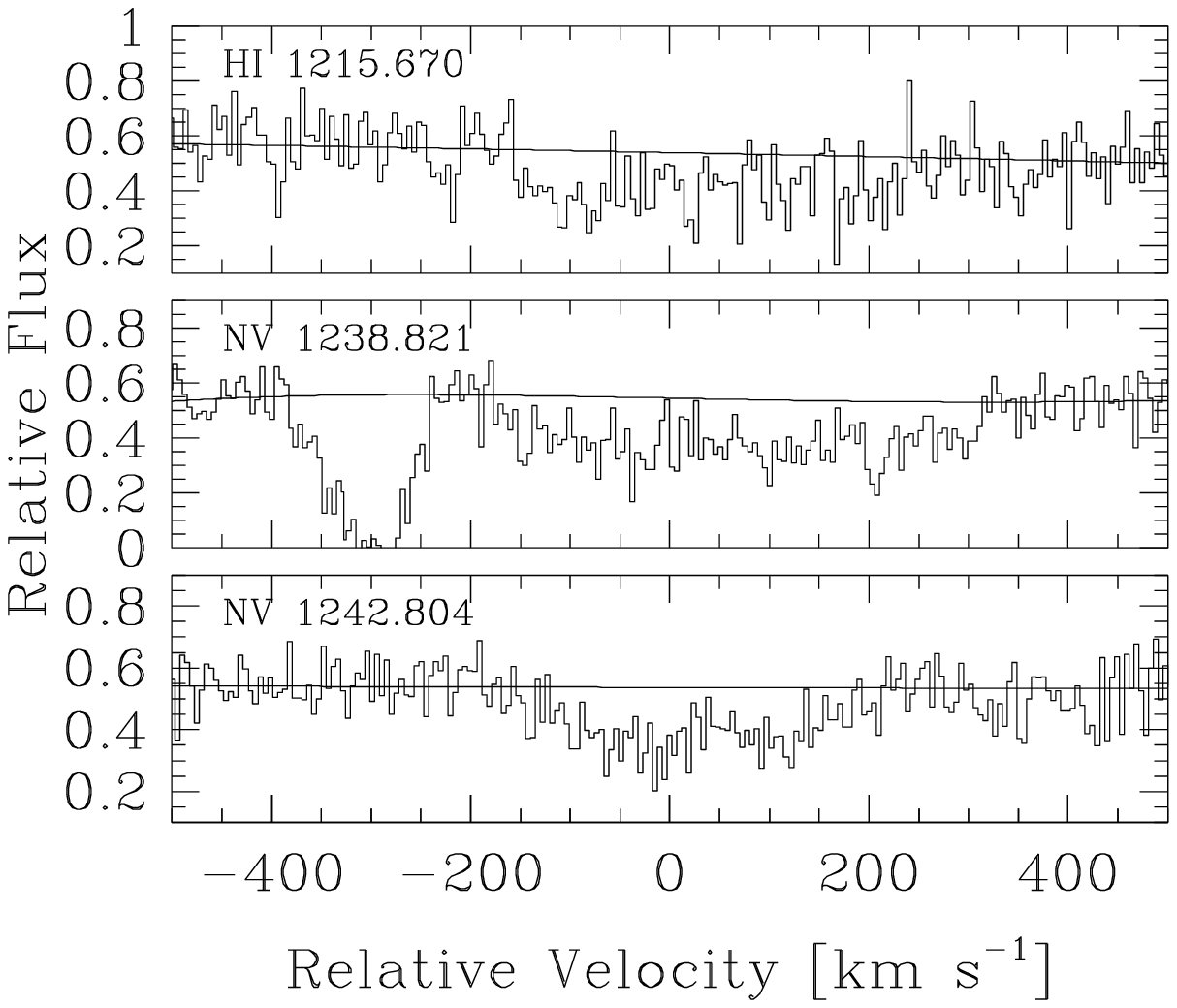}
\includegraphics[width=0.99\columnwidth]{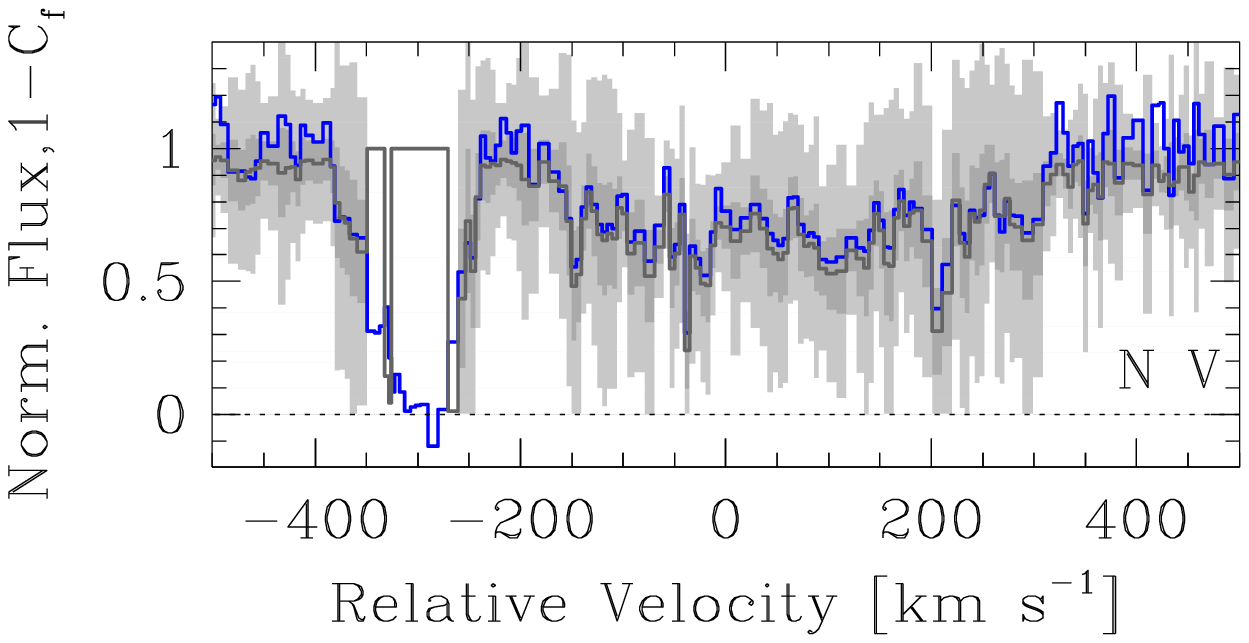}
\protect\caption[pg1206 system plot]{On the left, we present velocity-aligned absorption profiles of detected transitions for the absorption line system at $\zabs$=1.0280 observed toward PG\,1206+459.(The quoted redshift is the velocity zero-point; wavelength increases to the right.) On the right, we show the results of the partial coverage test on the \nv\ doublet. The normalized flux profile of the stronger transition is shown as a black histogram (blue histogram in color version). To improve signal-to-noise, the \nv\ profile is shown with a sampling of one bin per resolution element. The nominal velocity-dependent coverage fraction is shown as black points (black histogram in color version) with the grey shadings indicating the $1\sigma$\ (dark grey) and $3\sigma$\ (lighter grey) confidence limits.} \label{fig:pg1206}
\end{figure*}
\begin{figure*}
\includegraphics[width=0.99\columnwidth]{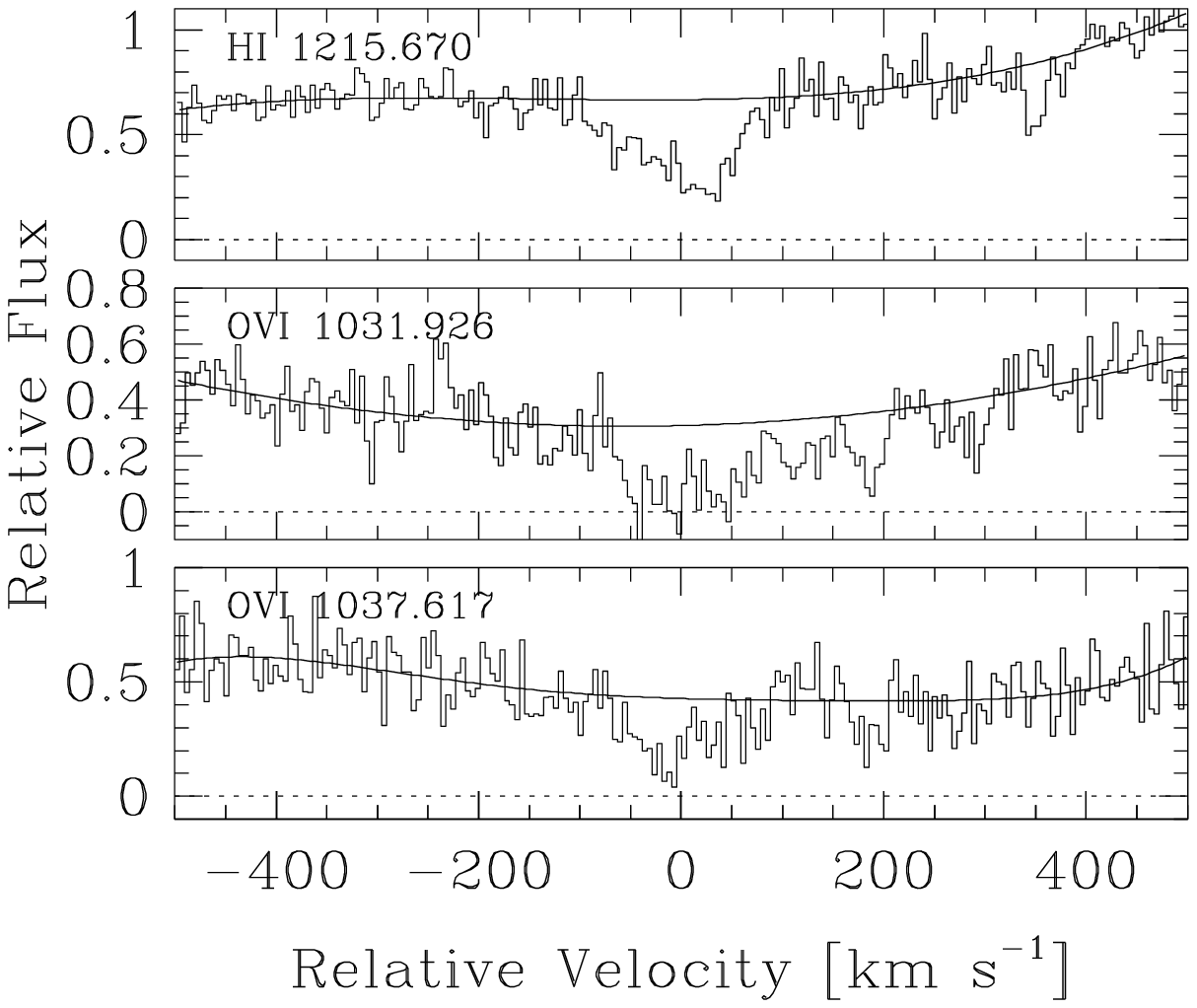}
\includegraphics[width=0.99\columnwidth]{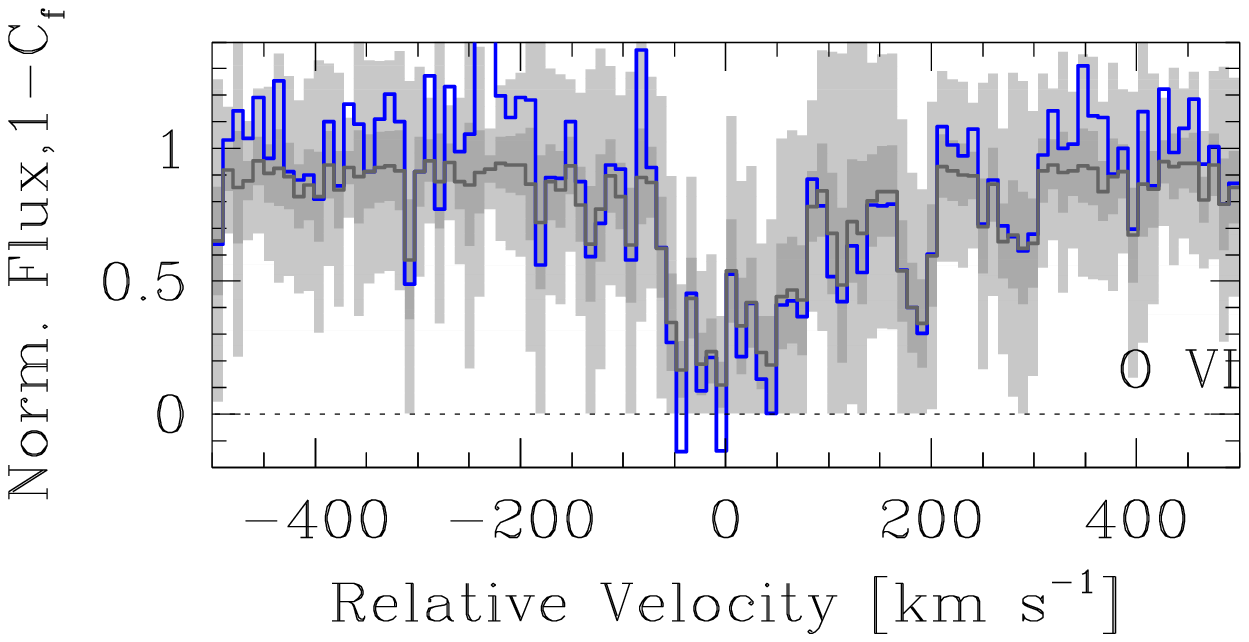}
\protect\caption[PG\,1241 system plot]{On the left, we present velocity-aligned absorption profiles of detected transitions for the associated absorption line system at $\zabs=1.2127$\ observed toward PG\,1241+176. (The quoted redshift is the velocity zero-point.) On the right, we show the results of the partial coverage test on the \ovi\ doublet. The normalized flux profile of the stronger transition is shown as a black histogram (blue histogram in color version). To improve signal-to-noise, the \ovi\ profile is shown with a sampling of one bin per resolution element. The nominal velocity-dependent coverage fraction is shown as black points (black histogram in color version) with the grey shadings indicating the $1\sigma$\ (dark grey) and $3\sigma$\ (lighter grey) confidence limits.} \label{fig:pg1241}
\end{figure*}
\begin{figure*}
\includegraphics[width=0.99\columnwidth]{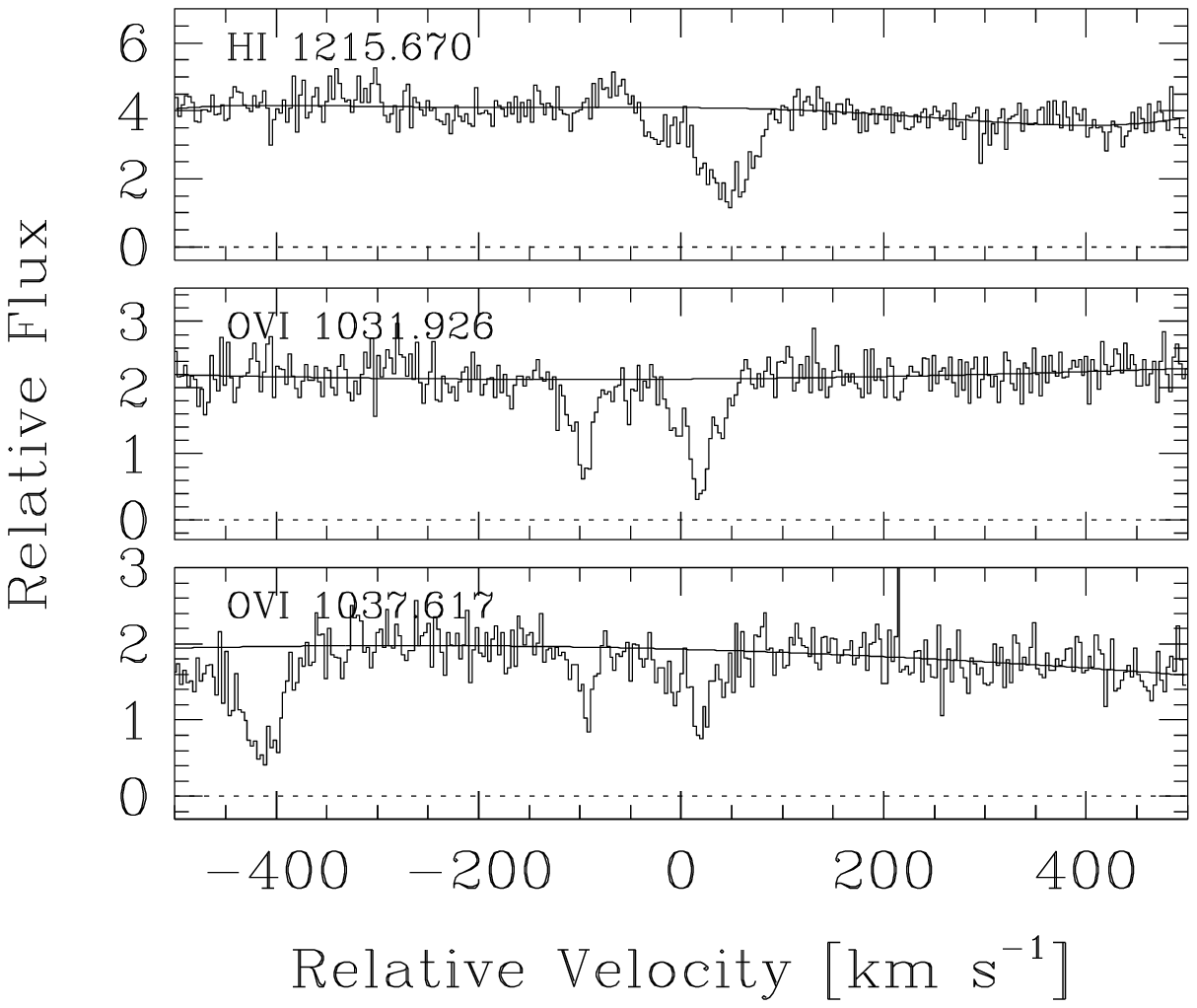}
\includegraphics[width=0.99\columnwidth]{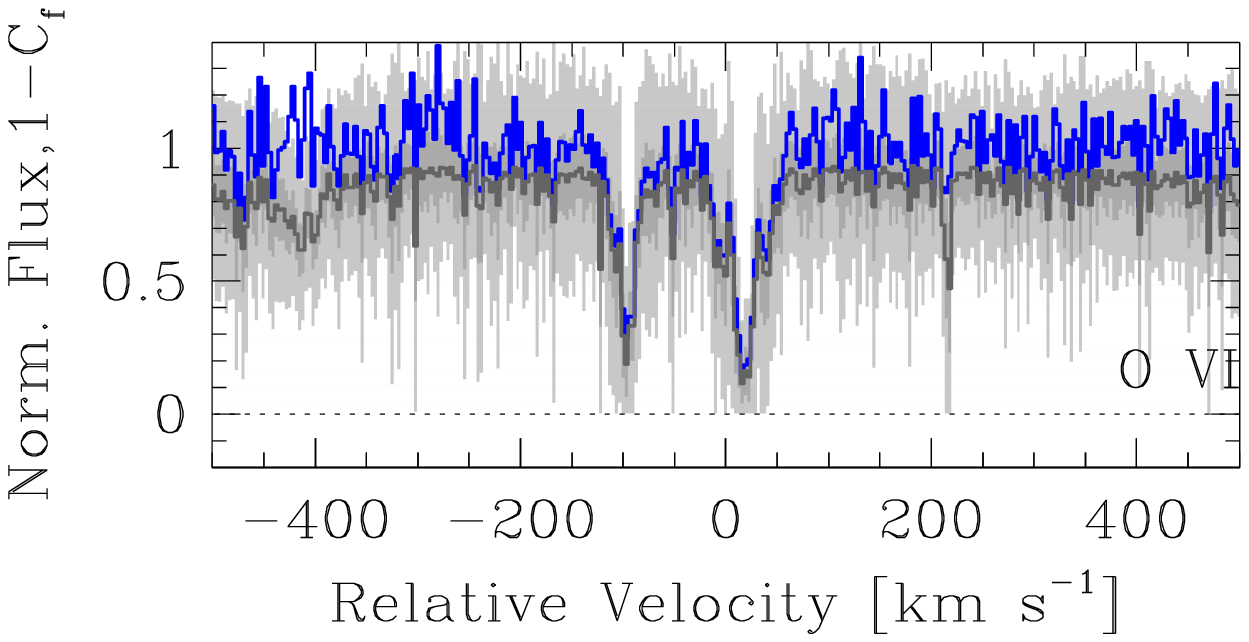}
\protect\caption[PG\,1444 system plot]{On the left, we present velocity-aligned absorption profiles of detected transitions for the associated absorption line system at $\zabs=0.2673$\ observed toward PG\,1444+407. (The quoted redshift is the velocity zero-point.) On the right, we show the results of the partial coverage test on the \ovi\ doublet. The normalized flux profile of the stronger transition is shown as a black histogram (blue in color version). The nominal velocity-dependent coverage fraction is shown as black points (black histogram in color version) with the grey shadings indicating the $1\sigma$\ (dark grey) and $3\sigma$\ (lighter grey) confidence limits.} \label{fig:pg1444}
\end{figure*}
\begin{figure*}
\includegraphics[width=0.99\columnwidth]{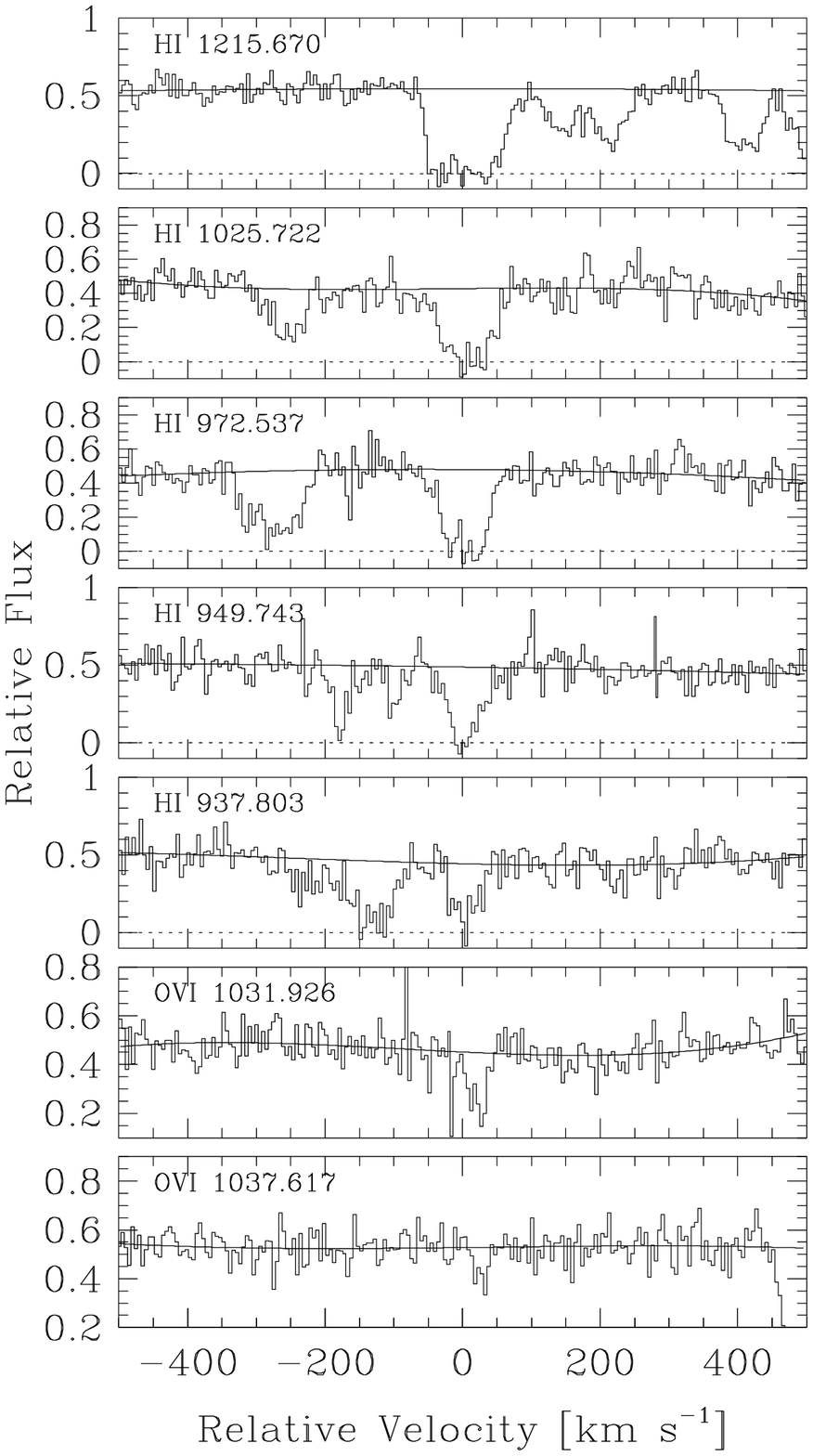}
\includegraphics[width=0.99\columnwidth]{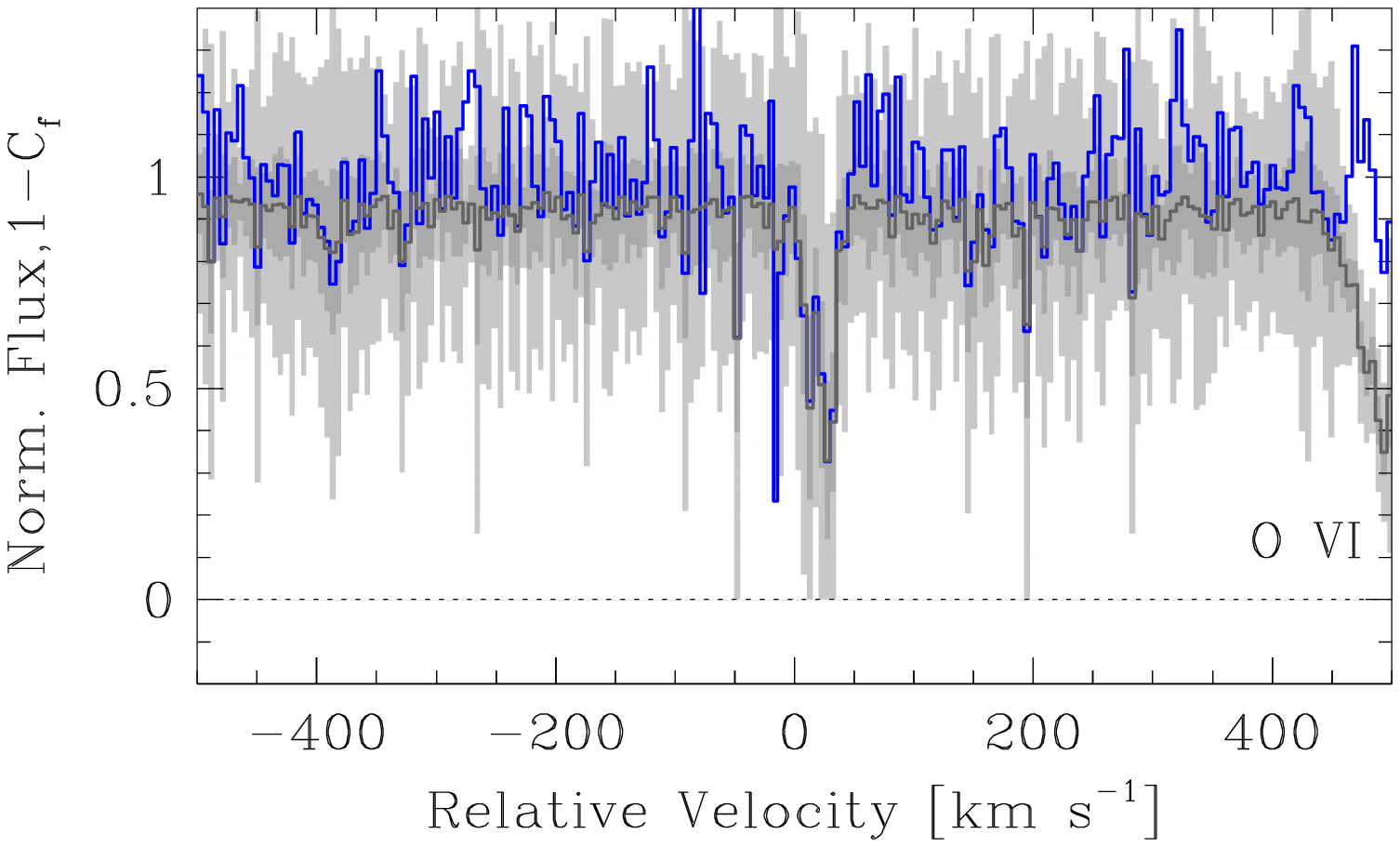}
\protect\caption[PG\,1630 system plot]{On the left, we present velocity-aligned absorption profiles of detected transitions for the associated absorption line system at $\zabs=1.4333$\ observed toward PG\,1630+377. (The quoted redshift is the velocity zero-point.) On the right, we show the results of the partial coverage test on the \ovi\ doublet. The normalized flux profile of the stronger transition is shown as a black histogram (blue histogram in color version). The nominal velocity-dependent coverage fraction is shown as black points (black histogram in color version) with the grey shadings indicating the $1\sigma$\ (dark grey) and $3\sigma$\ (lighter grey) confidence limits.} \label{fig:pg1630}
\end{figure*}
\begin{figure*}
\includegraphics[width=0.99\columnwidth]{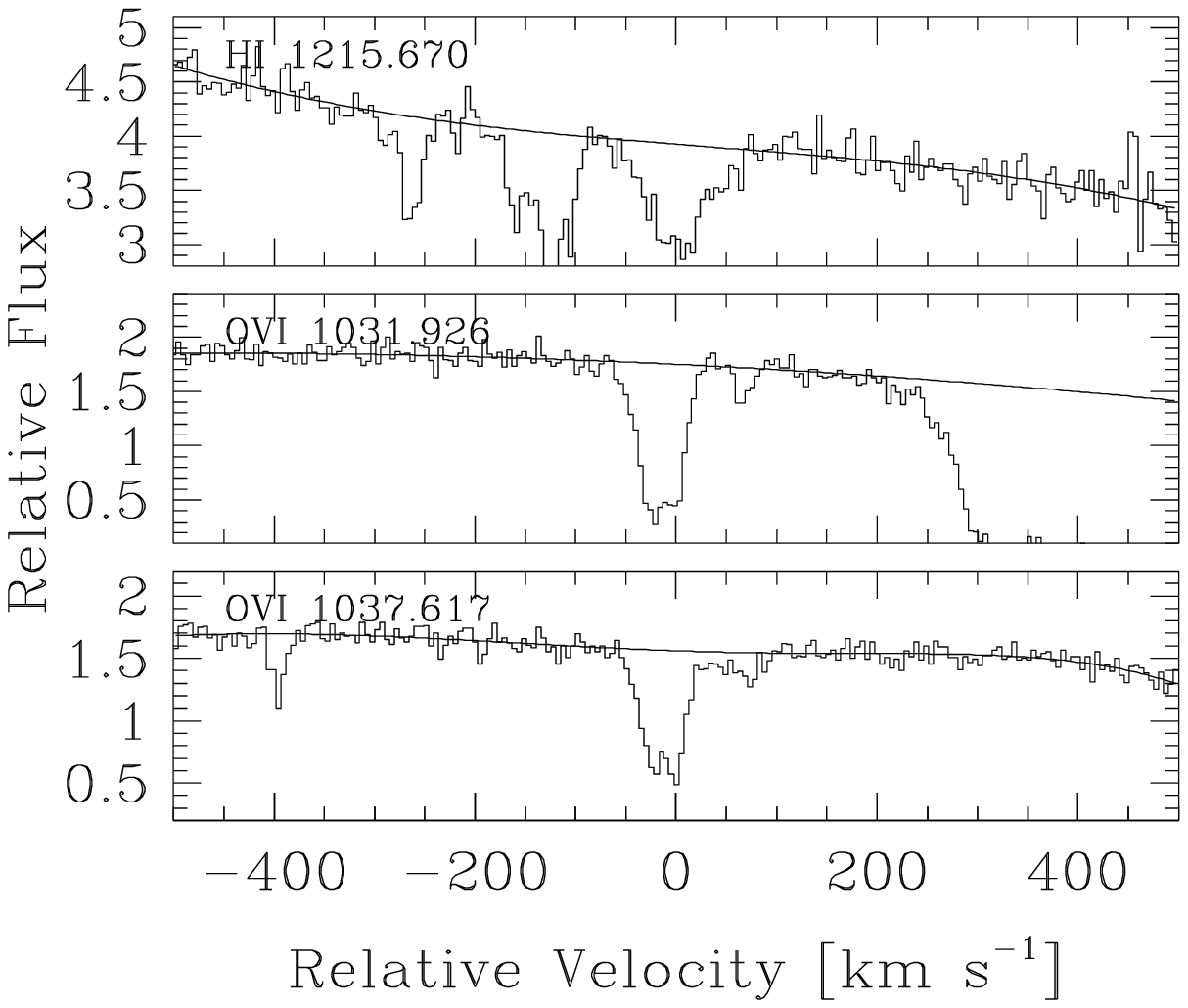}
\includegraphics[width=0.99\columnwidth]{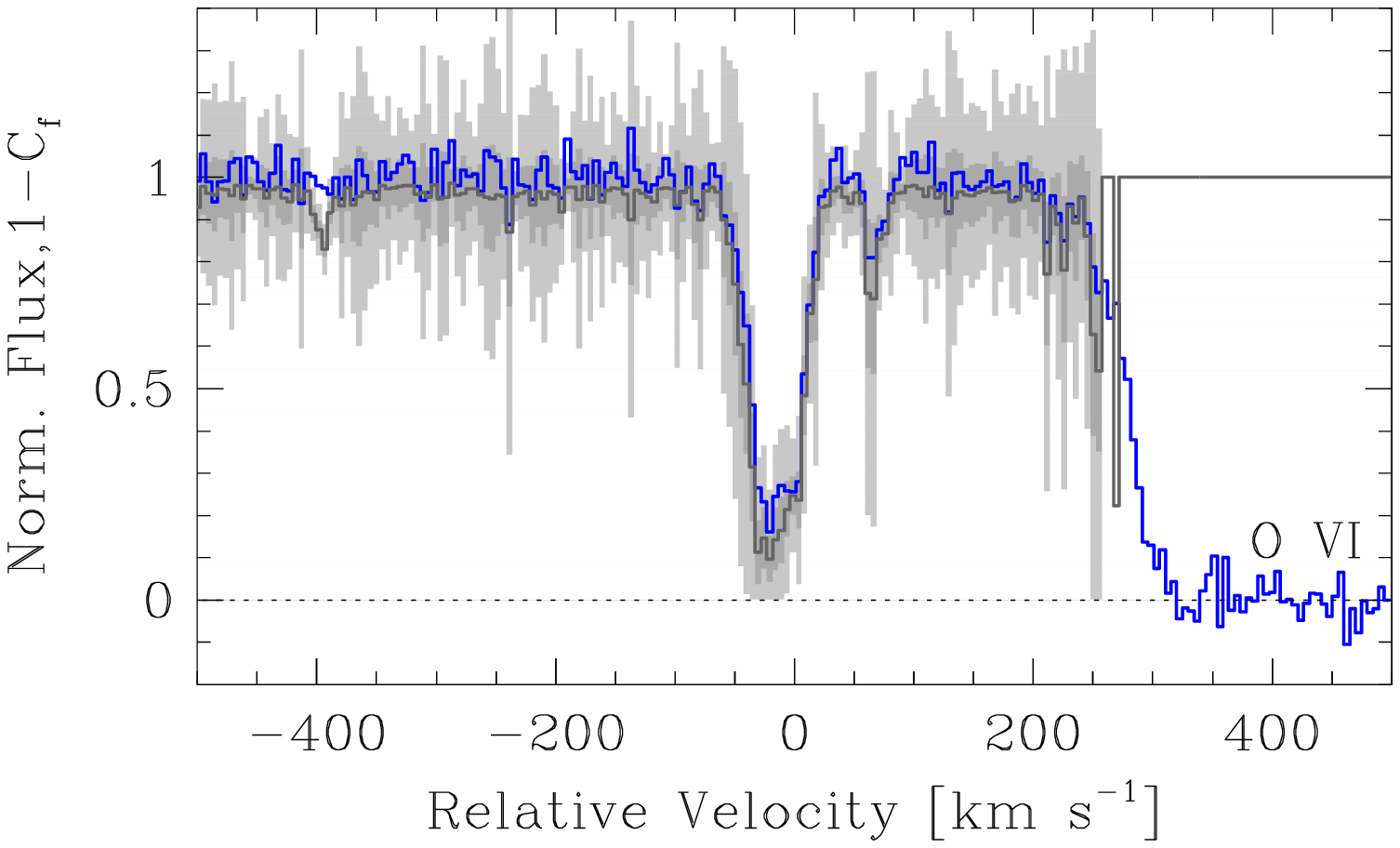}
\protect\caption[PG\,1634 system plot]{On the left, we present velocity-aligned absorption profiles of detected transitions for the associated absorption line system at $\zabs=1.3415$\ observed toward PG\,1634+706. (The quoted redshift is the velocity zero-point.) On the right, we show the results of the partial coverage test on the \ovi\ doublet. The normalized flux profile of the stronger transition is shown as a black histogram (blue histogram in color version). The nominal velocity-dependent coverage fraction is shown as black points (black histogram in color version) with the grey shadings indicating the $1\sigma$\ (dark grey) and $3\sigma$\ (lighter grey) confidence limits.} \label{fig:pg1634}
\end{figure*}
\clearpage
\begin{figure*}
\includegraphics[width=0.99\columnwidth]{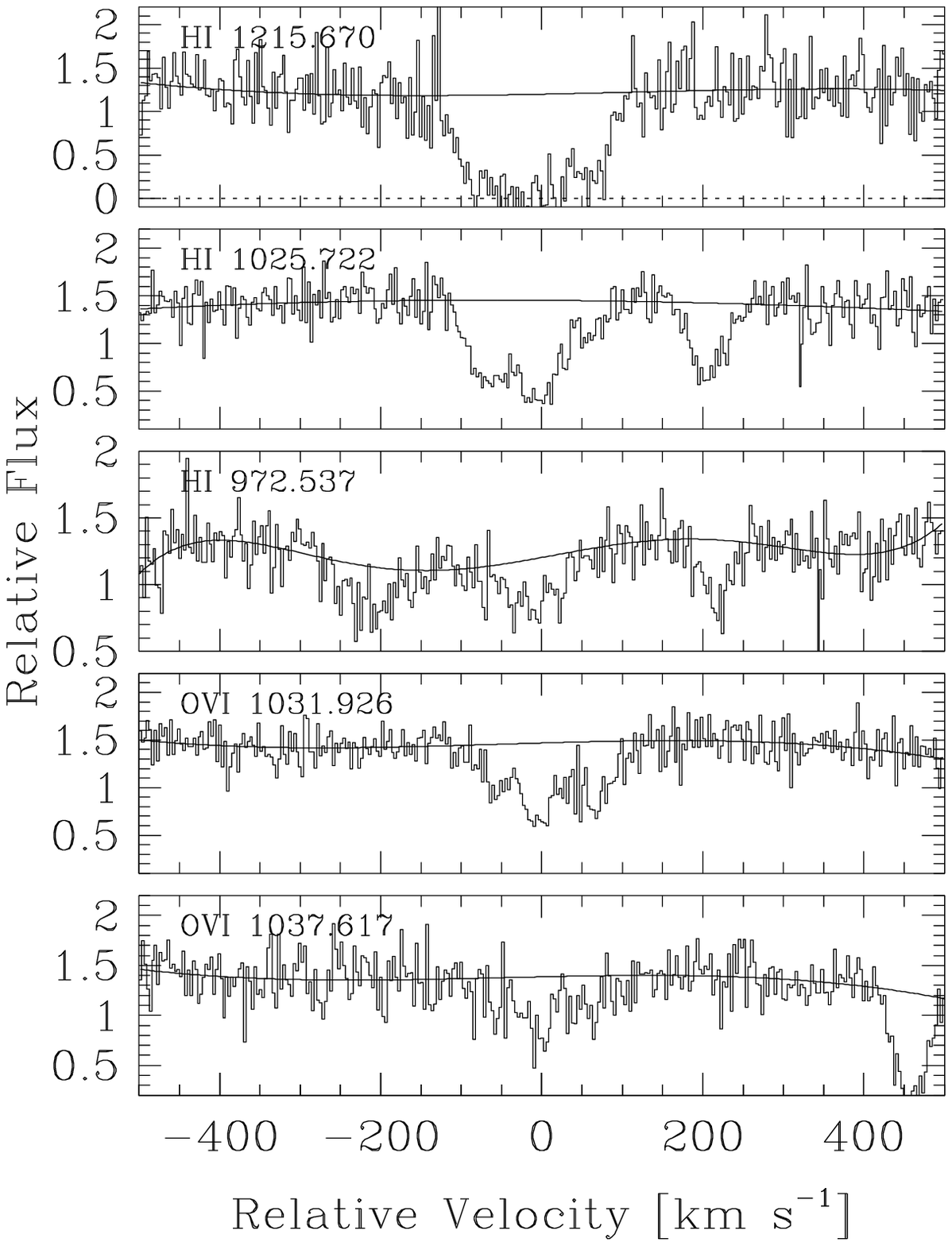}
\includegraphics[width=0.99\columnwidth]{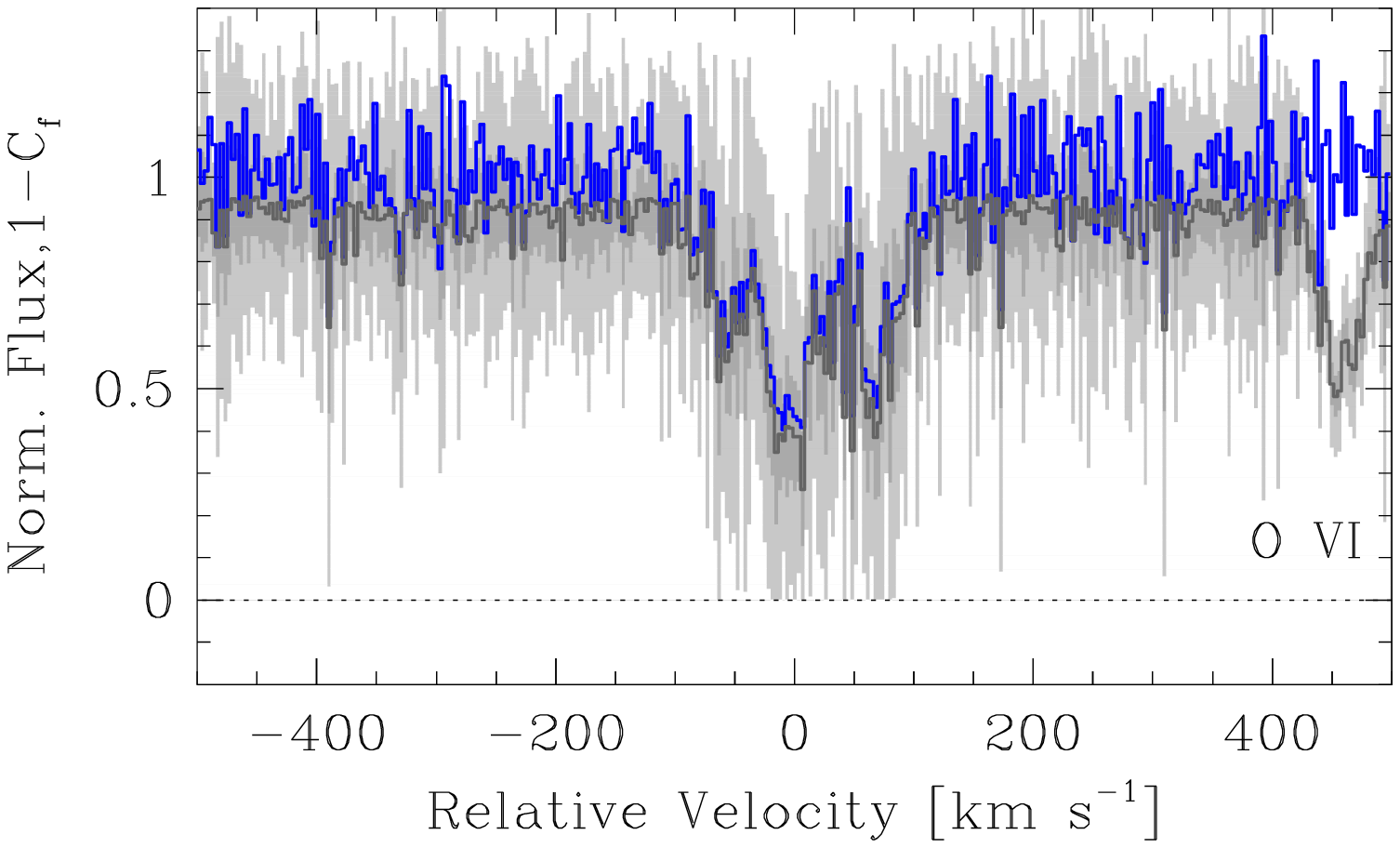}
\protect\caption[3C\,351 system plot]{On the left, we present velocity-aligned absorption profiles of detected transitions for the associated absorption line system at $\zabs=0.3166$\ observed toward 3C\,351. (The quoted redshift is the velocity zero-point.) On the right, we show the results of the partial coverage test on the \ovi\ doublet. The normalized flux profile of the stronger transition is shown as a black histogram (blue histogram in color version). The nominal velocity-dependent coverage fraction is shown as black points (black histogram in color version) with the grey shadings indicating the $1\sigma$\ (dark grey) and $3\sigma$\ (lighter grey) confidence limits.} \label{fig:3c351}
\end{figure*}
\begin{figure*}
\includegraphics[width=0.99\columnwidth]{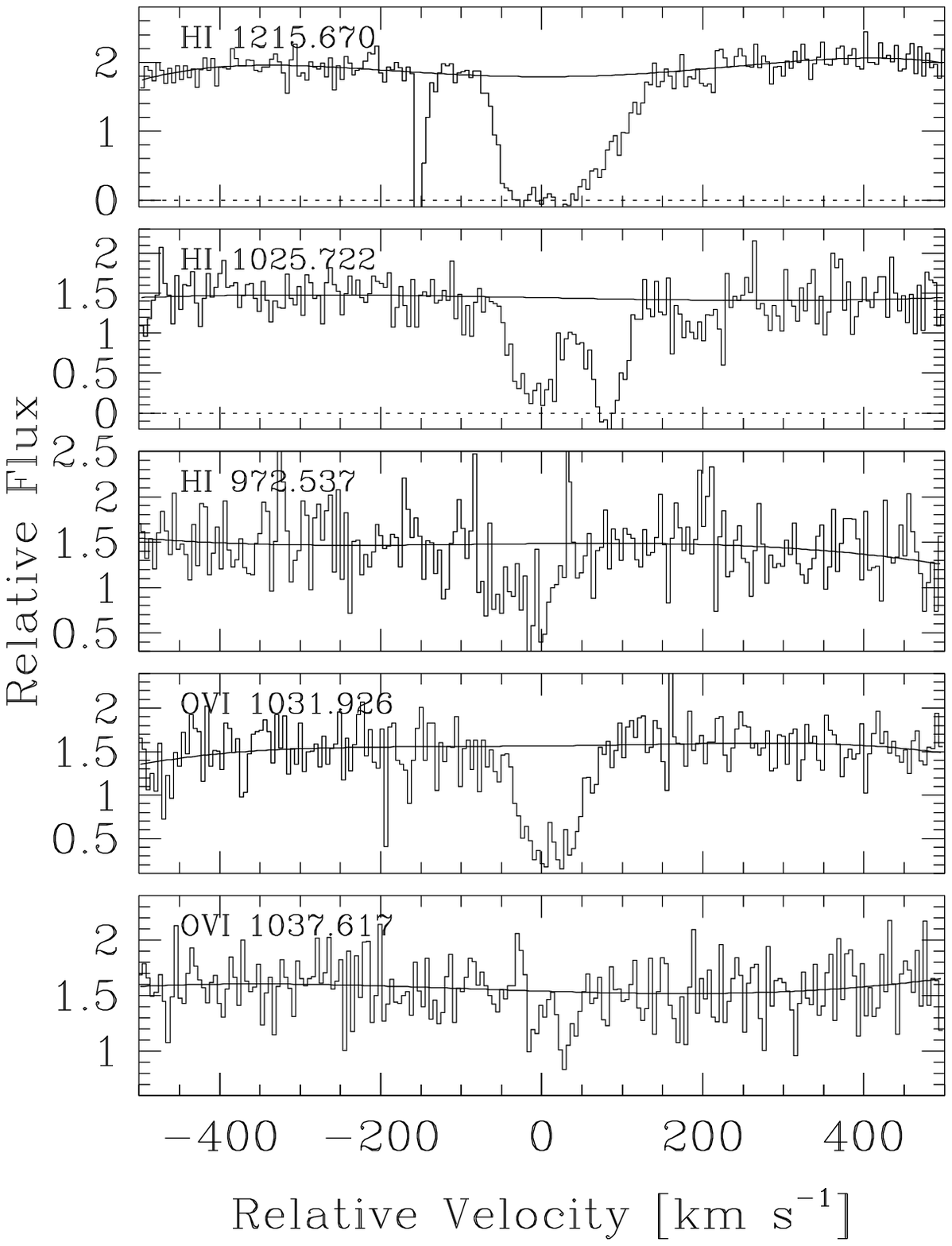}
\includegraphics[width=0.99\columnwidth]{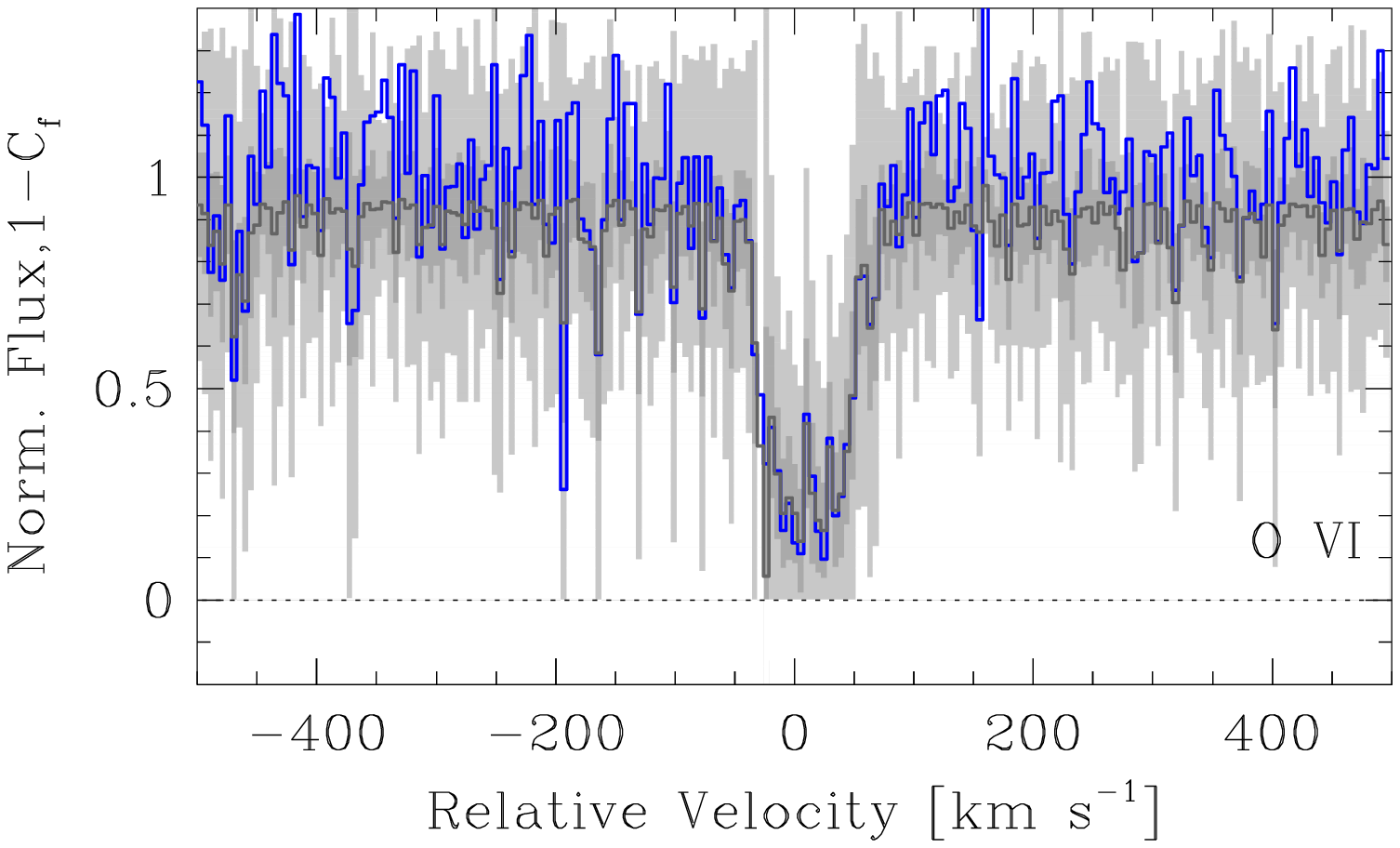}
\protect\caption[PG\,1718 system plot]{On the left, we present velocity-aligned absorption profiles of detected transitions for the associated absorption line system at $\zabs=1.0548$\ observed toward PG\,1718+481. (The quoted redshift is the velocity zero-point.) On the right, we show the results of the partial coverage test on the \ovi\ doublet. The normalized flux profile of the stronger transition is shown as a black histogram (blue histogram in color version). The nominal velocity-dependent coverage fraction is shown as black points (black histogram in color version) with the grey shadings indicating the $1\sigma$\ (dark grey) and $3\sigma$\ (lighter grey) confidence limits.} \label{fig:pg1718a}
\end{figure*}

\begin{figure*}
\includegraphics[width=0.99\columnwidth]{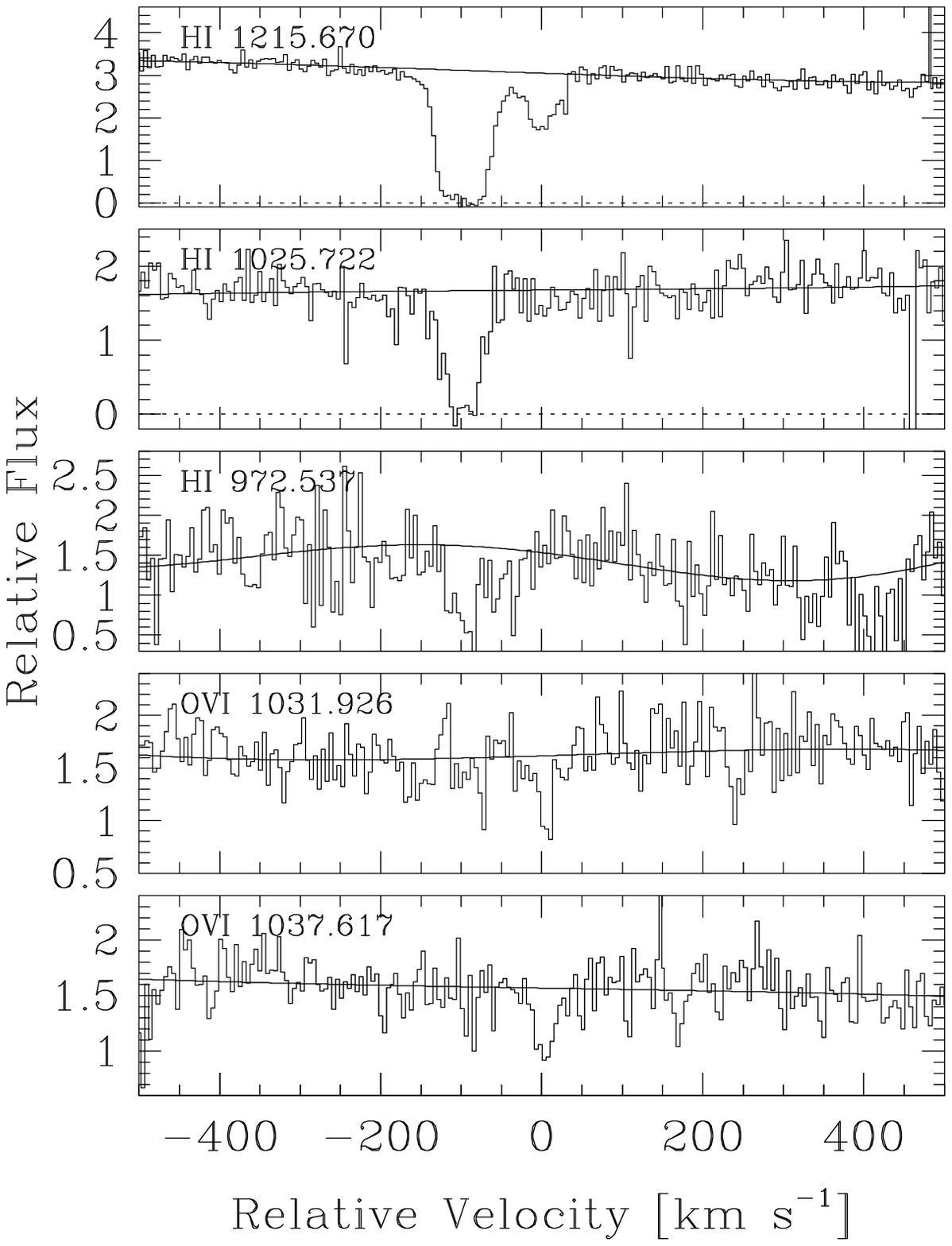}
\includegraphics[width=0.99\columnwidth]{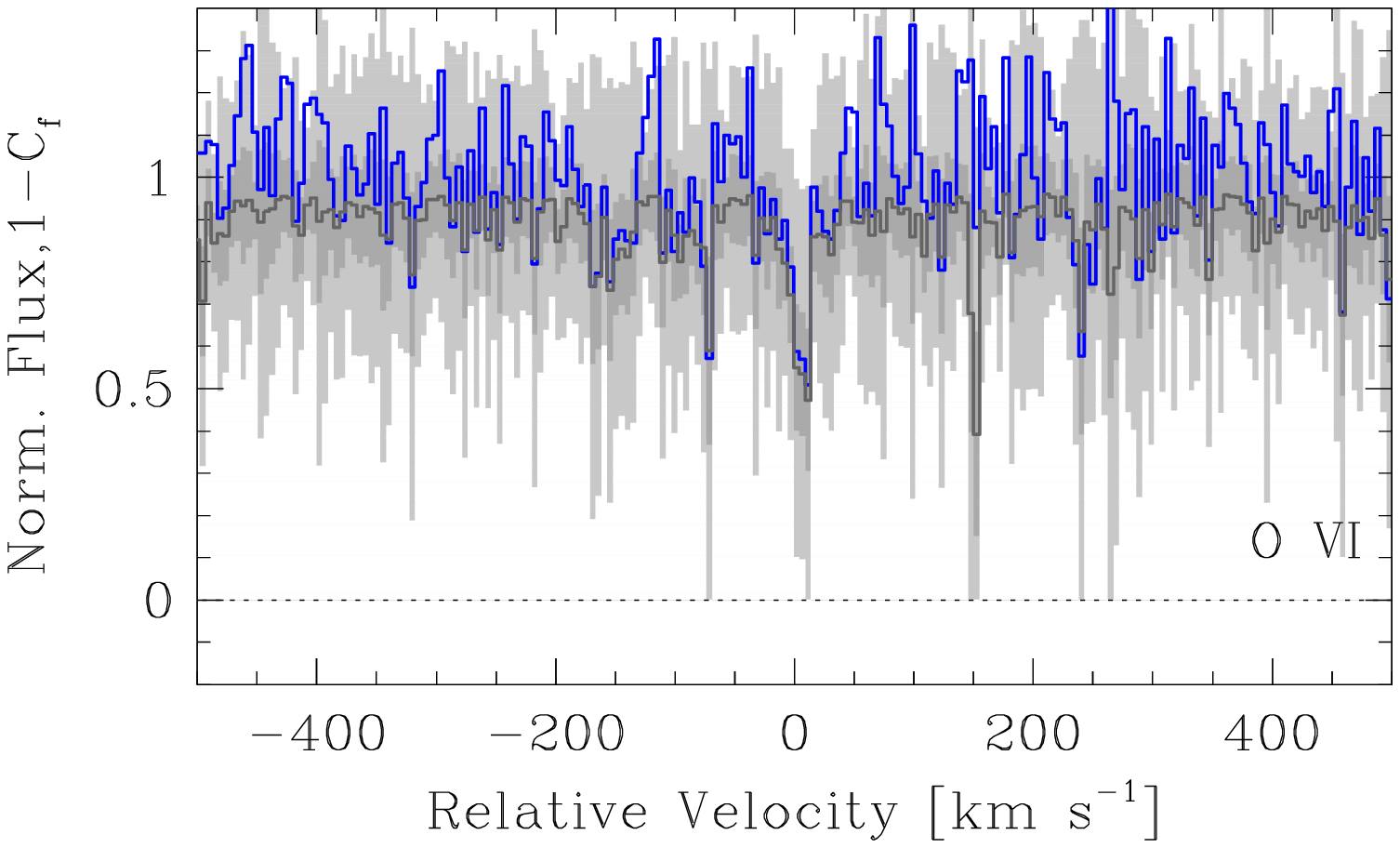}
\protect\caption[PG\,1718 system plot]{On the left, we present velocity-aligned absorption profiles of detected transitions for the associated absorption line system at $\zabs=1.0874$\ observed toward PG\,1718+481. (The quoted redshift is the velocity zero-point.) On the right, we show the results of the partial coverage test on the \ovi\ doublet. The normalized flux profile of the stronger transition is shown as a black histogram (blue histogram in color version). The nominal velocity-dependent coverage fraction is shown as black points (black histogram in color version) with the grey shadings indicating the $1\sigma$\ (dark grey) and $3\sigma$\ (lighter grey) confidence limits.} \label{fig:pg1718a}
\end{figure*}
\begin{figure*}
\includegraphics[width=0.99\columnwidth]{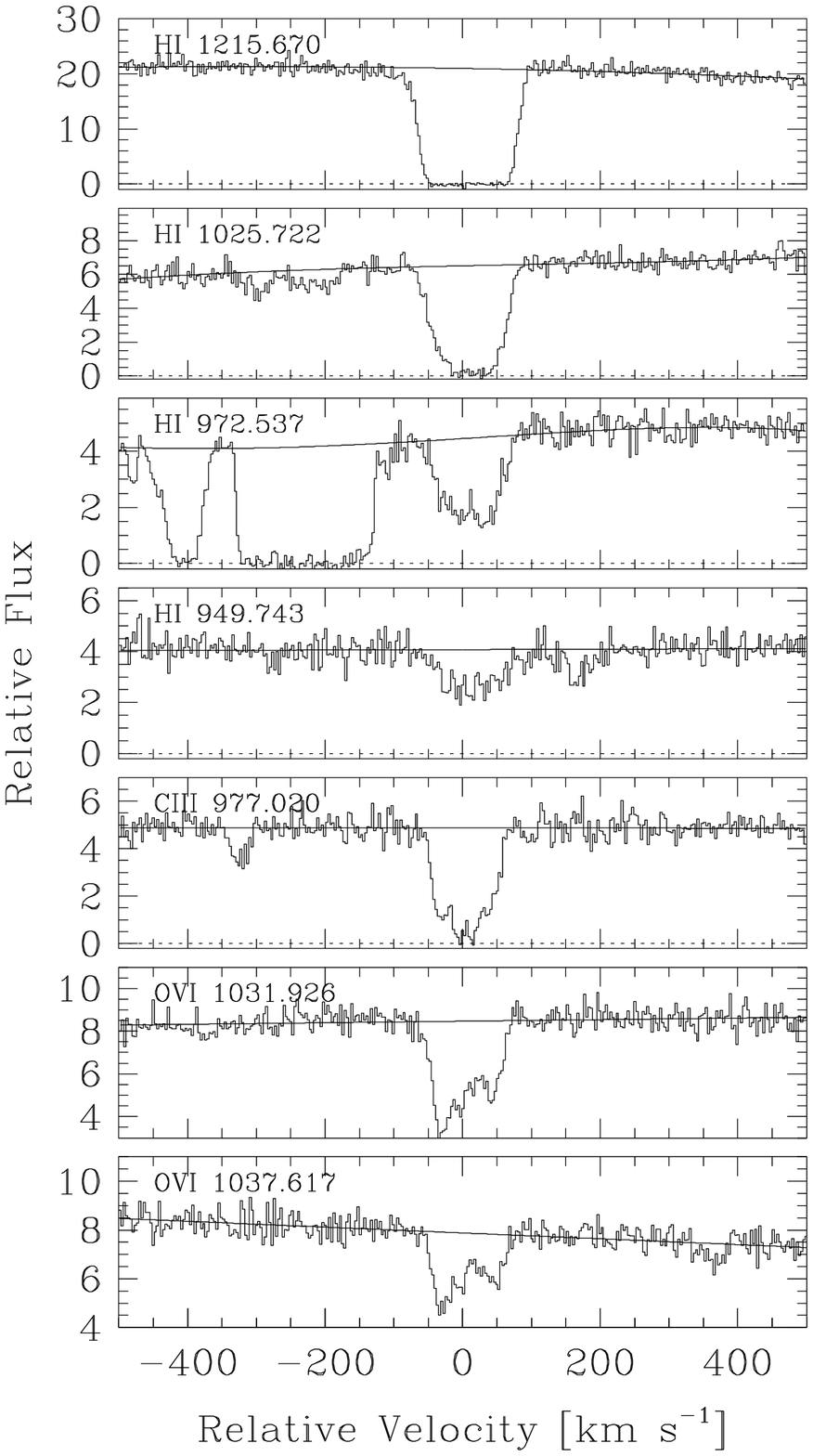}
\includegraphics[width=0.99\columnwidth]{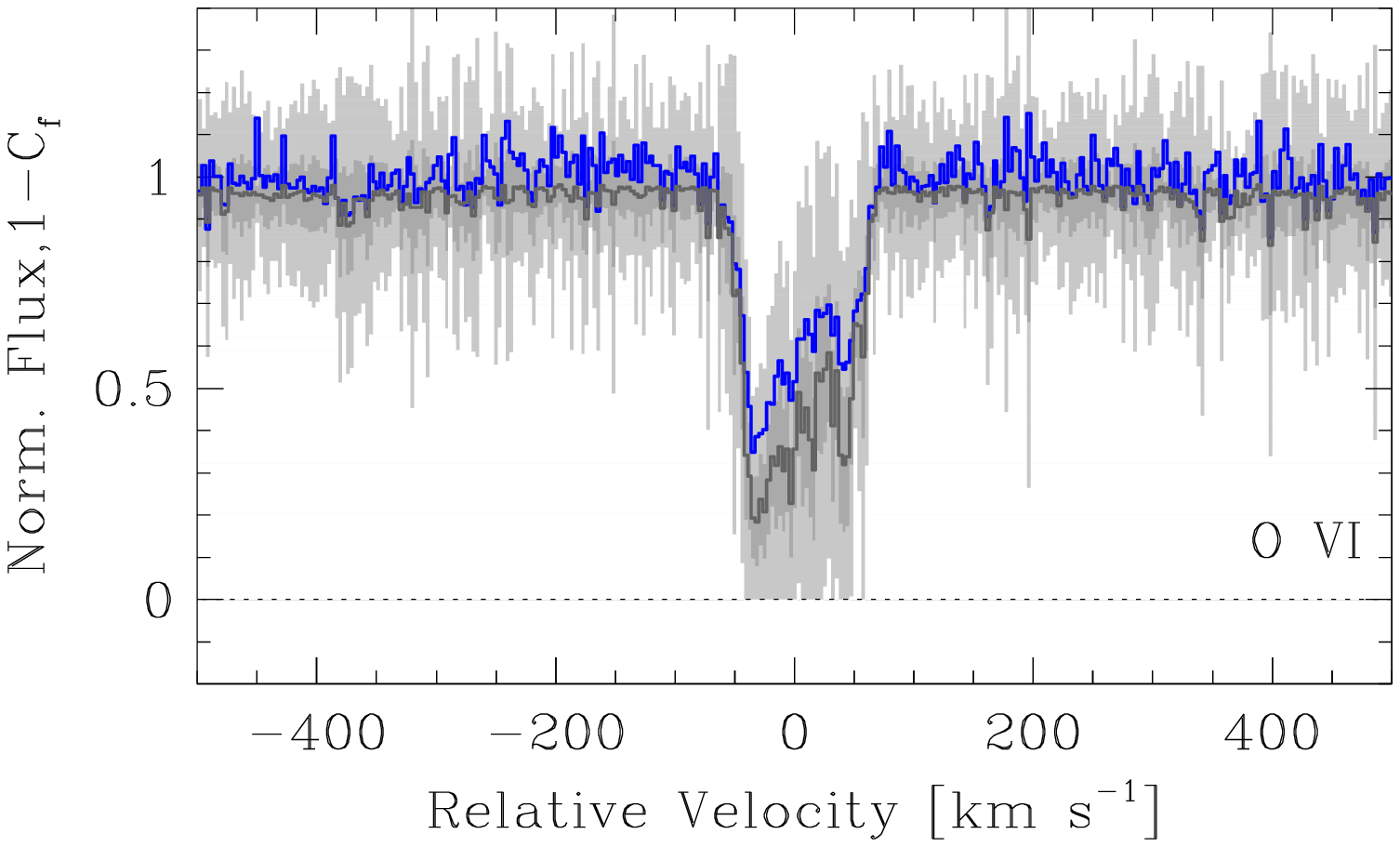}
\protect\caption[H\,1821 system plot]{On the left, we present velocity-aligned absorption profiles of detected transitions for the associated absorption line system at $\zabs=0.2967$\ observed toward H\,1821+643. (The quoted redshift is the velocity zero-point.) On the right, we show the results of the partial coverage test on the \ovi\ doublet. The normalized flux profile of the stronger transition is shown as a black histogram (blue histogram in color version). The nominal velocity-dependent coverage fraction is shown as black points (black histogram in color version) with the grey shadings indicating the $1\sigma$\ (dark grey) and $3\sigma$\ (lighter grey) confidence limits.} \label{fig:h1821}
\end{figure*}
\begin{figure*}
\includegraphics[width=0.99\columnwidth]{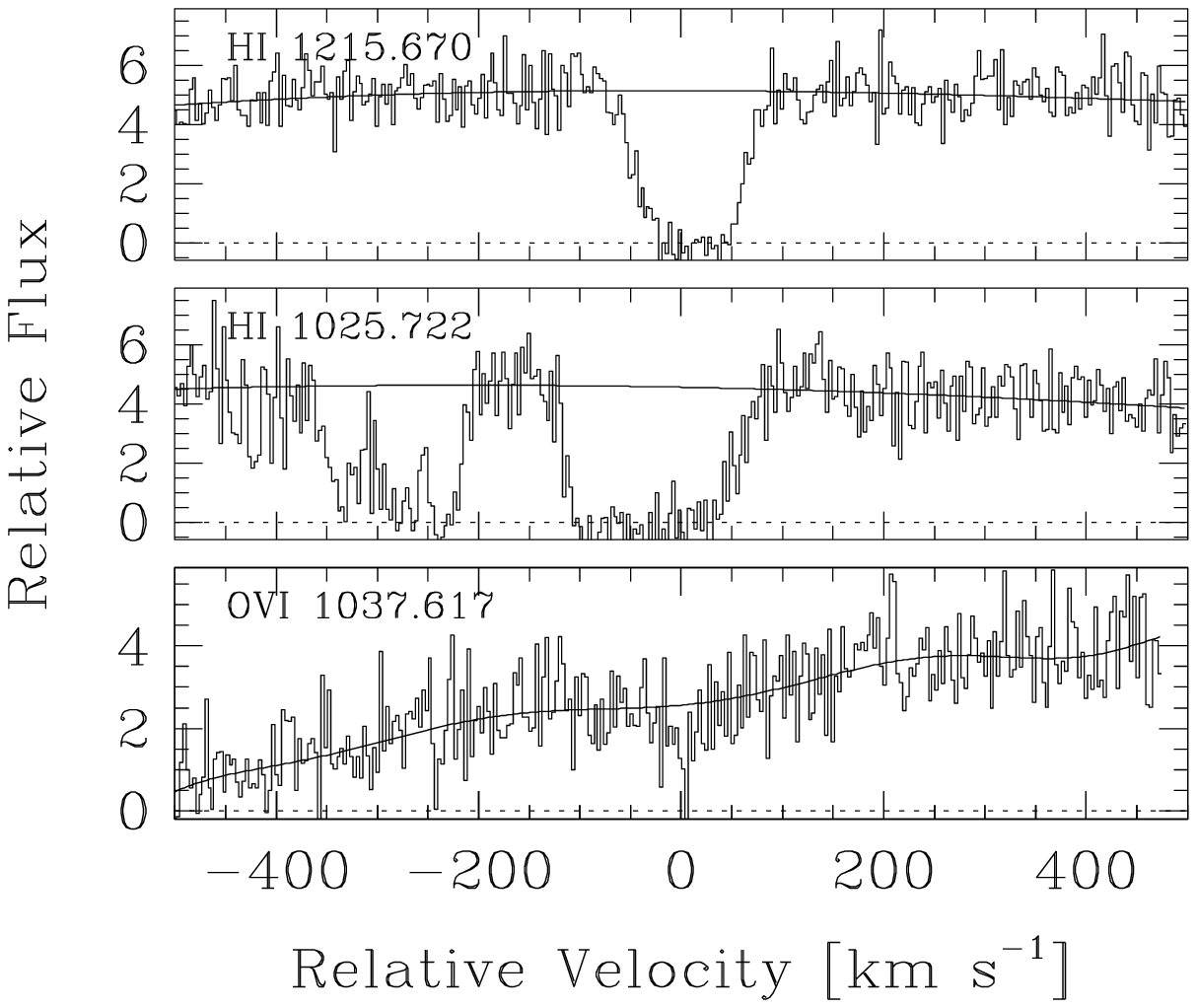}
\protect\caption[PHL\,1811 system plot]{We present velocity-aligned absorption profiles of detected transitions for the associated absorption line system at $\zabs=0.1765$\ observed toward PHL\,1811. (The quoted redshift is the velocity zero-point.)} \label{fig:phl1811a}
\end{figure*}

\begin{figure*}
\includegraphics[width=0.99\columnwidth]{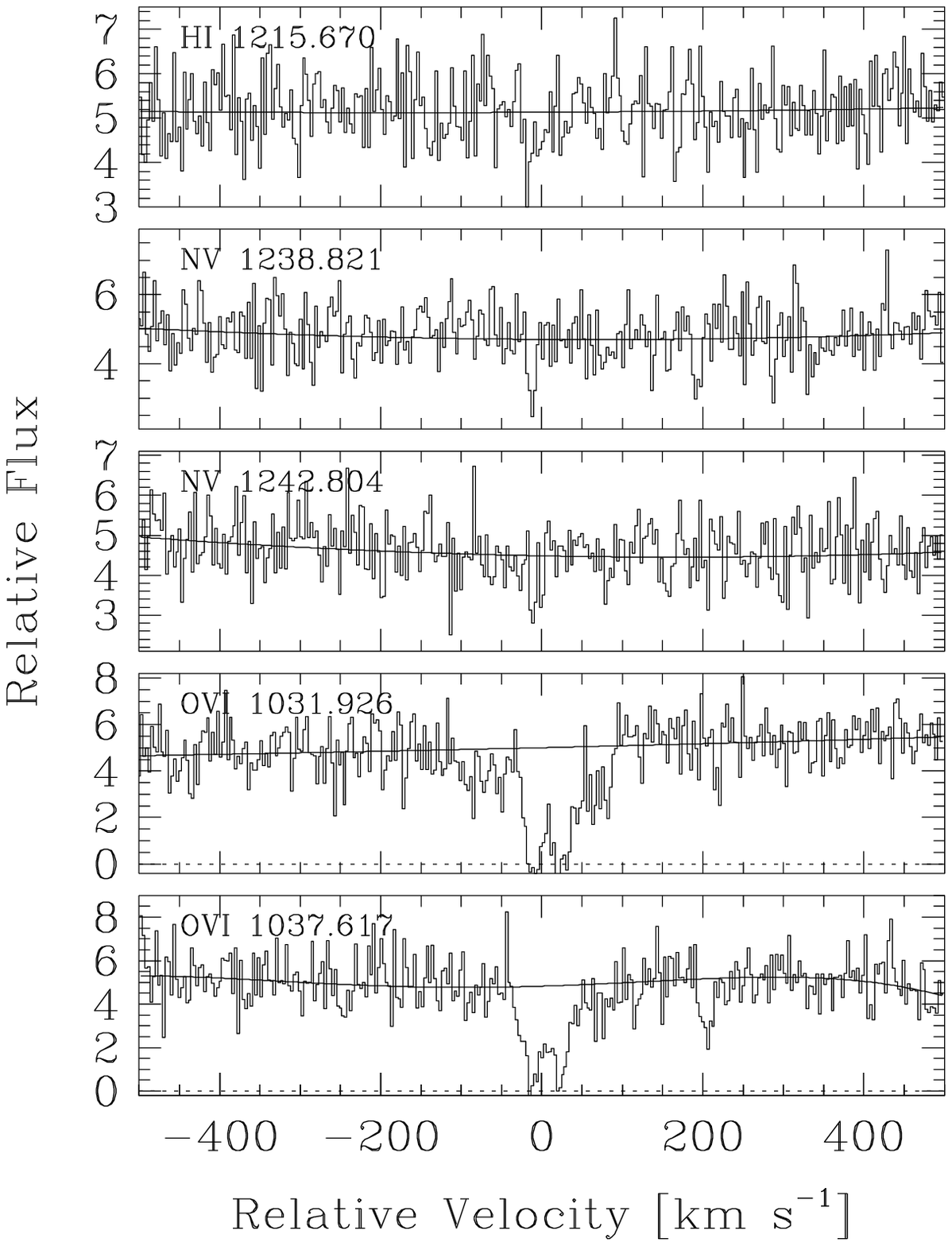}
\includegraphics[width=0.99\columnwidth]{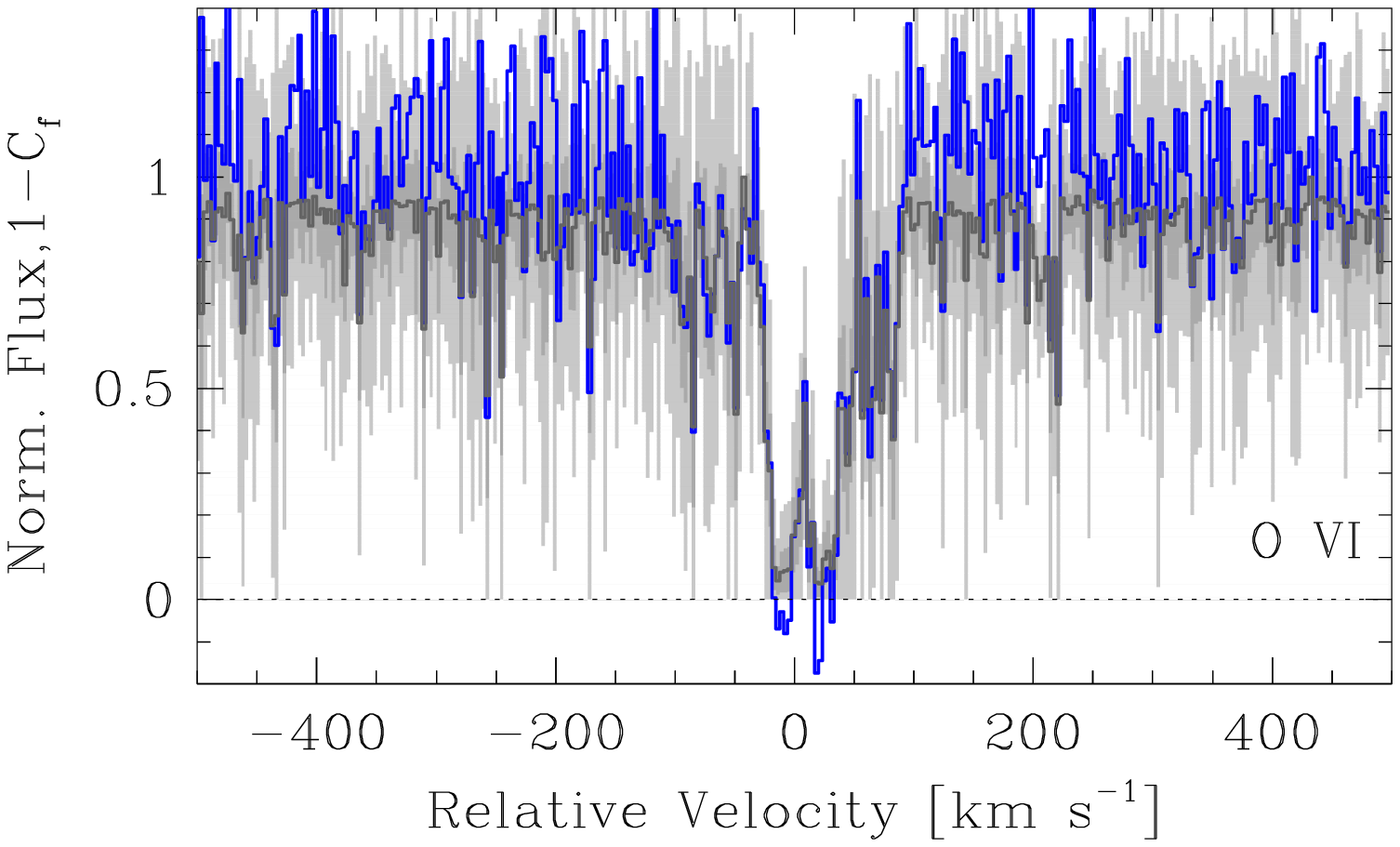}
\protect\caption[PHL\,1811 system plot]{On the left, we present velocity-aligned absorption profiles of detected transitions for the associated absorption line system at $\zabs=0.1919$\ observed toward PHL\,1811. (The quoted redshift is the velocity zero-point.) On the right, we show the results of the partial coverage test on the \ovi\ doublet. The normalized flux profile of the stronger transition is shown as a black histogram (blue histogram in color version). The nominal velocity-dependent coverage fraction is shown as black points (black histogram in color version) with the grey shadings indicating the $1\sigma$\ (dark grey) and $3\sigma$\ (lighter grey) confidence limits.} \label{fig:phl1811b}
\end{figure*}

\end{document}